\documentclass[aps,twocolumn,superscriptaddress,citeautoscript,floatfix,reprint]{revtex4-1}

\pdfoutput=1
\usepackage{array}
\usepackage{multirow}
\usepackage{microtype}
\usepackage[english]{babel}
\usepackage[utf8]{inputenc}
\usepackage{amsmath}
\usepackage{graphicx}
\usepackage[dvipsnames]{xcolor}
\usepackage{amssymb}
\usepackage{subfigure}
\usepackage{anysize}
\usepackage{physics}
\usepackage{makeidx}
\usepackage{float}
\usepackage{xcolor,colortbl}
\usepackage{listings}
\usepackage{enumitem}
\usepackage[group-separator={,}]{siunitx}

\graphicspath{{./figures/}}

\definecolor{darkblue}{rgb}{0.1,0.2,0.6}
\definecolor{darkred}{rgb}{0.8,0.1,0.2}
\definecolor{Gray}{gray}{0.9}
\usepackage[colorlinks,citecolor=darkblue,linkcolor=darkred,urlcolor=darkblue]{hyperref}	
\usepackage[all]{hypcap} 
\usepackage[top=2.5cm,bottom=2.5cm,left=2cm,right=2cm]{geometry}

\providecommand{\newoperator}[2]{\newcommand*{#1}{\mathop{\mathrm{#2}}\nolimits}}
\newoperator{\sgn}{sgn}
\newoperator{\arctanh}{arctanh}
\newoperator{\argmax}{argmax}
\newoperator{\diag}{diag}

\def\bra#1{\langle#1|}
\def\ket#1{|#1\rangle}
\def\braket#1#2{\langle#1|#2\rangle}
\def\rsim#1{\mbox{[Run#1]}}

\makeatletter
\def\@gobblesubsection#1\csname thesubsection\endcsname{\Alph{subsection}}
\def\@gobblesubsubsection#1\csname thesubsubsection\endcsname{\Alph{subsection}.\arabic{subsubsection}}
\renewcommand{\p@subsection}{\@gobblesubsection}
\renewcommand{\p@subsubsection}{\@gobblesubsubsection}
\makeatother

\def\Bristlecone#1{\mbox{Bristlecone-#1}}

\begin{document}

\title{A flexible high-performance simulator for verifying and benchmarking quantum circuits implemented on real hardware}

\def\urbana{
	Institute for Condensed Matter Theory and Department of Physics, 
	University of Illinois at Urbana-Champaign, Urbana, IL 61801, USA
}

\def\quail{
	Quantum Artificial Intelligence Lab. (QuAIL), 
    NASA Ames Research Center, Moffett Field, CA 94035, USA
}

\def\usra{
	USRA Research Institute for Advanced Computer Science (RIACS),
	615 National, Mountain View, California 94043, USA
}

\def\google{
	Google Inc., Venice, CA 90291, USA
}

\def\asrc{
	ASRC Federal InuTeq, 
    7000 Muirkirk Meadows Drive, Suite 100, Beltsville, MD 20705
}

\def\sgt{
	Stinger Ghaffarian Technologies Inc., 
    7701 Greenbelt Rd., Suite 400, Greenbelt, MD 20770
}

\author{Benjamin Villalonga}
	\affiliation{\urbana}
	\affiliation{\quail}
	\affiliation{\usra}
    
\author{Sergio Boixo}
	\affiliation{\google}
    
\author{Bron Nelson}
	\affiliation{\quail}
	\affiliation{\asrc}
    
\author{Christopher Henze}
	\affiliation{\quail}
    
\author{Eleanor Rieffel}
	\affiliation{\quail}
    
\author{Rupak Biswas}
	\affiliation{\quail}
    
\author{Salvatore Mandr\`a}
	\email{salvatore.mandra@nasa.gov}
	\affiliation{\quail}
	\affiliation{\sgt}

\date{\today}

\begin{abstract}
Here we present qFlex, a flexible tensor network based quantum circuit
simulator. qFlex can compute both exact amplitudes, essential for the
verification of the quantum hardware, as well as low fidelity amplitudes, in
order to mimic sampling from Noisy Intermediate-Scale Quantum (NISQ) devices. In
this work, we focus on random quantum circuits (RQCs) in the range of sizes
expected for supremacy experiments. Fidelity $f$ simulations are performed at a
cost that is $1/f$ lower than perfect fidelity ones. We also present a technique
to eliminate the overhead introduced by rejection sampling in most tensor network
approaches. We benchmark the simulation of square lattices and Google's
Bristlecone QPU. Our analysis is supported by extensive simulations on NASA HPC
clusters Pleiades and Electra.  For our most computationally demanding
simulation, the two clusters combined reached a peak of 20 PFLOPS (single
precision), \emph{i.e.}, 64\% of their maximum achievable performance, which
represents the largest numerical computation in terms of sustained FLOPs and
number of nodes utilized ever run on NASA HPC clusters. Finally, we introduce a
novel multithreaded, cache-efficient tensor index permutation algorithm of
general application.
\end{abstract}

\maketitle

\section{Introduction}
\label{sec:intro}

Building a universal, noise-resistant quantum computer is to date a long-term
goal driven by the strong evidence that such a machine will provide large
amounts of computational power, beyond classical capabilities
\cite{NCbook,RPbook, shor1994algorithms, grover1996fast, feynman1982simulating,
aspuru-guzik_simulated_2005, babbush_low_2017, jiang_quantum_2018,
babbush_encoding_2018}.  An imminent milestone in that direction is represented
by Noisy Intermediate-Scale Quantum (NISQ) devices \cite{preskill_quantum_2018}
of about 50-100 qubits.  Despite the lack of error correction mechanisms to run
arbitrarily deep quantum circuits, NISQ devices may be able to perform tasks
which already surpass the capabilities of today's classical digital computers
within reasonable time and energy constraints
\cite{bremner_average-case_2016,harrow_quantum_2017, boixo_characterizing_2018},
thereby achieving quantum supremacy \cite{aaronson2011computational,
bremner_average-case_2016, preskill_quantum_2012, bremner_achieving_2017,
boixo_characterizing_2018, harrow_quantum_2017, aaronson2017complexity,
neill_blueprint_2018, bouland_quantum_2018, harrow_approximate_2018,
movassagh_efficient_2018}.

Quantum circuit simulation plays a dual role in demonstrating quantum supremacy.
First, it establishes a classical computational bar that quantum computation
must pass to demonstrate supremacy. Indeed, formal complexity proofs related to
quantum supremacy are asymptotic, and therefore assume an arbitrarily large
number of qubits~\cite{aaronson2011computational, bremner_average-case_2016,
preskill_quantum_2012, bremner_achieving_2017, boixo_characterizing_2018,
harrow_quantum_2017, aaronson2017complexity, neill_blueprint_2018,
bouland_quantum_2018, harrow_approximate_2018, movassagh_efficient_2018}. This
is only possible with a fault tolerant quantum
computer~\cite{kalai2014gaussian,arkhipov2015bosonsampling,
rahimi2016sufficient,boixo_characterizing_2018,bremner_achieving_2017,yung2017can,boixo2017fourier,
gao2018efficient} (Note that the polynomial approximation algorithms in
Refs.~\cite{bremner_achieving_2017,yung2017can,gao2018efficient} never produce
a better approximation than trivially sampling bit-strings uniformly at random,
as shown in Ref.~\cite{boixo2017fourier}.), and therefore a near term practical
demonstration of quantum supremacy must rely on a careful comparison with
highly
optimized classical algorithms on state-of-the-art supercomputers.  Second, it
also provides verification that the quantum hardware is indeed performing as
expected up to the limits of classical computational capabilities. 

The leading near-term proposal for a quantum supremacy experiment on NISQ
devices is based on the sampling of bit-strings from a random quantum circuit
(RQC)~\cite{aaronson2017complexity, bouland_quantum_2018,
movassagh_efficient_2018, boixo_characterizing_2018}. Indeed, under reasonable
assumptions, sampling from large RQCs is classically unfeasible
\cite{aaronson2011computational, bremner_average-case_2016,
bremner_achieving_2017, boixo_characterizing_2018, aaronson2017complexity,
bouland_quantum_2018, movassagh_efficient_2018}. Further, these quantum circuits
appear to become difficult to simulate at relatively small sizes and within
error tolerances that are expected to be implementable on early NISQ hardware
\cite{boixo_characterizing_2018}. Here, we present a flexible simulator that
both raises the bar for quantum supremacy demonstrations and provides expanded
verification of quantum hardware through sampling.

It is important to emphasize the difference between the two tasks at hand: the
verification of a NISQ device and the computational task proposed for quantum
supremacy, as well as the role that a classical simulator plays in both of them. 

On the one hand, the fidelity of NISQ devices can be estimated by computing the
cross-entropy difference (cross-entropy benchmarking, or XEB) between the actual
output from the hardware given an RQC, and the corresponding exact output of
that particular RQC using classical simulators, as proposed in Boixo et
al.~\cite{boixo_characterizing_2018}. Note that this calculation requires the
sampling of about one million bit-strings from the device, and the computation
of their exact amplitudes using a classical simulator. For quantum circuits
beyond the ability to compute amplitudes classically, XEB can no longer be
performed. Alternatively, close correspondence between experiments, numerics,
and theory up to that point, for a variety of circuits with combinations of
fewer qubits, shallower depth, or simpler-to-simulate circuits (\emph{e.g.},
more Clifford gates) or architectures (see Methods~\ref{sec:cutting}) of the
same size, may suggest by extrapolation that the hardware is performing
correctly at a particular fidelity.

On the other hand, the computational task proposed for supremacy experiments
consists of sampling a million amplitudes from either the NISQ device or its
classical competitor at the same fidelity, for example 0.5\%. A quantum computer
performing this sampling task beyond the capabilities of the best
state-of-the-art algorithms in supercomputers would therefore achieve practical
quantum supremacy.

\begin{figure}[t!]
\includegraphics[width=1.00\columnwidth]{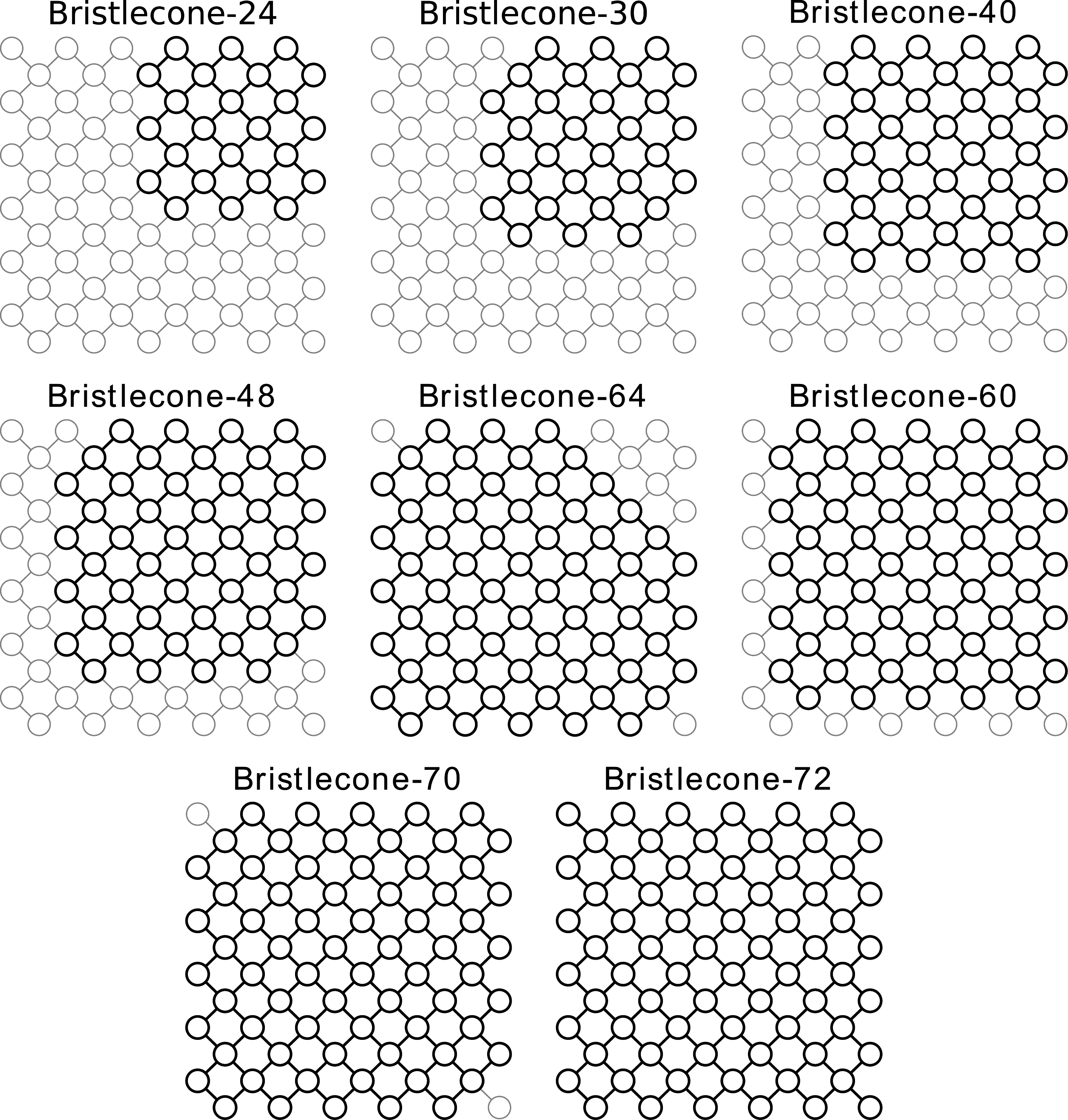}
\caption{\label{fig:bris_x} 
Sub-lattices of interest of the full \Bristlecone{72} (bottom right), ordered by
increasing hardness for a given depth. Note that \Bristlecone{72} (entire lattice)
is not harder to simulate than \Bristlecone{70}, since the two corner tensors can
be contracted trivially at a negligible cost (see Methods~\ref{sec:simulator}).
Note also that \Bristlecone{64} is similar in hardness to \Bristlecone{48}, and
substantially easier to simulate than \Bristlecone{60}, as is discussed in
Methods~\ref{sec:simulator} and Results. We identify a family of
sub-lattices of Bristlecone, namely \Bristlecone{24}, \mbox{-30}, \mbox{-40}, \mbox{-48}, \mbox{-60} and \mbox{-70},
that are hard to simulate classically, while keeping the number of qubits as low
as possible.}
\end{figure}

Here, we propose a flexible quantum circuit simulator (qFlex) that raises the
bar in the classical simulation of RQCs, including the simulation of the Google
Bristlecone QPU.  By design, our simulator is ``blind'' to the randomness in the
choice of single-qubit gates of the RQCs.  Therefore, it presents no
fluctuations in performance from one RQC to another. Moreover, by expanding on a
technique introduced in \cite{markov_quantum_2018}, including introducing
fine-grained ``cuts'' that enable us to judiciously balance memory requirements
with number of independent computations that can be done in parallel, our
simulator can output $1/f$ amplitudes with a target fidelity $f$ at the same
computational cost to compute a single perfect-fidelity amplitude; furthermore,
we present an alternative technique to simulate RQC sampling with target
fidelity $f$ with the same speedup factor of $1/f$. 

In the last few years, many different simulators have been proposed, either
based on the direct evolution of the quantum wave-function
\cite{de_raedt_massively_2007, smelyanskiy_qhipster:_2016, haner20170,
pednault_breaking_2017, boixo_characterizing_2018, markov_quantum_2018,
de_raedt_massively_2018, li_quantum_2018}, Clifford + $T$ gate
sets~\cite{bravyi_improved_2016}, and tensor network
contraction~\cite{markov_simulating_2008, boixo_simulation_2017,
chen_64-qubit_2018, chen_classical_2018}.  Tensor network contraction based
simulators have been particularly successful in simulating RQCs for sizes close
to the quantum supremacy regime.  Some recent simulators exploited
\cite{chen_64-qubit_2018, li_quantum_2018, chen_classical_2018} weaknesses in
the design of the RQCs presented in \cite{boixo_characterizing_2018}, and even
introduced small changes in the circuits that make them significantly easier to
simulate. These designs have been revised to remove these weaknesses (see
Ref.~\cite{markov_quantum_2018} and Methods~\ref{sec:RQCs}). Making RQCs as
difficult as possible to simulate is a key point in the route towards quantum
supremacy. At the same time, a thorough exploration of optimizations that make
classical simulators as efficient as possible is essential so that supremacy is
not overclaimed when a NISQ device surpasses classical capabilities. It is also
important to reiterate that the quantum supremacy computational task of interest
consists of producing a sample of bit-strings within some variational distance
of the output distribution defined by a quantum
circuit~\cite{boixo_characterizing_2018, aaronson2017complexity,
bouland_quantum_2018, movassagh_efficient_2018}.  This is very different from
computing a single output amplitude, as done in Ref.~\cite{chen_classical_2018}
(see Methods~\ref{sec:sampling}).

Among the proposed classical approaches, it is worth mentioning Markov et al.'s
simulator \cite{markov_quantum_2018}.
Their method is based on splitting $I\times J$ grids of qubits in halves, which
are then independently simulated \cite{chen_64-qubit_2018}.  To make the
simulator more competitive, Markov et al. introduce checkpoint states and reuse
them for different branches of a tree where internal nodes represent Schmidt
decompositions of cross-gates and leaves represent simulation results for each
tree path.
The number of independent circuits to simulate is exponential in the number of
projected ${\rm CZ}$-gates that cross from one half to the other.  As part of
their study, the authors propose for the first time a technique to ``match'' the
target fidelity $f$ of the NISQ device, which actually reduces the classical
computation cost by a factor $f$. By matching the fidelity of a realistic
quantum hardware ($f = 0.51\%$), Markov et al.~\cite{markov_simulating_2008}
were able to simulate $7\times 7$ and $7\times 8$ grids with depth $1+40+1$ by
numerically computing $10^6$ amplitudes in respectively \num{582000} hours and
\num{1407000} hours on single cores. However, the algorithm
in~\cite{markov_quantum_2018} becomes less efficient than our algorithm for
grids beyond $8\times 8$ qubits because of memory requirements.  Moreover, it is
not well suited for the simulation of the Google Bristlecone QPU. Indeed, as we
show here, the Google Bristlecone QPU implements circuit topologies with a large
diameter, which increases the run time exponentially.  In both cases, one could
mitigate the memory requirements by either using distributed memory protocols
like \texttt{MPI}, or by partitioning the RQCs in more sub-circuits. However,
the aforementioned approaches introduce a non-negligible slow-down that make
them unpractical (see SI~C for more details).
\\

To summarize, our tensor network based simulator relies on four different points of strength:
\begin{itemize}[label={}, leftmargin=\parindent]
  \item \textbf{Robustness.} RQCs are mapped onto regular tensor networks, where
        each tensor corresponds to a block of the circuit enclosing several
        gates; consequently, 2D grids of qubits, including the Bristlecone
        architecture, are mapped onto 2D grids of tensors. Since the blocking
        operation removes any randomness in the resulting tensor network
        topology (the only randomness left is in the tensor entries themselves),
        our simulator is robust against fluctuations from RQC to RQC and to changes 
        of the rules to generate RQCs.
  \item \textbf{Flexibility.} By computing an appropriate fraction of ``paths'', it
        is possible to control the ``fidelity'' of the simulated RQCs, as first
        introduced in Ref.~\cite{markov_quantum_2018}.  Therefore, our simulator
        can output $1/f$ amplitudes with target fidelity $f$ with the same
        computational cost to compute one perfect amplitude, for almost any $f$.
        This property is very important to ``mimick'' the sampling from NISQ
        devices.
  \item \textbf{Scalability.} By carefully choosing which cuts to apply to
        the RQCs, we are able to control the maximum size of tensors seen during
        tensor contraction. Thanks to the regularity of the resulting tensor
        network, together with a better memory management and a novel
        cache-efficient tensor index permutation routine, we
        are able to simulate circuits of as many as 72 qubits and realistic
        circuit depths on NISQ architectures such as Bristlecone. 
  \item \textbf{Performance.} To the best of our knowledge, our tensor
        contraction engine is optimized beyond all the
        existing CPU-based alternatives for contracting the RQCs with the largest
        number of qubits studied in this work.
\end{itemize}

Our analyses are supported by extensive simulations on Pleiades (27th in the
November 2018 TOP500 list) and Electra (33rd in the November 2018 TOP500 list)
supercomputers hosted at NASA Ames Research Center.  

In total, we used over 3.2 million core-hours and ran six different numerical
simulations (see Fig.~\ref{fig:bris_x} for nomenclature of Google Bristlecone):
\begin{enumerate}[label={\rsim{\arabic*}}, leftmargin=4\parindent]
	\item \Bristlecone{64} (1+32+1): 1.2M amplitudes with target fidelity 0.78\%,
    \item \Bristlecone{48} (1+32+1): 1.2M amplitudes with target fidelity 0.78\%,
    \item \Bristlecone{72} (1+32+1): 10 amplitudes with perfect fidelity,
    \item \Bristlecone{72} (1+24+1): 43K amplitudes with target fidelity 12.5\%,
    \item \Bristlecone{72} (1+24+1): 6000 amplitudes with perfect fidelity,
    \item \Bristlecone{60} (1+32+1): 1.15M amplitudes with target fidelity 0.51\%.
\end{enumerate}

For the most computationally demanding simulation we have run, namely sampling
from a $60$-qubit sub-lattice of Bristlecone, the two systems combined reached a
peak of 20 PFLOPS (single precision), that is $64\%$ of their maximum achievable
performance, while running  on about 90\% of the nodes. To date, this is the
largest computation run on NASA HPC clusters in terms of peak PFLOPS and number
of nodes used.  All Bristlecone simulation data are publicly
available (see Data Availability) and we plan to open source our simulator in the
near future.\\

This paper is structured as follows. Our results -- with an emphasis on our
ability to both simulate the computational task run on the quantum computer, as
well as to compute perfect fidelity amplitudes for the verification of the
experiments -- and discussion are presented in their respective sections. In
Methods~\ref{sec:RQCs} we review the rules for generating the revised RQCs
\cite{markov_quantum_2018}, which are based on the constraints of the quantum
hardware, while attempting to make classical simulations hard.  The hardness of
the revised RQCs motivates in part our simulator's approach, which is explained
in Methods~\ref{sec:simulator}, where both conceptual and implementation details
are discussed; here, we also introduce a novel, cache-efficient algorithm for
tensor index permutation which takes advantage of multithreading. In
Methods~\ref{sec:sampling} we discuss two methods to classically sample from an
RQC mimicking the fidelity $f$ of the output of a real device, while achieving a
speedup in performance of a factor of $1/f$ (see
Ref.~\cite{markov_quantum_2018}); in addition, we present a method to speedup
the classical sampling by a factor of about $10\times$ that, under reasonable
assumptions, is well suited to tensor network based simulators. We also discuss
the implications of classically sampling from a non fully-thermalized RQC.
Methods~\ref{sec:hardness} discusses the hardness of simulating RQCs implemented
on the Bristlecone QPU as compared to those implemented on square grids of
qubits.

\section{Results}
\label{sec:results}

In this section we review the performance and the numerical results obtained by
running our simulations \rsim{1-6} on the NASA HPC clusters Pleiades and
Electra.

In the time of exclusive access to large portions of the NASA HPC clusters, we
were able to run for over 3.2 million core-hours. Although most of the
computation ran on varying portions of the supercomputers, for a period of time
we were able to reach the peak of 20 PFLOPS (single precision), that corresponds
to $64\%$ of the maximum achievable performance for Pleiades and Electra
combined. For a comparison, the peak for the LINPACK benchmark is ~23 PFLOPS
(single precision, projected), which is only $15\%$ larger than the peak we
obtained with our simulator. This is to date the largest simulation (in terms of
number of nodes and FLOPS rate) run on the NASA Ames Research Center HPC
clusters. This is not a surprise since both LINPACK and our simulation do the
majority of work in MKL routines (\texttt{dgemm} or \texttt{cgemm} and similar),
in our case due in part to the fact that our cache-efficient memory reordering
routines lower the tensor indexes permutation bottleneck to a minimum.\\

\begin{figure*}[t!]
\includegraphics[width=1.00\textwidth]{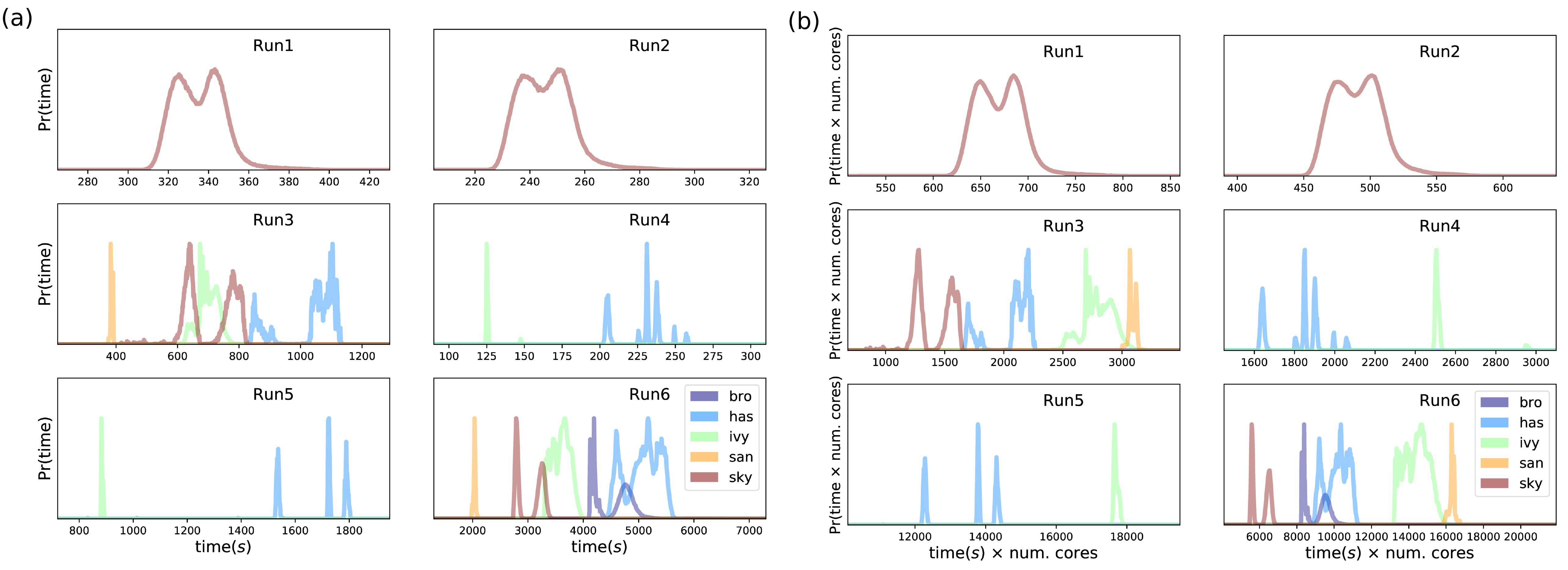}
\caption{\label{fig:times} (a) Distribution of the runtimes for a single
instance of each of the six simulations \rsim{1-6} run on different node
architectures.  An instance refers to a certain number of paths for a particular
number of amplitudes (output bit-strings); see Methods~\ref{sec:cutting} and
Table~\ref{table:times} for more details.  For clarity, all distributions have
been normalized so that their maxima are all at the same height. The nodes used
on NASA HPC clusters Pleiades and Electra are: Broadwell (bro), Intel Xeon
E5-2680v4 ; Haswell (has), Intel Xeon E5-2680v3; Ivy Bridge (ivy), Intel Xeon
E5-2680v2; Sandy Bridge (san), Intel Xeon E5-2670; Skylake (sky), 2 $\times$
20-core Intel Xeon Gold 6148 processors per node. (b) Same
distribution as above, but the runtimes are multiplied by the number of cores
per job on a single node, to provide a fairer comparison.  As one can see,
Skylake nodes provide generally the best performance, and belong to Electra, an
energy efficient HPC cluster.  The split of runtimes into groups is discussed
in SI~D.}
\end{figure*}

Fig.~\ref{fig:times} reports the distribution of the runtimes for a single
instance of each of the six simulations \rsim{1-6} for both Pleiades and
Electra.  Interestingly, we observe a split in the distribution of runtimes
(see SI~D for further details).  For our simulations run on
Pleiades, we used all the four available node architectures:
\begin{itemize}
\item 2016 Broadwell (bro) nodes: Intel Xeon E5-2680v4, 28 cores, 128GB per node.
\item 2088 Haswell (has) nodes: Intel Xeon E5-2680v3, 24 cores, 128GB per node.
\item 5400 Ivy Bridge (ivy) nodes: Intel Xeon E5-2680v2, 20 cores, 64GB per node.
\item 1936 Sandy Bridge (san) nodes: Intel Xeon E5-2670, 16 cores, 32GB per node.
\end{itemize}
For the Electra system, we used its two available node architectures:
\begin{itemize}
\item 1152 Broadwell (bro) nodes: same as above.
\item 2304 Skylake (sky) nodes: 2 $\times$ 20-core Intel Xeon Gold 6148, 40 cores, 192GB per node.
\end{itemize}
Note that the Skylake nodes at Electra form a much smaller machine than
Pleiades, but substantially more efficient, both time and energy-wise.

In Table~\ref{table:times} we report runtime, memory footprint, and number of
cores (threads) used for all six cases run on NASA Pleiades and Electra HPC
clusters. As we describe in Methods~\ref{sec:simulator}, instances (which
involve a certain number of paths given a cut prescription, as well as a batch
size $N_c$, as introduced in Methods~\ref{sec:fast}) can be collected for a
large number of low fidelity amplitudes or for a smaller number of high fidelity
amplitudes at the same computational cost. \rsim{1-6} were performed sharing the
Pleiades and Electra clusters on maintenance period, which made nodes on both
supercomputers become available and unavailable for our simulations without
prior notice. For this reason, the ZeroMQ software (more suited for this sort of
scheduling than MPI) was used for scheduling different instances of the
simulations; all instances were scheduled from a master task. In practice,
instances corresponding to different paths of the same amplitude were grouped
together onto a single instance and run sequentially on the same group of cores
of a single node (except for \rsim{3}, which requires a large number of paths,
whose computations were also parallelized across nodes); this provides some
advantage due to the reuse of tensors across paths (see SI~A). Due
to the inhomogeneous nature of our two clusters, with five different types of
nodes across two supercomputers, each job instance included an estimate of its
memory footprint (see Table~\ref{table:times}), and was scheduled on any
available node with enough available memory. Given that the number of instances
per node was always smaller than the number of cores, each instance was
multithreaded, using as many threads as the number of physical cores given;
cores were assigned proportionally to the memory footprint of the instance. Note
that both matrix multiplications and tensor index permutations take advantage of
multithreading (see Methods~\ref{sec:implementation}). All the numerical data gathered
during the simulations \rsim{1-6}, including all the amplitudes, are publicly
available (see Data Availability).

Note that, after running our
simulations on Pleiades and Electra we have identified for \Bristlecone{48} and
-70 a better contraction procedure (\rsim{2b} and \rsim{3b}, respectively). This
new contraction is is about twice as fast as the one used in \rsim{2-3}, which
was similar in approach to the contraction used for \Bristlecone{60} (see
SI~B for more details); we include the estimated
benchmark of these new contractions as well.

In Table~\ref{table:times_sampling} we estimate the effective runtime needed for
the computation of $10^6$ batches of amplitudes (\emph{i.e.}, sampling $10^6$
bit-strings) with a target fidelity close to $0.5\%$ on a single core, for
different node types. As one can see, the \Bristlecone{60} sub-lattice is almost
$10\times$ harder to simulate than the \Bristlecone{64} sub-lattice, while
\Bristlecone{64} is only $2\times$ harder than \Bristlecone{48}.\\

In the following, we report the (estimated) runtime and energy consumption for
both the tasks of verification and sampling for rectangular grids of qubits, up
to $8\times9$, as well as the full \Bristlecone{70} layout. The estimation is
obtained by computing a small percentage of the calculations required for the
full task.  We would like to stress that our simulator's runtimes are
independent of any particular RQC instance and, therefore, our estimations are
quite robust.

Table~\ref{table:verification} shows the estimated performance (runtimes and
energy consumption) of our simulator in computing perfect fidelity amplitudes of
output bit-strings of an RQC (rectangular lattices and \Bristlecone{70}), for both
Pleiades and Electra.  Runtimes are estimated assuming that fractions of the
jobs are assigned to each group of nodes of the same type in a way that they all
finish simultaneously, thus reducing the total real time of the run.  The power
consumption of Pleiades is 5MW, and a constant power consumption per core,
regardless of the node type, is assumed for our estimations.  For Electra, the
2304 Skylake nodes have an overall power consumption of 1.2MW, while the 1152
Broadwell nodes have an overall power consumption of 0.44MW.

Classically sampling bit-strings from the output state of an RQC involves the
computation of a large number (approximately one million) of low-fidelity (about
$0.5\%$) batches of probability amplitudes, as better described in Methods~\ref{sec:low}.
Table~\ref{table:classical} shows the estimated performance of our simulator
in this task, with runtimes and energy consumption requirements on the two HPC
clusters, Pleiades and Electra.\\

Finally, we compare our approach to the two leading previously existing
simulators of RQCs, introduced in Ref.~\cite{chen_classical_2018} (Alibaba) and
Ref.~\cite{markov_quantum_2018} (MFIB) (see also Table~\ref{table:igor}), as
well as to the recently proposed simulation methods of
Ref.~\cite{chen_quantum_2019} (Teleportation-Inspired Algorithm, or TIA) and
Ref.~\cite{guo2019general} (General-purpose quantum circuit simulator, or GPQS). 

\begin{figure}[t]
\includegraphics[width=1.00\columnwidth]{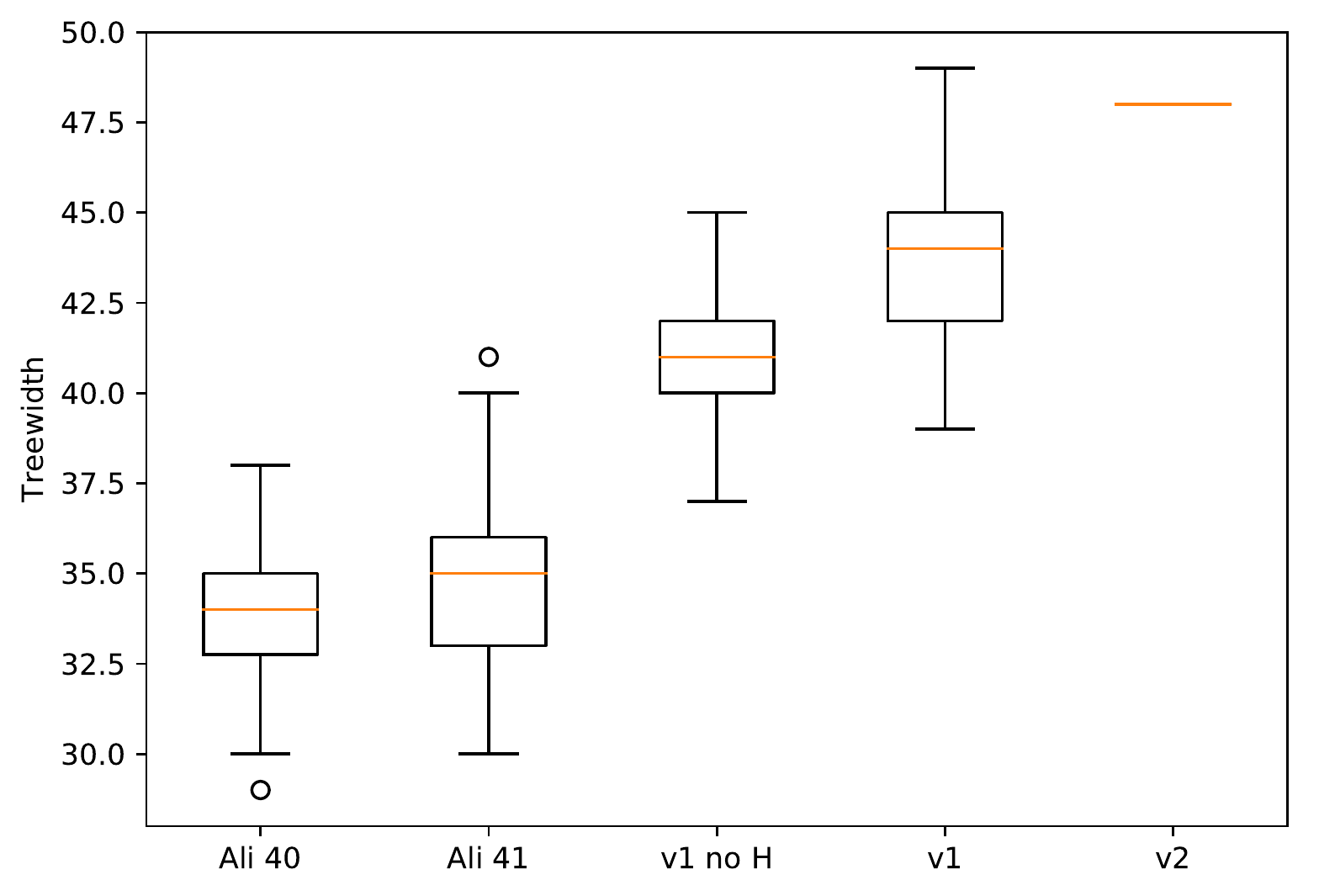}
\caption{\label{fig:treewidths} From left to right, upper bounds of treewidths
of RQCs on a $7\times 7$ square lattice simulated in: [Ali 40]
Ref.~\cite{chen_classical_2018} with depth (1+40) (note that no layer of
Hadamards is added at the end of the circuit); [Ali 41]
Ref.~\cite{chen_classical_2018} with depth (1+41); [v1 no H] old prescription of
the RQCs~\cite{boixo_characterizing_2018} without the final layer of Hadamards
and depth (1+41); [v1] old prescription of the
RQCs~\cite{boixo_characterizing_2018} with the final layer of Hadamards and
depth (1+40+1); [v2] revised prescription of the RQCs~\cite{markov_quantum_2018}
with depth (1+40+1).  Note that in all cases the tree width of the RQCs is
substantially larger than that ones simulated in
Ref.~\cite{chen_classical_2018}, making the simulations about $2^{13}\times$ or
$2^{14}\times$ harder (on average). Moreover, fluctuations in the treewidth for
the revised prescriptions of RQCs are completely absent. The upper bounds were
obtained by running {\tt quickbb}~\cite{gogate2004complete} with settings {\tt
--time 60 --min-fill-ordering}.}
\end{figure}

Compared to Ref.~\cite{chen_classical_2018}, our simulator is between
$3.6\times$ and $100\times$ slower (see SI~D for complementary
details), depending on the case.  However, it is important to stress that
Ref.~\cite{chen_classical_2018} reports the computational cost to simulate a
class of RQCs which is much easier to simulate than the class of RQCs reported
in Ref.~\cite{boixo_characterizing_2018}. Indeed, Chen et al. fail to include
the final layer of Hadamards in their RQCs and use more ${\rm T}$ gates at the
beginning of the circuit. For these reasons, we estimate that such class is
about $1000\times$ easier to simulate than the new prescription of RQCs we
present in this work.  The computational cost of simulating a circuit using
Alibaba's simulator scales as $2^{\rm TW}$, where ${\rm TW}$ is the treewidth of
the undirected graphical model of the circuit~\cite{boixo_simulation_2017}. We
show in Fig.~\ref{fig:treewidths} the treewidths of the circuits simulated in
Ref.~\cite{chen_classical_2018}, the old prescription of the
circuits~\cite{boixo_characterizing_2018} (with and without the final layer of
Hadamards), and the revised prescription, for RQCs on a $7\times 7\times
(1+40+1)$ square grid.  Note that with this number of qubits and depth, the
circuits simulated in Ref.~\cite{chen_classical_2018} are (on average)
$1000\times$ easier or more than the revised ones. Here we are comparing the
treewidth of the circuits to simulate, while Alibaba's simulator applies first a
preprocessing algorithm that projects a well chosen subset of variables in the
undirected graphical model of the circuit; this generates a number of
simulations that is exponential in the number of projected variables, but that
decreases the runtime of each of these simulations, which also allows for
parallelization. In our comparison, we are assuming that the trade-off between
the number of simulations generated after the projections and the decrease in
runtime for each simulation is comparable between different versions of the
circuits, and hence the treewidth is directly a good measure of the hardness of
the simulation of the circuits using the Alibaba simulator.  It is worth noting
that the revised RQCs have no variation in treewidth from one instance to
another.  In addition, note that Ref.~\cite{chen_classical_2018} reports
runtimes corresponding to the 80 percentile best results, excluding the worst
runtimes. On the contrary, our runtimes have little fluctuations and are RQC
independent.  Finally, in the absence of an implementation of the fast sampling
technique introduced in Methods~\ref{sec:fast}, this simulator would suffer from
a multiplicative runtime overhead when using rejection sampling (see
Methods~\ref{sec:sampling}).

Compared to Ref.~\cite{markov_quantum_2018}, our simulator is $7\times$ less
efficient to compute $10^6$ amplitudes with fidelity $0.51\%$ for $7\times7$
grids of qubits with depth $1+40+1$, using the new prescription of RQCs.
However, it is important to note that the runtime of MFIB's simulator and our
simulator scale in completely different ways.  Indeed, MFIB's approach has the
advantage to compute a large number of amplitudes with a small cost overhead. On
the contrary, our approach performs much better in the computation of a smaller
subset of amplitudes; both methods use comparable resources when computing about
$10^5$ amplitudes of a $7\times 7\times (1+40+1)$ RQC.  Note also that MFIB's
approach is limited by memory usage, and it scales unfavorably compared to our
simulator for circuits with a large number of qubits (\emph{e.g.}, beyond
$8\times 8$ rectangular grids), with a large diameter (\emph{e.g.},
\Bristlecone{60} and \mbox{-70}), or both.  For instance, \Bristlecone{70} would
require 825GB per node, which is currently unavailable for most of HPC clusters.
To mitigate the memory requirements, one could either partition the RQCs in more
sub-circuits, or use distributed memory protocols like \texttt{MPI}.  However,
both approaches introduce a non-negligible slow-down that make them unpractical
(see SI~C for more details on the impact further partitions have on the runtime,
as well as Ref.~\cite{jones2018quest} for insight on the strong and weak scaling
of distributed wave function simulators).

Sometime after this manuscript appeared on the arXiv, two other simulators have
been posted: TIA~\cite{chen_quantum_2019} and GPQS~\cite{guo2019general}.  TIA
is used to compute single amplitudes of RQCs. The most challenging case
benchmarked using TIA largest number of logical qubits is the simulation
involving the computation of an amplitude for \Bristlecone{72}, with depth
$1+32+1$; \Bristlecone{72} is trivially equivalent to \Bristlecone{70} (see
Methods~\ref{sec:contraction}), and therefore a comparison to our computation
times of \Bristlecone{70} is in place. TIA takes 14.1 minutes to compute one
amplitude on 16384 Sunway SW26010 260C nodes, with 256 cores each, and a
theoretical peak performance of 3.05 TFLOPS per node.  Our simulator computes an
amplitude in 4121.49 core-hours using Skylake nodes (see
Table~\ref{table:times}).  A direct comparison of core-hours between both
simulators running on their respective architectures results on our simulator
being $239\times$ faster than TIA.  However, Taihu Sunlight uses a different
architecture, with slower cores compared to Electra's cluster of Skylake nodes.
Therefore, for a fairer comparison, we use both node's theoretical peak
performance: 3.05 TFLOPS for Sunway SW26010 260C nodes, and 3.07 TFLOPS for
Electra's Skylake nodes.  Both node types deliver a similar performance, which
leads to the estimation of our simulator being $37\times$ more efficient than
TIA for the circuit considered in this comparison, \emph{i.e.},
\mbox{\Bristlecone{70} (72)} with depth $1+32+1$. Note that, while potentially
adaptable to this simulator, in the absence of an implementation of the fast
sampling technique, this simulator needs of the computation of a few amplitudes
in order to sample each bitstring, with the corresponding multiplicative runtime
overhead (see Methods~\ref{sec:sampling} and Ref.~\cite{chen_quantum_2019}).

GPQS is also used to compute amplitudes one at a time, although could
potentially implement the fast sampling technique, and is closely related to our
simulator in that it first contracts the circuit tensor network in the time
direction. However, it then opts for a distributed contraction across several
nodes of a supercomputer using the Cyclops Tensor Framework, as opposed to
performing cuts to allow for single-node contractions. On the $7\times 7\times
(1+40+1)$ computation of single perfect fidelity amplitudes, which can be
directly compared with our equivalent computation, GPQS takes 31 minutes using a
fraction of the Tianhe-2 supercomputer, delivering a theoretical peak of 1.73
PFLOPS (double precision), as reported by the authors. Our simulator takes
$1.16\times 10^{-2}$ hours to compute a single perfect fidelity amplitude on
Electra (see Table~\ref{table:verification}), which delivers a theoretical peak
of 8.32 PFLOPS. From a direct comparison between both simulators, we estimate
that our simulator is $9.26\times$ more efficient than GPQS for the
aforementioned circuit instance. However, GPQS performs calculations using
double precision. If this simulator were to be adapted for single precision
calculations, we estimate ours would still be $4.63\times$ more efficient. This
comparison gives meaningful insight on the cost of relying on inter-node
communication, instead of cuts, for tensor network contractions of the sizes
relevant to supremacy circuit sizes.

\section{Discussion}
\label{sec:conclusions}

In this work, we introduced a flexible simulator, based on tensor contraction
(qFlex), to compute both exact and noisy (with a given target
fidelity~\cite{markov_quantum_2018}) amplitudes of the output wave-function of a
quantum circuit.  While the simulator is general and can be used for a wide
range of circuit topologies, it is well optimized for quantum circuits with a
regular design, including rectangular grids of qubits and the Google Bristlecone
QPU. To test the performance of our simulator, we focused on the benchmark of
random quantum circuits (RQCs) presented in
Refs.~\cite{boixo_characterizing_2018,markov_quantum_2018} for both the 2-D
grids ($7\times7$, $8\times8$ and $8\times9$) and the Google Bristlecone QPU
($24$, $48$, $60$, $64$, and $70$ qubits). Compared to some existing
methods~\cite{li_quantum_2018,chen_64-qubit_2018,chen_classical_2018}, our
approach is more robust to modifications in the class of circuits to simulate
and performs well on the redesigned, harder class of RQCs. While other
benchmarks exploit~\cite{li_quantum_2018}, and sometimes
introduce~\cite{chen_64-qubit_2018,chen_classical_2018}, weaknesses in
particular ensembles of random quantum circuits that affect their reported
performance significantly, our runtimes are directly determined by the number of
full lattices of two-qubit gates at a given depth (see
Fig.~\ref{fig:3D_to_2D_grid}). 

Our performance analyses are supported by extensive simulations on Pleiades
(24th in the November 2018 TOP500 list) and Electra (43rd in the November 2018
TOP500 list) supercomputers hosted at NASA Ames Research Center. Among other
``diamond-shaped'' lattices of qubits benchmarked, which are likely to be used
for supremacy experiments, our simulator is able to compute exact amplitudes for
the benchmark of RQCs using the full Google Bristlecone QPU with depth 1+32+1 in
less than $(f \cdot 4200)$ hours on a single core, with $f$ the target fidelity.
This corresponds to 210 hours in Pleiades or 264 hours in Electra for $10^6$
amplitudes with fidelity close to 0.5,\% a computation needed to perform the RQC
sampling task.  All our data are publicly available to use
(see Data Availability).\\

At first sight, compared to Alibaba's simulator~\cite{chen_classical_2018}, our
simulator is between $3.6\times$ and $100\times$ slower, depending on the case.
However, Alibaba's simulator heavily exploits the structure of RQCs and its
performance widely varies from one RQC instance to another. Indeed,
Ref.~\cite{chen_classical_2018} reports only runtimes corresponding to the 80th
percentile best results, excluding the worst runtime. In contrast, our runtimes
have little variation in performance between instances and are independent of
RQC class.  Moreover, Ref.~\cite{chen_classical_2018} fails to include the final
layer of Hadamards and uses fewer non-diagonal gates at the beginning of the
circuit which, we estimate, makes the corresponding circuits much easier to
simulate: approximately $1000\times$ easier for the $7\times 7\times (1+40+1)$
circuit. We would like to encourage the reporting of benchmarking against the
circuit instances publicly available in~\cite{instances} in order to arrive
at meaningful conclusions.

Compared to Ref.~\cite{markov_quantum_2018}, our simulator is $7\times$ less
efficient (on Electra Skylake nodes) to compute $10^6$ amplitudes with fidelity
$0.51\%$ for $7\times7$ grids of qubits with depth $1+40+1$.  However, compared
to Ref.~ \cite{markov_quantum_2018} our simulator scales better on grids beyond
$8\times8$ and on circuits with a large number of qubits and diameter, including
the Bristlecone QPU and its sub-lattices \Bristlecone{60} and \mbox{-70}.

Compared to Ref.~\cite{chen_quantum_2019}, our simulator is $37\times$ more
efficient in computing an amplitude of \Bristlecone{70} at depth 1+32+1, which
is equivalent in hardness to \Bristlecone{72} (see
Methods~\ref{sec:contraction}).

Compared to Ref.~\cite{guo2019general}, our simulator is more than $9\times$
more efficient in computing an amplitude of a $7\times 7$ circuit of depth
$1+40+1$.\\

In addition, we were able to simulate (\emph{i.e.}, compute over $10^6$
amplitudes) RQCs on classically hard sub-lattices of the Bristlecone of up to 60
qubits with depth (1+32+1) and fidelity comparable to the one expected in the
experiments (around 0.50\%) in effectively well below half a day using both
Pleiades and Electra combined.  We also discussed the classical hardness in
simulating sub-lattices of Bristlecone as compared to rectangular grids with the
same number of qubits. Our theoretical study and numerical analyses show that
simulating the Bristlecone architecture is computationally more demanding than
rectangular grids with the same number of qubits and we propose a family of
sub-lattices of Bristlecone to be used in experiments that make classical
simulations hard, while keeping the number of qubits and gates involved as small
as possible to increase the overall fidelity.\\

As a final remark, we will explore using our approach and extensions to simulate
different classes of quantum circuits, particularly those with a regular
structure, including quantum circuits for algorithms with potential applications
to challenging optimization and machine learning problems arising in
aeronautics, Earth science, and space exploration, as well as to simulate
many-body systems for applications in material science and chemistry.

\section{Methods}

\subsection{Revised set of Random Quantum Circuits}
\label{sec:RQCs}

In this section, we review the prescription to generate RQCs proposed originally
by Google \cite{boixo_characterizing_2018}, and its revised version
\cite{markov_quantum_2018}. This prescription can be used to generate RQCs for
2D square grids, including the Bristlecone architecture (which is a diamond
shaped subset of a 2D square grid).  The circuit files used for the numerical
simulations in this paper are publicly available in~\cite{instances}.

Given a circuit depth and circuit topology of $n$ qubits, Google's
RQCs~\cite{boixo_characterizing_2018,markov_quantum_2018} are an ensemble of
quantum circuits acting on a Hilbert space of dimension $N=2^n$. The
computational task consists of sampling bit-strings as defined by the final
output.

\begin{figure}[t]
\includegraphics[width=1.00\columnwidth]{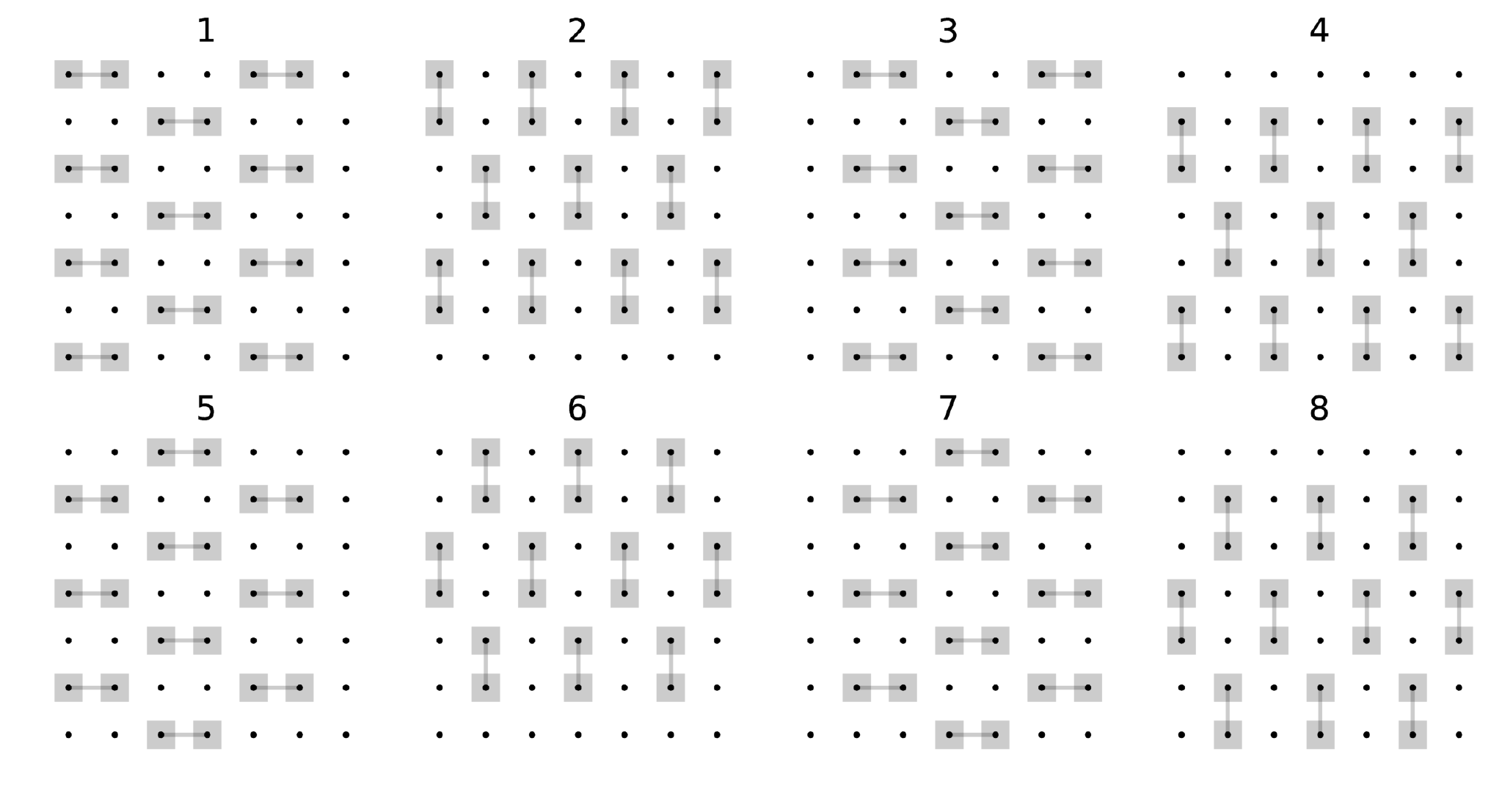}
\caption{\label{fig:coverings} Layout of two-qubit gates and the corresponding
cycle order (from 1 to 8). This layout can be tiled over 2D square grids of
arbitrary size. The Bristlecone architecture is a diamond shaped subset of such
a 2D grid. For our simulations, we use ${\rm CZ}$ gates as the two-qubit gate.}
\end{figure}

Due to the limitation of the current technology and the constraints imposed by
the quantum hardware, circuits are randomly generated using the following
prescription:
\begin{enumerate}
  \item[(1)] Apply a first layer of Hadamard (${\rm H}$) gates to all the qubits.
  \item[(2)] After the initial layer (1), subsequent layers of two-qubit gates
             are applied. There are 8 different layers, which are cycled through
             in a consistent order (see Fig.~\ref{fig:coverings}).
  \item[(3)] Within these layers, for each qubit that is not being acted upon by
             a two-qubit gate in the current layer, and such that a two-qubit
             gate was acting on it in the previous layer, randomly apply (with
             equal probability) a gate in the set $\{{\rm X^{1/2}},\,{\rm Y^{1/2}}\}$.
  \item[(4)] Within these layers, for each qubit that is not being acted upon by
             a two-qubit gate in the current layer, and was acted upon by a gate in
             the set $\{{\rm X^{1/2}},\,{\rm Y^{1/2}},\,{\rm H}\}$ in the
             previous layer,  apply a $T$ gate.
  \item[(5)] Apply a final layer of ${\rm H}$ gates to all the qubits.
\end{enumerate}

The depth of a circuit will be expressed as $1+t+1$, where the prefix and
suffix of $1$ explicitly denote the presence of an initial and a final layer of
Hadamard gates.\\

For our simulations, as was done in prior RQC works, we use the ${\rm CZ}$ gate
as our two-qubit gate.  One of the differences between the original prescription
\cite{boixo_characterizing_2018} and this new prescription
\cite{markov_quantum_2018} for the generation of RQCs is that we now avoid
placing ${\rm T}$ gates after ${\rm CZ}$ gates.  If a ${\rm T}$ gate follows a
${\rm CZ}$ gate, this structure can be exploited to effectively reduce the
computational cost to simulate the RQCs, as was done in
\cite{chen_64-qubit_2018, li_quantum_2018, chen_classical_2018}. The revised RQC
formulation ensures that each ${\rm T}$ gate is preceded by a $\{{\rm
X^{1/2}},\,{\rm Y^{1/2}},\,{\rm H}\}$ gate, which foils this exploit.  In
addition, the layers of two-qubit gates have been reordered, in order to avoid
consecutive ``horizontal'' or ``vertical'' layers, which is known to make
simulations easier.  Finally, it is important to keep the final layer of ${\rm
H}$ gates, as otherwise multiple two-qubit gates at the end of the circuit can
be simplified away, making the simulation
easier~\cite{boixo_characterizing_2018}.

Replacing ${\rm CZ}$ gates with ${\rm iSWAP} = \left(
\left|00\right\rangle\!\!\left\langle00\right|+
\left|11\right\rangle\!\!\left\langle11\right|+
i\left|01\right\rangle\!\!\left\langle10\right|+
i\left|10\right\rangle\!\!\left\langle01\right|\right)$ gates is known to make the
circuits yet harder to simulate.  More precisely, an RQC of depth $1+t+1$ with
${\rm CZ}$ gates is equivalent, in terms of simulation cost, to an RQC of depth
$1+t/2+1$ with ${\rm iSWAP}$s.  In future work, we will benchmark our approach
on these circuits as well.

\subsection{Overview of the simulator}
\label{sec:simulator}

\begin{figure}[t]
\includegraphics[width=1.00\columnwidth]{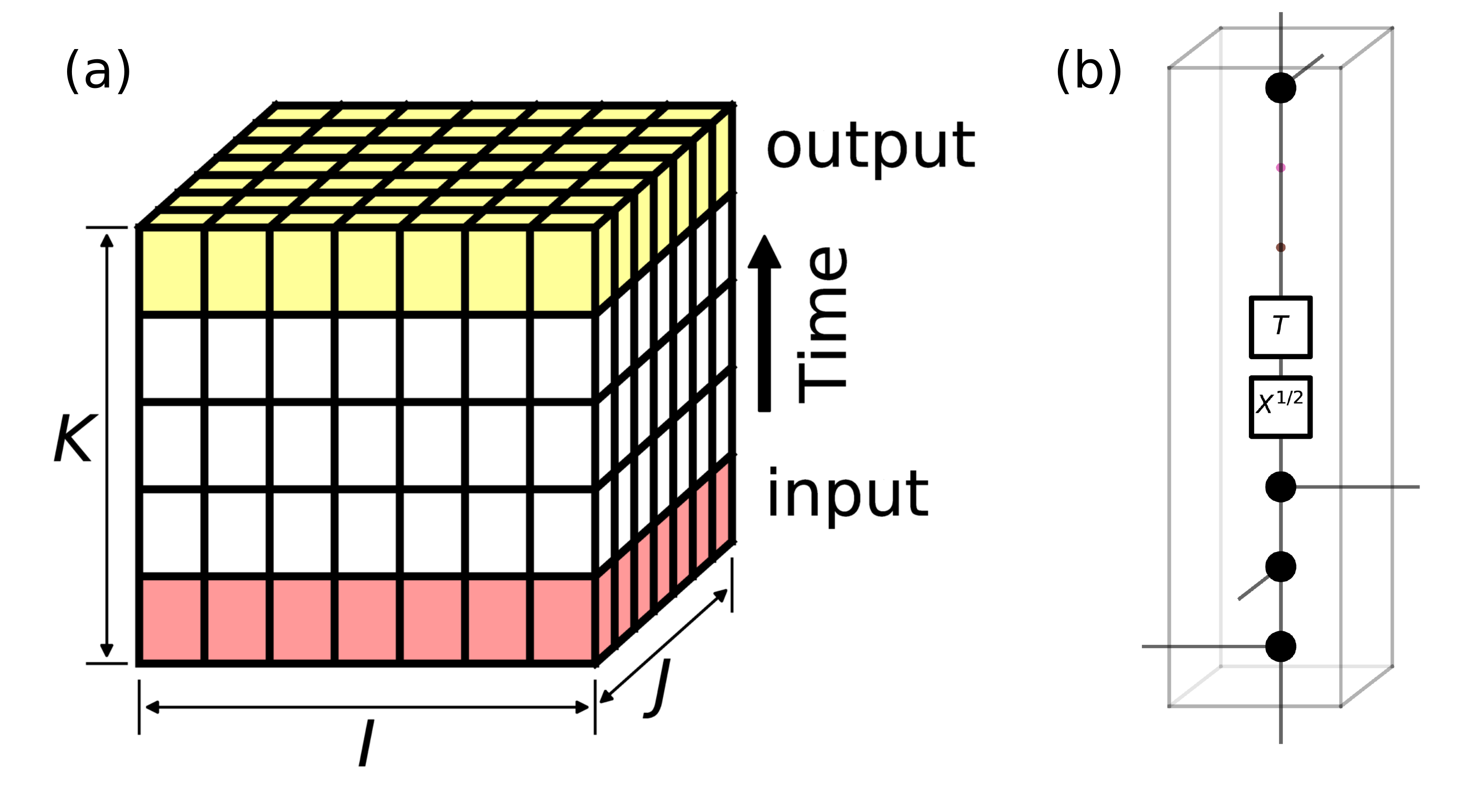}
\caption{\label{fig:3D_grid} (a) 3D grid of tensors obtained by
contracting 8 consecutive layers of ${\rm CZ}$ gates, including the single qubit
gates. (b) Example of a typical block of 8 layers of gates on a
single qubit; note that the qubit shares one ${\rm CZ}$ gate with each of its
four neighbors per block.}
\end{figure}

A given quantum circuit can always be represented as a tensor network, where
one-qubit gates are rank-2 tensors (tensors of 2 indexes with dimension 2 each),
two-qubit gates are rank-4 tensors (tensors of 4 indexes with dimension 2 each),
and in general $n$-qubit gates are rank-$2n$ tensors.  The computational and
memory cost for the contraction of such networks is exponential with the number
of open indexes and, for large enough circuits, the network contraction is
unpractical; nonetheless, it is always possible to specify input and output
configurations in the computational basis through rank-1 Kronecker deltas over
all qubits, which can vastly simplify the complexity of the tensor network.
This representation of quantum circuits gives rise to an efficient simulation
technique, first introduced in Ref.~\cite{markov_simulating_2008}, where the
contraction of the network gives amplitudes of the circuit at specified input
and output configurations.

Our approach allows the calculation of amplitudes of RQCs through the
contraction of their corresponding tensor networks, as discussed above, but with
an essential first step, which we now describe.  One of the characteristics of
the layers of ${\rm CZ}$ gates shown in Fig.~\ref{fig:coverings} is that it
takes 8 cycles for each qubit to share one, and only one, ${\rm CZ}$ gate with
each of its neighbors.  This property holds for all subsets of a 2D square grid,
including the Bristlecone architecture.  Therefore, it is possible to contract
every 8 layers of the tensor network corresponding to an RQC of the form
described in Methods~\ref{sec:RQCs} onto an $I\times J$ two-dimensional grid of
tensors, where $I$ and $J$ are the dimensions of the grid of qubits.  While in
this work we assume that the number of layers is a multiple of $8$, our
simulator can be trivially used for RQCs with a depth that is not a multiple of
$8$.  The bond dimensions between each tensor and its neighbors are the Schmidt
rank of a ${\rm CZ}$ gate, which (as for any diagonal two-qubit gate) is equal
to 2 (note that for iSWAP the Schmidt rank is equal to 4, thus effectively
doubling the depth of the circuit as compared to the ${\rm CZ}$ case).  After
contracting each group of 8 layers in the time direction onto a single, denser
layer of tensors, the RQC is mapped onto an $I\times J\times K$
three-dimensional grid of tensors of indexes of bond dimension 2, as shown in
Fig.~\ref{fig:3D_grid}, where $K=t/8$, and $1+t+1$ is the depth of the circuit
(see Methods~\ref{sec:RQCs}).  Note that the initial (final) layer of Hadamard
gates, as well as the input (resp.~output) delta tensors, can be trivially
contracted with the initial (resp.~final) cycle of 8 layers of gates.  At this
point, the randomness of the RQCs appears only in the entries of the tensors in
the tensor network, but not in its layout, which is largely regular, and whose
contraction complexity is therefore independent of the particular RQC instance
at hand.  This approach contrasts with those taken in
Refs.~\cite{boixo_simulation_2017, li_quantum_2018, chen_classical_2018}, which
propose simulators that either benefit from an approach tailored for each random
instance of an RQC, or take advantage of the particular layout of the ${\rm CZ}$
layers.

The contraction of the resulting 3D tensor network described above (see
Fig.~\ref{fig:3D_grid}) in order to compute the amplitude corresponding to
specified initial and final bit-strings is described in the following
Methods~\ref{sec:contraction}.

\subsubsection{Contraction of the 3D tensor network}
\label{sec:contraction}

\begin{figure}[t]
\includegraphics[width=1.00\columnwidth]{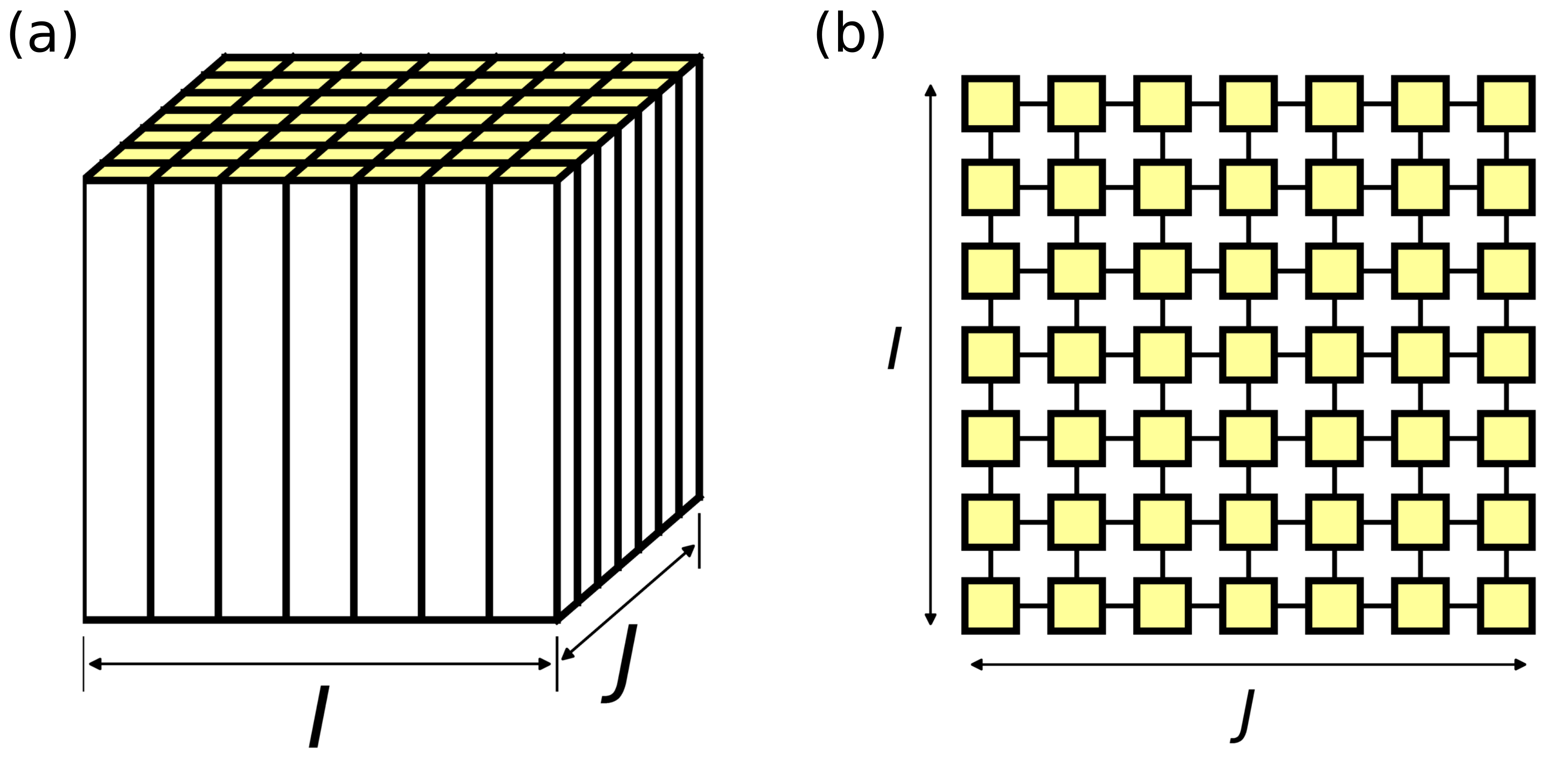}
\caption{\label{fig:3D_to_2D_grid} (a) Contraction of the 3D grid of tensors
(see Fig.~\ref{fig:3D_grid}) in the time direction 
to obtain (b) a 2D grid of tensors.}
\end{figure}

\begin{figure}[t]
\includegraphics[width=1.00\columnwidth]{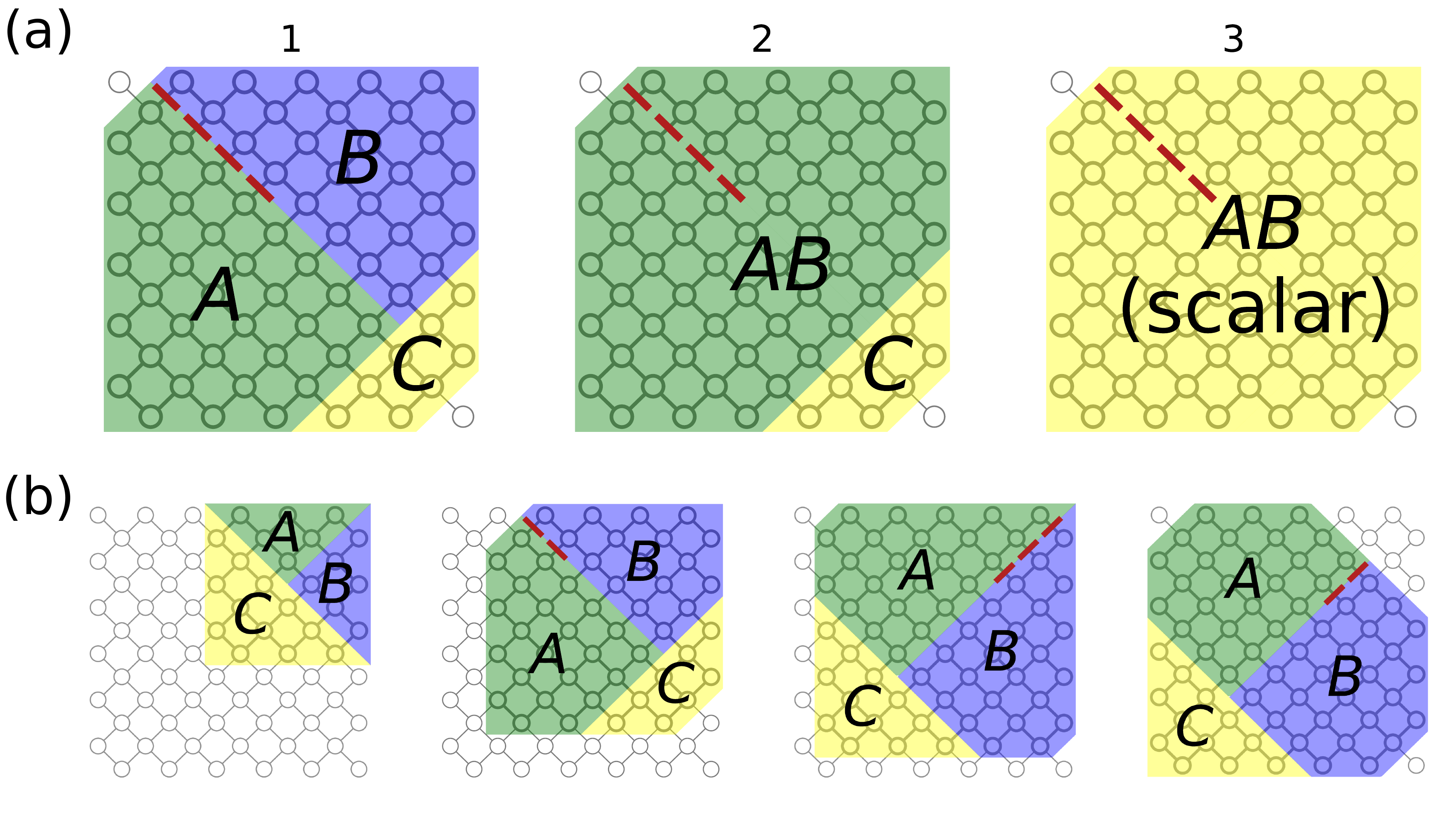}
\caption{\label{fig:bris_x_C} (a) Sketch of the contraction procedure
followed to obtain one path of one amplitude of the \Bristlecone{70} with depth
(1+32+1). We first make four cuts of dimension $2^4$ each, leaving us with
$2^{16}$ paths; for each path, we contract all tensors on region $A$, and all
tensors on region $B$; then tensors $A$ and $B$ are contracted together;
finally, tensor $C$ (which is independent of chosen path, and can in addition be
computed very efficiently) is contracted with $AB$, which obtains the
contribution of this path to this particular amplitude. (b)
Corresponding regions $A$, $B$, and $C$ for the \Bristlecone{24}, \mbox{-48}, \mbox{-60}, and
\mbox{-64}. Note that both the \Bristlecone{48} and the \Bristlecone{64} need 2 cuts of
dimension $2^4$ each, while the \Bristlecone{60} needs three of such cuts, making
it a factor of $2^4$ times harder than \Bristlecone{64}, even though it has 4
qubits less.}
\end{figure}

In this section, we describe the contraction procedure followed for the
computation of single perfect-fidelity output amplitudes for the 3D grid of
tensors described in the previous section. 

Starting from the 3D grid of tensors of Fig.~\ref{fig:3D_grid}, we first
contract each vertical ($K$ direction) column of tensors onto a single tensor of
at most 4 indexes of dimension $2^K$ each (see left panel of
Fig.~\ref{fig:3D_to_2D_grid}).  Note that for the circuit sizes and depths we
simulate, $K$ is always smaller than $I$ and $J$, and so this contraction is
always feasible in terms of memory, fast, and preferable to a contraction in
either the direction of $I$ or $J$.  This results in a 2D grid of tensors of
size $I\times J$, where all indexes have dimension $2^K$ (see right panel of
Fig.~\ref{fig:3D_to_2D_grid}).  Note that contracting in the time direction
first is done at a negligible cost, and reduces the number of high complexity
contractions to only the ones left in the resulting 2D grid.

While we have focused so far on the steps leading to the 2D square grid tensor
network of Fig.~\ref{fig:3D_to_2D_grid}, it is easy to see that the Bristlecone
topology (see \Bristlecone{72} in Fig.~\ref{fig:bris_x}) is a sub-lattice of a
square grid or qubits, and so all considerations discussed up to this point are
applicable.  Even though Bristlecone has 72 qubits, the top-left and
bottom-right qubits of the network can be contracted trivially with their
respective only neighbor, adding no complexity to our classical simulation of
RQCs.  For this reason, without loss of generality, we ``turn off'' those two
qubits from the Bristlecone lattice, and work only with the resulting
sub-lattice, which we call \Bristlecone{70} (see Fig.~\ref{fig:bris_x}).  For the
remainder of this section, we will focus on \Bristlecone{70} and other
sub-lattices of Bristlecone (see sub-lattices considered in
Fig.~\ref{fig:bris_x}), and we will refer back to square grids of qubits in
later sections.\\

\label{sec:cutting}
\noindent\textit{Cutting indexes and the contraction of the 2D tensor network ---}
From Fig.~\ref{fig:bris_x}, it is easy to see that it is not possible to
contract the \Bristlecone{70} tensor network without generating tensors of rank
11, where each index has dimension $2^K$.  For a circuit of depth $1+32+1$ and
$K=4$, the dimension of the largest tensors is $2^{11\times 4}$, which needs
over 140 TB of memory to be stored using single precision floating point complex
numbers, far beyond the RAM of a typical HPC cluster node (between 32 GB and 512
GB).  Therefore, to avoid the memory bottleneck, we decompose the contraction of
the \Bristlecone{70} tensor network into independent contractions of several
easier-to-compute sub-networks.  Each sub-network is obtained by applying
specific ``cuts'', as is described below.

Given a tensor network with $n$ tensors and a set of indexes to contract
$\{i_l\}_{l=1,\ldots}$, $\sum_{i_1,i_2,\ldots} T^1 T^2 \ldots T^n$, we define a
cut over index $i_k$ as the explicit decomposition of the contraction into
$\sum_{i_k}\left(\sum_{\{i_l\}_{l=1,\ldots}-\{i_k\}} T^1 T^2 \ldots T^n\right)$.
This implies the contraction of ${\rm dim}(i_k)$ many tensor networks of lower
complexity, namely each of the $\sum_{\{i_l\}_{l=1,\ldots}-\{i_k\}} T^1 T^2
\ldots$ networks, where tensors involving index $i_k$ decrease their rank by 1,
fixing the value of $i_k$ to the particular value given by the term in
$\sum_{i_k}$.  This procedure, equivalent to the ones used in
Refs.~\cite{chen_64-qubit_2018,li_quantum_2018,chen_classical_2018,markov_quantum_2018},
reduces the complexity of the resulting tensor network contractions to
computationally manageable tasks (in terms of both memory and time), at the
expense of creating exponentially many contractions.  The resulting tensor
networks can be contracted independently, which results in a computation that is
embarrassingly parallelizable. It is possible to make more than one cut on a
tensor network, in which case $i_k$ refers to a multi-index; the contribution to
the final sum of each of the contractions (each of the values of the multi-index
cut) is called a ``path'', and the final value of the contraction is the sum of
all path contributions.

For the \Bristlecone{70} example with depth $(1+32+1)$, making four cuts, as shown
in Fig.~\ref{fig:bris_x_C} (a), decreases the size of the maximum tensor
stored during the contraction from $2^{11\times 4}$ to $2^{7\times4}$ entries,
at the price of $2^{4 \times 4}$ contractions to be computed.  At the same time,
the choice of cuts aims at lowering the number of high complexity contractions
needed per path, as well as lowering the number of largest tensors held
simultaneously in memory.  Note that for \Bristlecone{60}, tensors $A$ and $B$ are
both equally large, and that the number of high complexity contractions is
larger than for a single path of \Bristlecone{70}.

After making these cuts, the contraction of each path is carried out in the
following way (see Fig.~\ref{fig:bris_x_C}): first, we contract all tensors
within region $A$ onto a tensor of rank 7 (tensor $A$); we do the same for
tensor $B$; then tensors $A$ and $B$ are contracted onto a rank-6 tensor, $AB$;
finally, tensor $C$ is contracted, which does not depend on the particular path
at hand, followed by the contraction of $AB$  with $C$ onto a scalar.
In Fig.~\ref{fig:bris_x_C} (b) we depict the corresponding $A$, $B$,
and $C$ regions for the sub-lattices of Bristlecone we use in our simulations,
as well as the cuts needed to contract the resulting tensor networks using the
described method, in particular for \Bristlecone{48}, \mbox{-60}, and \mbox{-64}.  Note that
that \Bristlecone{48} and \mbox{-64} need both two cuts of depth 4, making them similar
to each other in complexity, while \Bristlecone{60} needs three cuts, making it
substantially harder to simulate.

We identify a family of sub-lattices of Bristlecone, namely \Bristlecone{24},
\mbox{-30}, -40, \mbox{-48}, \mbox{-60} and \mbox{-70}, that are hard to
simulate classically, while keeping the number of qubits and gates as low as
possible. Indeed, the fidelity of a quantum computer decreases with the number
of qubits and gates involved in the experiment~\cite{boixo_characterizing_2018},
and so finding classically hard sub-lattices with a small number of qubits is
essential for quantum supremacy experiments.  It is interesting to observe that
\Bristlecone{64} is an example of a misleadingly large lattice that is easy to
simulate classically (see Results for our numerical results).

Note that the rules for generating RQCs cycle over the layers of two-qubit gates
depicted in Fig.~\ref{fig:coverings}.  In the case that the cycles or the layers
are perturbed, our simulator can be trivially adapted.  In particular: 1) if the
layers are applied in a different order, but the number of two-qubit gates
between all pairs of neighbors is the same, then the 2D grid tensor network of
Fig.~\ref{fig:3D_to_2D_grid} still holds, and the contraction method can be
applied as described; 2) if there is a higher count of two-qubit gates between
some pairs of neighbors than between others, then the corresponding anisotropy
in the bond dimensions of the 2D tensor network can be exploited through
different cuts.\\

\label{sec:remarks}
\noindent\textit{Remarks on the choice of cuts and contraction ordering ---}
The cost of contracting a tensor network depends strongly on the contraction
ordering chosen and is a topic covered in the literature (see
Ref.~\cite{markov_simulating_2008,boixo_simulation_2017}); determining the
optimal contraction ordering is an NP-hard problem. Given an ordering, the
leading term to the time complexity of the contraction is given by the most
expensive contraction between two tensors encountered along the contraction of
the full network; the optimal ordering given this cost model is closely related
to the treewidth of the line-graph of the tensor network.  However, a more
practical approach to determining the cost of a particular contraction ordering
is the FLOP count of the contraction of the entire network (\emph{i.e.}, the
number of scalar additions and multiplications needed); this accounts for cases
where a single high-complexity contraction is preferable to a large number of
low-complexity ones. The latter cost model is commonly used in order to
determine the optimal ordering for the contraction of tensor networks
(\emph{e.g.} in Ref.~\cite{chen_classical_2018}). Note that, although the choice
of a contraction ordering affects also its memory complexity, by what we mean
the memory required to perform the contraction, this is usually a smaller
concern as compared to time complexity.

An even more realistic approach needs to consider the performance efficiency of
the different tensor contractions, \emph{i.e.}, the delivered FLOP/s of each
contraction.  In particular, modern computing architectures benefit from high
arithmetically intensive contractions, \emph{i.e.}, contractions with a large
ratio between the FLOP count and the number of reads and writes from and to
memory. Highly unbalanced matrix multiplications, for example, present low
arithmetic intensity, while the multiplication of large squared matrices shows
high arithmetic intensity, and therefore achieves a performance very close to
the theoretical peak FLOP/s of a particular CPU. This principle (which
prioritizes time-to-solution over FLOP-count-to-solution) is at the heart of our
choice to contract the quantum circuits first into blocks and subsequently along
the ``time'' direction. This choice, beyond ``regularizing'' the tensor network,
serves two purposes, aimed at decreasing the overall time-to-solution in
practice: 1) the vast majority of contractions are performed in these first
steps, and are done in a negligible amount of time, leaving most of the
computation to a \emph{small number} of subsequent high-complexity contractions;
and 2) the remaining high-complexity contractions have \emph{high arithmetic
intensity} and achieve therefore high efficiency. In addition, contracting all
blocks in the time direction first reduces the memory requirement of the
subsequent contraction as compared to keeping the tensor network in its three
dimensional version.

The cost model discussed here becomes more intricate when cuts are considered.
Each choice of cuts not only has a different impact on the memory complexity of
the resulting, simplified tensor networks, but it can also substantially affect
their time complexity. As was explained in Methods~\ref{sec:cutting}, the main
purpose of the cuts is reducing the memory requirement of the resulting
contractions to fit the limits of each computation node; this is indeed the main
factor taken into account when choosing a set of cuts. However, the right choice
of cuts can also allow us to, again, have a small number of high arithmetic
intensity contractions. This is the case with the contractions depicted in
Fig.~\ref{fig:bris_x_C}, where the main bottleneck is given by the contraction
of $A$ with $B$, which is very arithmetically intensive; see also
Fig.~A1 for an example on a square lattice.

Finally, it is worth noting two more factors on the runtime of a contraction.
First, if the memory requirement of a contraction is well below the limits of a
computing node, then several contractions can be run in parallel. In these
cases, there is a trade-off between the number of cuts made and the number of
contractions run in parallel. Second, the choice of cuts can affect the number
of tensors reused across different paths (which is done at a memory cost), which
can have a substantial impact on the computation time of an amplitude, as can be
seen in the examples of SI~A.

Although a cost model involving all the factors discussed above can be well
characterized, automatically optimizing the cuts chosen and the contraction
ordering, together with the tensors reused across paths, is beyond the scope of
this work. In practice, we study different configurations ``by hand''. Note
that, once the tensor networks have a certain level of regularity, it is not
hard to make a choice that is close to optimal.

\subsubsection{Implementation of the simulator}
\label{sec:implementation}

We implemented our tensor network contraction engine for CPU-based
supercomputers using $C^{++}$. We have planned to release our tensor contraction
engine in the near future.  During the optimization, we were
able to identify two clear bottlenecks in the implementation of the
contractions: the matrix multiplication required for each index (or multi-index)
contraction, and the reordering of tensors in memory needed to pass the
multiplication to a matrix multiplier in the appropriate storage order (in
particular, we always use row-major storage).  In addition, to avoid
time-consuming allocations during the runs, we immediately allocate large enough
memory to be reused as scratch space in the reordering of tensors and other
operations.\\

\noindent\textit{Matrix multiplications with Intel\textregistered MKL ---}
For the multiplication of two large matrices that are not distributed over
several computational nodes, Intel's MKL library is arguably the best performing
library on Intel CPU-based architectures.  We therefore leave this essential
part of the contraction of tensor networks to MKL's efficient, hand-optimized
implementation of the BLAS matrix multiplication functions.  Note that, even
though we have used MKL, any other BLAS implementation could be used as well,
and other linear algebra libraries could be used straightforwardly.\\

\label{sec:reorderings}
\noindent\textit{Cache-efficient index permutations}
The permutation of the indexes necessary as a preparatory step for efficient
matrix multiplications can be very costly for large tensors, since it involves
the reordering of virtually all entries of the tensors in memory; similar issues
have been an area of study in other contexts~\cite{lokhmotov2007optimal,
weng2009practical, knittel2011qtib}.  In this section we describe our novel
cache-efficient implementation of the permutation of tensor indexes.

Let $A_{i_0,\ldots,i_k}$ be a tensor with $k$ indexes.  In our implementation,
we follow a row-major storage for tensors, a natural generalization of matrix
row-major storage to an arbitrary number of indexes.  In the tensor network
literature, a permutation of indexes formally does not induce any change in
tensor $A$.  However, given a storage prescription (\emph{e.g.}, row-major), we
will consider that a permutation of indexes induces the corresponding reordering
of the tensor entries in memory.  A naive implementation of this reordering
routine will result in an extensive number of cache misses, with poor
performance.

\begin{figure}[t] 
\includegraphics[width=1.00\columnwidth]{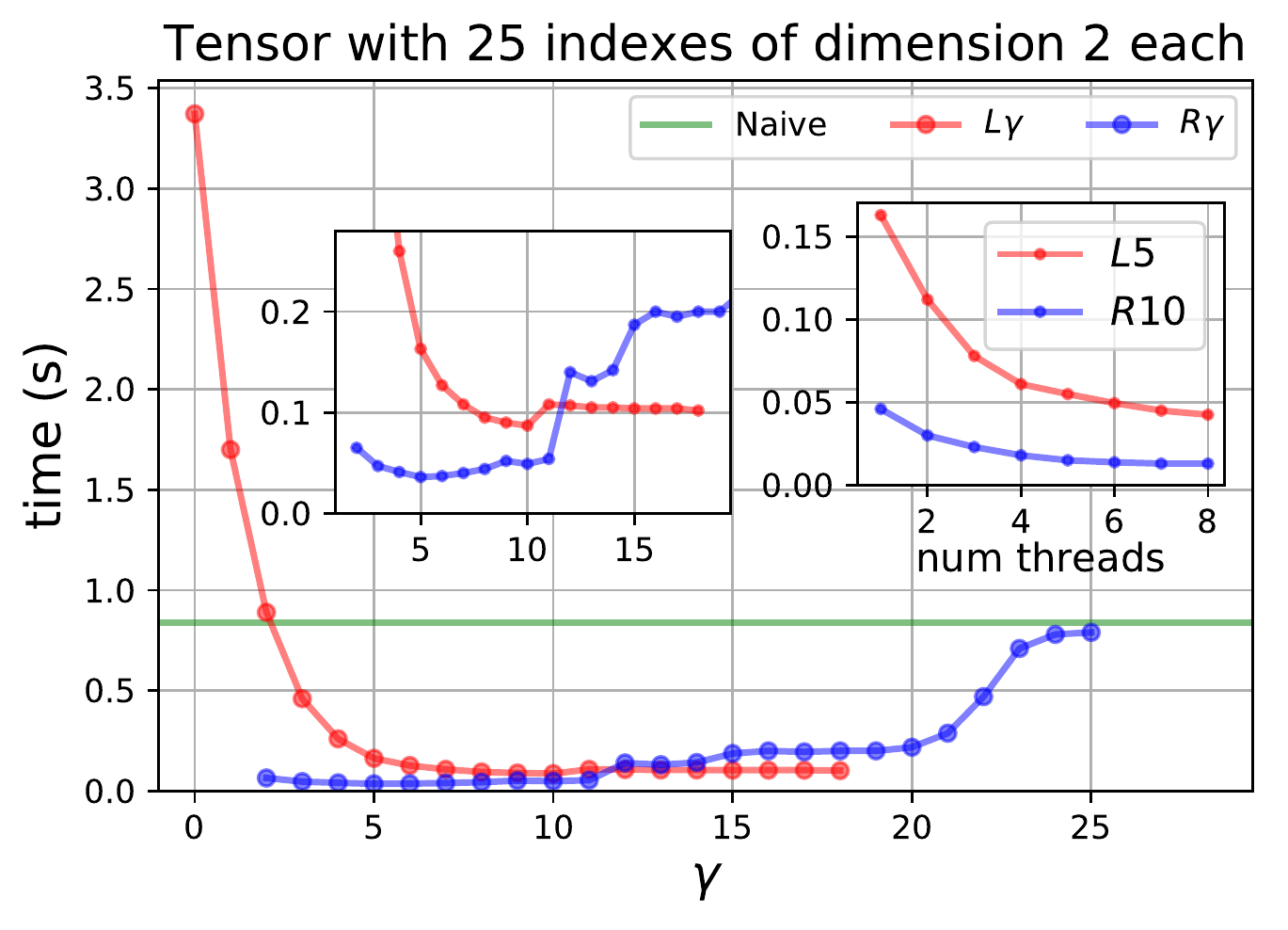}
\caption{\label{fig:LR} Single thread computation times on Broadwell nodes on
Pleiades for an arbitrary permutation of the indexes of a tensor of single
precision complex entries (and 25 indexes of dimension 2 each) following an
optimized, naive implementation of the reordering (green), an arbitrary
$L\gamma$ move (red), and an arbitrary $R\gamma$ move (blue).  The optimized,
naive approach performs comparably to python's numpy implementation of the
reordering.  Note that, for a wide range of $\gamma$, left and right moves are
very efficient.  \emph{Left inset:} zoomed version of the main plot.  For
$\gamma\in[5,10]$, both right and left moves are efficient.  \emph{Right inset:}
computation times for $L5$ and $R10$ (used in practice) as a function of the
number of threads used.}
\end{figure}

We implement the reorderings in a cache-efficient way by designing two
reordering routines that apply to two special index permutations.  Let us divide
a tensor's indexes into a left and a right group:
$A_{\underbrace{i_0,\ldots,i_{j}}\underbrace{i_{j+1},\ldots,i_{k}}}$.  If a
permutation involves only indexes in the left (right) group, then the
permutation is called a left (resp. right) move.  Let $\gamma$ be the number of
indexes in the right group.  We will denote left (resp. right) moves with
$\gamma$ indexes in the right group by $L\gamma$ (resp. $R\gamma$).  The
importance of these moves is that they are both cache-efficient for a wide range
of values of $\gamma$ and that an arbitrary permutation of the indexes of a
tensor can be decomposed into a small number of left and right moves, as will be
explained later in this section.  Let $d_\gamma$ be the dimension of all
$\gamma$ right indexes together. Then left moves involve the reordering across
groups of $d_\gamma$ entries of the tensor, where each group of $d_\gamma$
entries is contiguous in memory and is moved as a whole, without any reordering
within itself, therefore largely reducing the number of cache misses in the
routine.  On the other hand, right moves involve reordering within all of
$d_\gamma$ entries that are contiguous in memory, but involves no reordering
across groups, hence greatly reducing the number of cache misses, since all
reorderings take place in small contiguous chunks of memory.  Fig.~\ref{fig:LR}
shows the efficiency of $R\gamma$ and $L\gamma$ as compared to a naive (but
still optimized) implementation of the reordering that is comparable in
performance to python's numpy implementation.  A further advantage of the left
and right moves is that they can be parallelized over multiple threads and
remain cache-efficient in each of the threads. This allows for a very efficient
use of the computation resources, while the naive implementation does not
benefit from multi-threading. 

\begin{figure}[t] 
\includegraphics[width=1.00\columnwidth]{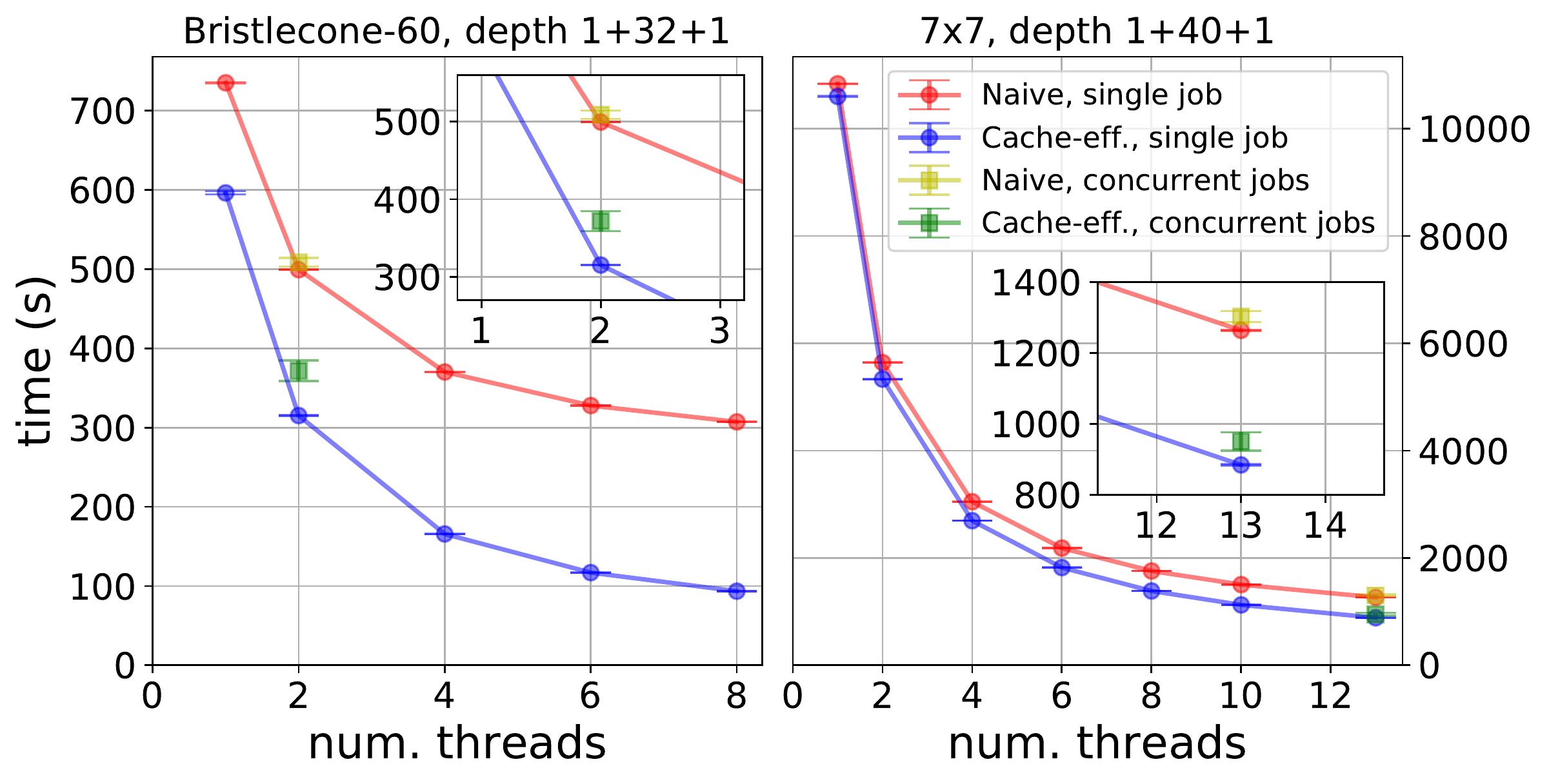}
\caption{\label{fig:reordering_advantage} Comparison of the runtime of the
computation of an amplitude using a naive index permutation implementation
(comparable to python's numpy implementation) and the cache-efficient
implementation described in Methods~\ref{sec:reorderings}, using Skylake nodes
on Electra. \emph{Left:} computation of three paths of an amplitude of
\Bristlecone{60} with depth $1+32+1$, corresponding to a target fidelity
$f=3/2^{12}$ (see Results). \emph{Right:} computation of three
paths of an amplitude of a grid of size $7\times 7$ with depth $1+40+1$,
corresponding to a target fidelity $f=3/2^{10}$ (see also
Results). Runtimes using a varying number of threads (one
threads per physical core) in an otherwise idle node are presented (single job),
as well as runtimes for the case where several amplitudes are computed on the
same node, fitting as many concurrent computations as possible (concurrent jobs,
also shown in the insets for clarity), a number that is constrained by memory
requirements; the latter is the case for the simulations presented in
Results. Cache-efficient runs show always better runtimes. The
scaling of runtime with number of threads is better for the cache-efficient
runs, given multithreading; note that the matrix-matrix multiplication part of
the contraction is always multithreaded. Concurrent jobs suffer from contention,
which is larger for the cache-efficient runs, presumably due to a larger use of
bandwidth; however, runtimes are still substantially improved in this case.}
\end{figure}

Let us introduce the decomposition of an arbitrary permutation into left and
right moves through an example.  Let $A_{abcdefg}$ be a tensor with 7 indexes of
dimension $d$ each.  Let $abcdefg\rightarrow cfeadgb$ be the index permutation
we wish to perform.  Furthermore, let us assume that it is known that $L2$ and
$R4$ are cache-efficient.  Let us also divide the list of 7 indexes of this
example in three groups: the last two (indexes 6 and 7), the next group of two
indexes from the right (indexes 4 and 5), and the remaining three indexes on the
left (1, 2, and 3). We now proceed as follows.  First, we apply an $L2$ move
that places all indexes in the left and middle groups that need to end up in the
rightmost group in the middle group; in our case this is index $b$, and the $L2$
we have in mind is $\underbrace{abc|de}_{L2}|fg\rightarrow
\underbrace{cae|bd}_{L2}|fg$; note that if the middle group is at least as big
as the rightmost group, then it is always possible to do this.  Second, we apply
an $R4$ move that places all indexes that need to end up in the rightmost group
in their final positions; in our case, that is
$cae|\underbrace{bd|fg}_{R4}\rightarrow cae|\underbrace{fd|gb}_{R4}$; note that,
if the first move was successful, then this one can always be done.  Finally, we
take a final $L2$ move that places all indexes in the leftmost and middle groups
in their final positions, \emph{i.e.}, $\underbrace{cae|fd}_{L2}|gb\rightarrow
\underbrace{cfe|ad}_{L2}|gb$.  We have decomposed the permutation into three
cache-efficient moves, $L\mu-R\nu-L\mu$, with $\mu=2, \nu=4$.  Note that it is
essential for this particular decomposition that the middle group has at least
the same dimension as the rightmost group.  It is also crucial that both $L\mu$
and $R\nu$ are cache-efficient.

In practice, we find that (beyond the above example, where $\mu=2$ and $\nu=4$)
for tensors with binary indexes, $\mu=5$ and $\nu=10$ are good choices for our
processors (see Fig.~\ref{fig:LR}).  If the tensor indexes are not binary, this
approach can be generalized: if all indexes have a dimension that is a power of
2, then mapping the reordering onto one involving explicitly binary indexes is
trivial; in the case where indexes are not all powers of 2, then different
values of $\mu$ and $\nu$ could be found, or decompositions more general than
$L\mu-R\nu-L\mu$ could be thought of.  In our case, we find good results for the
$L5-R10-L5$ decomposition.  Note also that in many cases a single $R$ or a
single $L$ move is sufficient, and sometimes a combination of only two of them
is enough, which can accelerate contractions by a large factor.\\

We apply a further optimization to our index permutation routines. A reordering
of tensor entries in memory (either a general one or some of $R\gamma$ or
$L\gamma$ moves) involves two procedures: generating a map between the old and
the new positions of each entry, which has size equal to the dimension of all
indexes involved, and applying the map to actually move the entries in memory.
The generation of the map takes a large part of the computation time, and so
storing maps that have already been used in a look-up table (memoization), in
order to reuse them in future reorderings, is a desirable technique to use.
While the size of such maps might make this approach impractical in general, for
left and right moves memoization becomes feasible, since the size of the maps is
now exponentially smaller than in the general case due to left and right moves
only involving a subset of indexes.  In the contraction of regular tensor
networks we work with maps reappear often, and so memoization proves very
useful.\\

The implementation of the decomposition of general permutations of indexes into
left and right moves, with all the details discussed above, give us speedups in
the contractions that range from under $5\%$ in single-threaded contractions
that are dominated by matrix multiplications, to well over $50\%$ in
multithreaded contractions that are dominated by reorderings. A detailed
comparison of runtimes using a naive approach and the cache-efficient approach
discussed is shown in Fig.~\ref{fig:reordering_advantage}. The contraction
corresponding to the simulation of \Bristlecone{60} with depth $1+32+1$ (presented
in Methods~\ref{sec:cutting}) is dominated by reorderings, since a large part of
the runtime is spent in building tensor A, which involves a large number of
``unbalanced'' contractions, where the reordering of a large tensor is followed
by the multiplication of a large matrix with a small one. The contraction
corresponding the a $7\times 7$ grid of depth $1+40+1$ (presented in
SI~A) is dominated by a few ``balanced'' contractions of
large tensors, and so the overall runtime is dominated by matrix-matrix
multiplication. In this case, larger speedups are still achieved using a large
number of threads, due to multithreading, and the speedup using a single threads
is appreciable but small; in practice, we use 13 threads per job on Skylake
nodes, where we can only fit three jobs per node due to memory constraints, and
a larger number of threads per job on other node types. Finally, it is worth
mentioning that runtimes are still robust when using cache-efficient,
multithreaded index permutations, showing small variation among runs.

\subsection{Fast sampling of bit-strings from low fidelity RQCs}
\label{sec:sampling}

While the computation of perfect fidelity amplitudes of output bit-strings of
RQCs is needed for the verification of quantum supremacy
experiments~\cite{boixo_characterizing_2018}, classically simulating sampling
from low fidelity RQCs is essential in order to benchmark the performance of
classical supercomputers in carrying out the same task as a noisy quantum
computer.  Indeed, present day quantum computers suffer from noise and errors in
each gate. In the commonly used digital error model~\cite{emerson_scalable_2005,
knill2008randomized, magesan_robust_2011, fowler2012surface,
barends_superconducting_2014, barends_digital_2015, boixo_characterizing_2018},
the total error probability for a RQC is the sum of the probability of error
from each gate. The fidelity, or probability of no errors, of a quantum computer
or of a classical algorithm for RQC sampling can be estimated with
XEB~\cite{boixo_characterizing_2018}. Therefore, we only require the same value
for the cross-entropy or fidelity in the classical algorithm as in the noisy
quantum computer. A superconducting quantum processor with present day
technology is expected to achieve a fidelity of around 0.5\% for circuits with
the number of gates considered here~\cite{boixo_characterizing_2018}.

In Methods~\ref{sec:low} we describe two methods to mimic the fidelity $f$ of
the output wave-function of the quantum computer with our simulator, providing a
speedup of a factor of $1/f$ to the simulation as compared to the computation of
exact amplitudes~\cite{markov_quantum_2018}. Both methods can be adjusted to
provide the same fidelity, and therefore the same cross
entropy~\cite{boixo_characterizing_2018}, as a noisy quantum computer. That is,
they result in an equivalent RQC sampling. In Methods~\ref{sec:fast} we describe
a way to reduce the computational cost of the sampling procedure on tensor
contraction type simulators by a factor of almost $10\times$, under reasonable
assumptions. Finally, in Methods~\ref{sec:no_convergence} we discuss the
implications of sampling from a Porter-Thomas distribution that has not fully
converged.

\subsubsection{Simulating low fidelity RQCs}
\label{sec:low}

Here, we describe two methods to reduce the computational cost of classically
sampling from an RQC given a target fidelity.

\label{sec:fraction_paths}
\noindent\textit{Summing a fraction of the paths ---}
This method, presented in Ref.~\cite{markov_quantum_2018}, exploits the fact
that, for RQCs, the decomposition of the output wave-function of a circuit into
paths $\ket{\psi} = \sum_{p\in\{paths\}} \ket{\psi_{p}}$ (see
Methods~\ref{sec:cutting}) leads to terms $\ket{\psi_p}$ that have similar norm and
that are almost orthogonal to each other. For this reason, summing only over a
fraction $f$ of the paths, one obtains a wave-function $\ket{\tilde{\psi}}$ with
norm $\braket{\tilde\psi}{\tilde\psi} = f$. Moreover, $\ket{\tilde{\psi}}$ has
fidelity $f$ as compared to $\ket{\psi}$, that is:
\begin{align}
\label{fidelity_1}
\left|\frac{\braket{\psi}{\tilde\psi}}{\sqrt{\braket{\tilde\psi}{\tilde\psi}}}\right|^2 = f.
\end{align}
Therefore, amplitudes of a fidelity $f$ wave-function can be computed at a cost
that is only a fraction $f$ of that of the perfect fidelity case.

\begin{figure*}[t]
\includegraphics[width=1.00\textwidth]{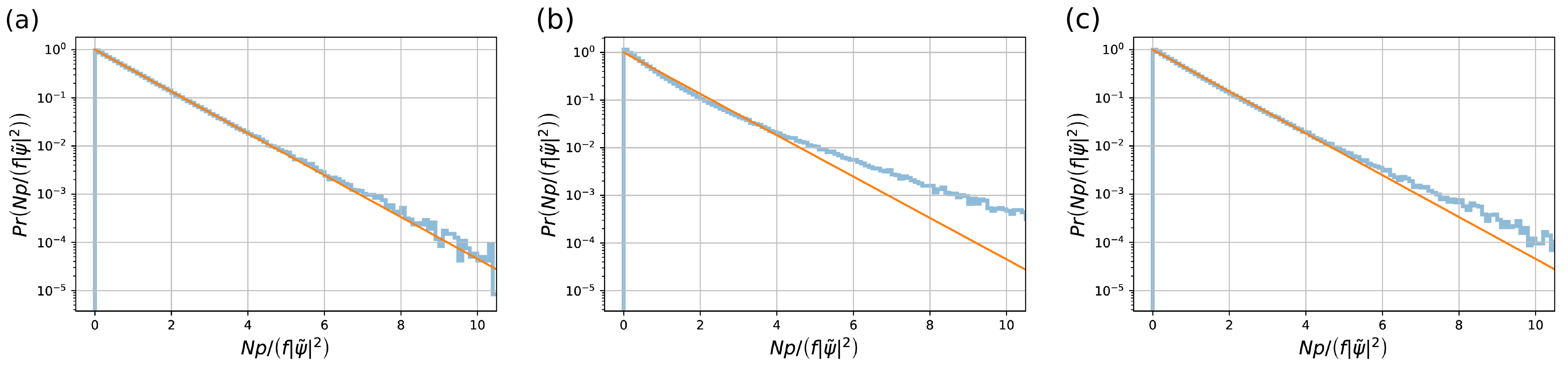}
\caption{\label{fig:PT_bris_x} Porter-Thomas distribution for the three
different sub-lattices of Bristlecone simulated with depth (1+32+1).
(a) \Bristlecone{64} (\rsim{1}); $1.2\times 10^6$ amplitudes with a
target fidelity $f=0.78\%$; $\braket{\tilde{\psi}}{\tilde{\psi}}=0.87f$.
(b) \Bristlecone{48} (\rsim{2}); $1.2\times 10^6$ amplitudes with
target fidelity $f=0.78\%$; $\braket{\tilde{\psi}}{\tilde{\psi}}=0.77f$.
(c) \Bristlecone{60} (\rsim{6}); $1.15\times 10^6$ amplitudes with a
target fidelity $f=0.51\%$; $\braket{\tilde{\psi}}{\tilde{\psi}}=1.38f$.  For
reference, the theoretical value of the Porter-Thomas distribution is plotted
(solid line).  For some simulations, the depth is not sufficient to fully
converge to a Porter-Thomas distribution.  Furthermore, summing a small number
of paths of low fidelity might lead to worse convergence than expected for a
particular depth (see Methods~\ref{sec:fraction_paths}).}
\end{figure*}

\begin{figure}[t!]
\includegraphics[width=1.00\columnwidth]{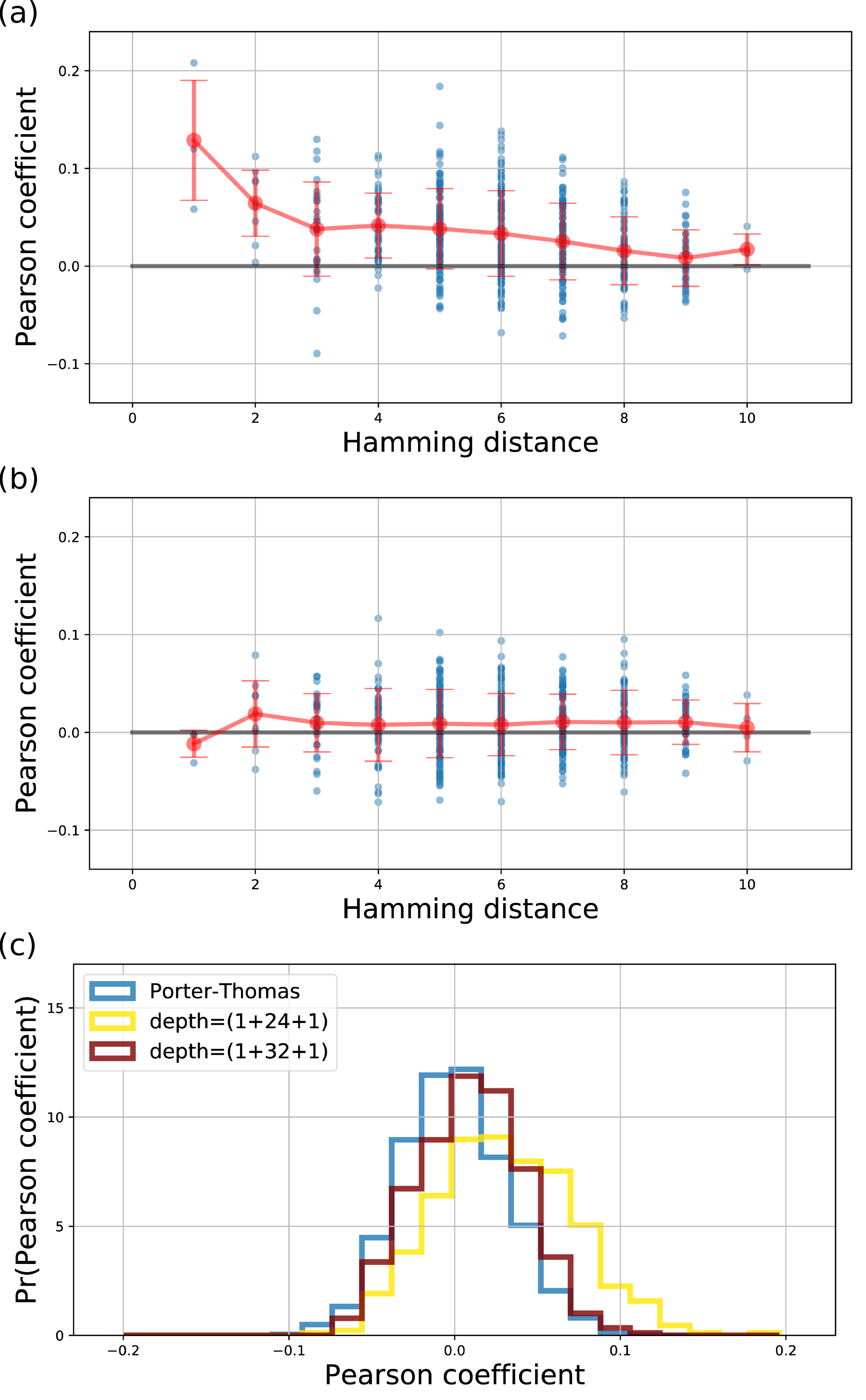}
\caption{\label{fig:pearson_vs_hamming} \emph{(a)-(b):} Pearson
coefficient as a function of Hamming distance for pairs generated at random on
subsystem $C$ of sub-lattice \Bristlecone{24}, for samples of size 1000 of random
strings on subsystem $A+B$.  All pairs between strings of a set of 32 random
strings on subsystem $C$ are considered.  The average and standard deviation
(error bars) for each Hamming distance is plotted with a solid line.  We can see
that, for depth (1+24+1) (a), the system has not thermalized, and the
correlation decreases with Hamming distance; for depth (1+32+1) (b),
correlations approach zero, and become Hamming distance independent (on
subsystem $C$).  \emph{(c)} we compare the distribution of Pearson
coefficients, obtained as described above, to the distribution of Pearson
coefficients obtained (numerically) from probability amplitudes with the same
sample size as in the simulations above, drawn from a Porter-Thomas
distribution. At large enough depth the system is expected to thermalize and the
two distributions match, meaning that the probability amplitudes obtained by
varying bit-strings only on subsystem $C$ are uncorrelated.}
\end{figure}

We find empirically that, while the different contributions $\ket{\psi_p}$
fulfill the orthogonality requirement (with a negligible overlap; \emph{e.g.},
in the \Bristlecone{60} simulation, the mutual fidelity between pairs out of 4096
paths is about $10^{-6}$), there is some non negligible variation in their norms
(see Results and Fig.~\ref{fig:PT_bris_x}), and thus the
fidelity achieved by $\ket{\psi_p}$ is equal to:
\begin{align}
\label{fidelity_2}
\left|\frac{\braket{\psi}{\psi_p}}{\sqrt{\braket{\psi_p}{\psi_p}}}\right|^2 = 
\braket{\psi_p}{\psi_p},
\end{align}
which is in general different than $(\# paths)^{-1}$.  If an extensive subset of
paths is summed over, then the variations on the norm and the fidelity are
suppressed, and the target fidelity is achieved.  This was the case in
Ref.~\cite{markov_quantum_2018}.  However, in this work we aim at minimizing the
number of cuts on the circuits, and so low fidelity simulations involve a small
number of paths (between 1 and 21 in the cases simulated).  In this case, some
``unlucky'' randomly selected paths might contribute with a fidelity that is
below the target, while others might achieve a higher fidelity than expected.

Finally, the low fidelity probability amplitudes reported in
Ref.~\cite{markov_quantum_2018}, obtained using the method described above,
follow a Porter-Thomas distribution as expected for perfect fidelity amplitudes.
Again, this is presumably true only in the case when a large number of paths is
considered.  In our case, we find distributions that have not fully converged to
a Porter-Thomas, but rather have a larger tail (see Results
and Fig.~\ref{fig:PT_bris_x}).  We attribute this phenomenon to the cuts in the
circuit acting as removed gates between qubits, thus increasing the effective
diameter of the circuit, which needs higher depth to fully thermalize.  We
discuss the implications of these tails for the sampling procedure in
Methods~\ref{sec:no_convergence}.

\label{sec:fraction_amplitudes}
\noindent\textit{Fraction of perfect fidelity amplitudes ---}
There exists a second method to simulate sampling from the output wave-function
$\ket{\psi}$ with a target fidelity $f$ that avoids summing over a fraction of
paths. 

The output density matrix of a random quantum circuit with fidelity $f$ can be
written as~\cite{boixo_characterizing_2018}
\begin{align}
\rho = f \ket{\psi}\bra{\psi} + (1-f) \frac{\openone}{N}\;.
\end{align}
This means that to produce a sample with fidelity $f$ we can sample from the
exact wave-function $\ket{\psi}$ with probability $f$ or produce a random
bit-string with probability $1-f$. The sample from the exact wave-function can
be simulated by calculating the required number of amplitudes with perfect
fidelity.

Note that the method presented in this section involves the computation of the
same number of paths as the one described in Methods~\ref{sec:fraction_paths}
for a given $f$, circuit topology, circuit depth, and set of cuts.  However,
this second method is more robust in achieving a target fidelity. Note that by
this argument the 6000 amplitudes of \rsim{5} are equivalent to 1.2M amplitudes at
0.5\% fidelity.

Note also that, even though this method and the one presented in
Methods~\ref{sec:fraction_paths} have the same computational cost for tensor
network based simulators, for Schr\"odinger-Feynman simulators like the one
presented in Ref.~\cite{markov_quantum_2018} it is preferable to consider a
fraction of paths as opposed to a fraction of perfect fidelity amplitudes. This
is due to the small cost overhead of computing an arbitrary number of amplitudes
using these simulators.

\subsubsection{Fast sampling technique}
\label{sec:fast}

While $10^6$ sampled amplitudes are necessary for cross entropy verification of
the sampling task~\cite{boixo_characterizing_2018}, the \emph{frugal rejection
sampling} proposed in Ref.~\cite{markov_quantum_2018} needs the numerical
computation of $10\times 10^6=10^7$ amplitudes in order to carry out the correct
sampling on a classical supercomputer.  This is due to the acceptance of $1/M$
amplitudes (on average) of the rejection sampling, where $M=10$ when sampling
from a given Porter-Thomas distribution with statistical distance $\epsilon$ of
the order of $10^{-4}$ (negligible).

In this section, we propose a method to effectively remove the $10\cross$
overhead in the sampling procedure for tensor network based simulators, which
normally compute one amplitude at a time. For the sake of clarity, we tailor the
proposed fast sampling technique to the Bristlecone architecture.  However, it
can be straightforwardly generalized to different architectures (see SI~A).
Given the two regions of the Bristlecone (and sub-lattices) $AB$ and $C$ of
Fig.~\ref{fig:bris_x_C}, and the contraction proposed (see
Methods~\ref{sec:cutting}), the construction of tensor $C$ and its subsequent
contraction with $AB$ are computationally efficient tasks done in a small amount
of time as compared to the full computation of the particular path.  This
implies that one can compute, for a given output bit-string on $AB$, $s_{AB}$, a
set of $2^{12}$ amplitudes generated by the concatenation of $s_{AB}$ with all
possible $s_C$ bit-strings on $C$ at a small overhead cost per amplitude.  We
call this set of amplitudes a ``batch'', we denote its size by $N_C$, and each
of the (concatenated) bit-strings by $s_{ABC}$.  In practice, we find that for
the \Bristlecone{64} and \mbox{-60} with depth (1+32+1), the computation of a
batch of 30 amplitudes is only around 10\% more expensive than the computation
of a single amplitude, while for the \Bristlecone{48} and \mbox{-70} with depth
(1+32+1), the computation of a batch of 256 amplitudes is around 15\% more
expensive than the computation of a single amplitude, instead of a theoretical
overhead of $30\times$ and $256\times$, respectively.

The sampling procedure we propose is a modification of the frugal rejection
sampling presented in Ref.~\cite{markov_quantum_2018} and proceeds as follows.
First, we choose slightly over $10^6$ (see below) random bit-strings on $AB$,
$s_{AB}$. For each $s_{AB}$, we choose $N_C$ bit-strings on $C$, $s_C$, at
random (without repetition).  We then compute the probability amplitudes
corresponding to all $s_{ABC}$ bit-strings on all (slightly over) $10^6$
batches.  We now shuffle each batch of bit-strings.  For each batch, we proceed
onto the bit-strings in the order given by the shuffle; we accept a bit-string
$s_{ABC}$ with probability $\min\left[1, p(s_{ABC})N/M\right]$, where
$p(s_{ABC})$ is the probability amplitude of $s_{ABC}$, and $N$ is the dimension
of the Hilbert space; once a bit-string is accepted, or the bit-strings of the
batch have been exhausted without acceptance, we proceed to the next batch.  By
accepting at most one bit-string per batch we avoid introducing spurious
correlations in the final sample of bit-strings.

Given an $M$ and a batch size $N_C$, the probability that a bit-string is
accepted from a batch is (on average) $1-(1-1/M)^{N_C}$.  For $M=10$ and
$N_C=30$, the probability of acceptance in a batch is $95.76\%$, and one would
need to compute amplitudes for $1.045\times 10^6$ batches in order to sample
$10^6$ bit-strings; for $M=10$ and $N_C=60$, the probability goes up to
$99.82\%$, and one only needs $1.002\times 10^6$ batches; for $M=10$ and
$N_C=256$, the probability of acceptance is virtually $100\%$, and $1.00\times
10^6$ batches are sufficient.  There is an optimal point, given by the overhead
in computing batches of different sizes $N_C$ and the probability of accepting a
bit-string on a batch given $N_C$, that minimizes the runtime of the algorithm.

There is a crucial condition for this sampling algorithm to work, namely the
absence of correlations between the probability amplitudes of the bit-strings
$\{s_{ABC}\}$ for fixed $s_{AB}$, so that they are indistinguishable from
probability amplitudes taken from random bit-strings over $ABC$.  We expect this
condition to be satisfied for chaotic systems that have converged to a
Porter-Thomas distribution.  In order to test this, we perform the following
test: for \Bristlecone{24}, we choose 1000 bit-strings over $AB$ ($s_{AB}$) at
random and for each of them we generate a batch of size $N_C=32$, where we
constrain the bit-strings $\{s_C\}$ to be the same across batches.  We now
compute the Pearson correlation coefficient between the two sets of 1000
amplitudes gotten for each pair of bit-strings in $C$, and we do this for all
$32\times 31/2$ pairs.  If the probability amplitudes of each batch are really
uncorrelated to each other, we expect the correlation coefficient to vanish.  We
show the coefficient as a function of Hamming distance between the pairs in
Fig.~\ref{fig:pearson_vs_hamming} (a) and (b). We can see that, for depth
(1+24+1) (a) there is a small but non negligible correlation, which in fact
decreases on average with Hamming distance.  For depth (1+32+1) (b), the
correlation is Hamming distance independent and approaches zero. In
Fig.~\ref{fig:pearson_vs_hamming} (c) we compare the distribution of
Pearson coefficients obtained for both depths analyzed to that one obtained from
pairs of sets of size 1000 sampled from a Porter-Thomas distribution.  While a
fairer comparison would involve  sampling from the distribution of the output
wave-function of the RQC, which might differ from the Porter-Thomas in the
absence of convergence, we still see a clear tendency of the distributions to
match for longer depth, \emph{i.e.}, closer to convergence.

\subsubsection{Sampling from a non fully-thermalized Porter-Thomas distribution}
\label{sec:no_convergence}

In Ref.~\cite{markov_quantum_2018} an error of the order of $10^{-4}$ is
computed for a frugal rejection sampling with $M=10$, assuming Porter-Thomas
statistics.  When the distribution has not converged to a Porter-Thomas, but
rather has a larger tail, we expect the error to increase.
We can estimate the
error in sampling numerically for the cases simulated here as the sum of the
probability amplitudes larger than $M/N$ with $N$ being the dimension of the
Hilbert space, multiplied by $N$ and divided by the number of amplitudes computed.
For $M=10$, we estimate an error $\epsilon=9.3\times 10^{-4}$
for \rsim1, $\epsilon=1.0\times 10^{-2}$ for \rsim2, and $\epsilon=2.5\times
10^{-3}$ for \rsim6, respectively.  If instead we consider $M=15$, this lowers
the error to $\epsilon=1.3\times 10^{-5}$ for \rsim1, $\epsilon=1.0\times
10^{-3}$ for \rsim2, and $\epsilon=1.15\times 10^{-4}$ for \rsim6, respectively.
Increasing $M$, in order to reduce the error in the frugal sampling, implies a
lower acceptance rate in the fast sampling, which is resolved by increasing the
size of the batches $N_C$, which is done at a small cost.

\subsection{Simulation of Bristlecone compared to rectangular grids}
\label{sec:hardness}

The diamond shape of the topology of Bristlecone and its hard sub-lattices (see
Fig.~\ref{fig:bris_x}: \Bristlecone{24}, \mbox{-30}, \mbox{-40}, \mbox{-48},
\mbox{-60}, and \mbox{-70}) makes them particularly hard to simulate classically
when compared to rectangular grids of the same (or smaller) number of qubits.
Indeed, these lattices are subsets of large rectangular grids, from which they
inherit their diameter; \emph{e.g.}, \Bristlecone{70} is a sublattice of a
$10\times 11$ grid.  When cutting the lattice (see Methods~\ref{sec:cutting}),
one has to apply several cuts in order to decrease the maximum size of the
tensors in the contraction to manageable sizes; in the case of \Bristlecone{70}
and depth (1+32+1), four cuts are needed in order to have tensors in the
contraction of at most dimension $2^{7\times 4}$, while for a rectangular
$8\times 9$ lattice (with $72$ qubits) only 3 cuts are needed.  Note that the
computational cost scales with the dimension of the indexes cut, \emph{i.e.},
exponentially with the number of cuts.

The same applies to a simulator based on a full split of the circuit into two
parts, as in Refs.~\cite{harrow_quantum_2017,chen_64-qubit_2018,markov_quantum_2018}.
For instance, the number of ${\rm CZ}$ gates for RQCs with depth
(1+32+1) which are cut when splitting \Bristlecone{60} in two halves is equal to
$40$. In comparison, $8\times8$ grids of qubits with the same depth have only
$32$ ${\rm CZ}$ gates cut. See SI~C for more details.

As was discussed in Methods~\ref{sec:simulator}, identifying topologies that are
hard to simulate classically, but that minimize the number of qubits involved,
increases the chances of success success of quantum supremacy experiments, due
to the decrease of the overall fidelity of the quantum computer with the number
of gates and qubits~\cite{boixo_characterizing_2018}.  For this reason, we find
that Bristlecone is a good setup for quantum supremacy experiments.

\section{Data Availability}

The simulation data that support the findings of this study are available
at~\url{https://data.nas.nasa.gov/quail/data.php?dir=/quaildata/quantum/qcSim}.
The circuit files used for the numerical simulations in this paper are publicly
available in~\cite{instances}.

\begin{acknowledgments}
The authors would like to thank Edward Farhi, Bryan A. O'Gorman, Alan Ho, Sergei
V. Isakov, Norman M. Tubman and Dmitry Lyakh for enlightening discussions and
Igor L. Markov for reviewing the manuscript. The authors also thank Orion
Martin, Alan Kao and David Yonge-Mallo for helping open sourcing qFlex code. We
thank the authors of Ref.~\cite{chen_classical_2018} for sharing the instances
used in that work. We are grateful for support from NASA Ames Research Center,
from the NASA Advanced Exploration Systems (AES) program, and the NASA
Transformative Aeronautics Concepts Program (TACP).  We also appreciate support
from the AFRL Information Directorate under grant number F4HBKC4162G001. We
acknowledge support of the NASA Advanced Division for providing access to the
NASA HPC systems, Pleiades and Electra, during dedicated downtime.\\

The views and conclusions contained herein are those of the authors and should
not be interpreted as necessarily representing the official policies or
endorsements, either expressed or implied, of the U.S.  Government. The U.S.
Government is authorized to reproduce and distribute reprints for governmental
purpose notwithstanding any copyright annotation thereon.
\end{acknowledgments}

\section{Author Contribution}
\label{sec:contrib}

B.V., S.B. and S.M. designed qFlex and the study; B.V, S.B and S.M devised new
improvements for approximated sampling; B.V. wrote the original qFlex code and
B.V, B.N. and C.H. further optimized it; B.V, B.N and C.H. ran simulations on
NASA Pleiades and Electra; B.V., S.B., E.R., R.B and S.M. performed the analysis
of data; all the authors contributed to the discussions and wrote the paper.

\begin{table*}
\centering
\begin{tabular}{ |c||c|c|c|c|ccccc| }
 \hline
 \multirow{2}{*}{\textbf{Run}} & \multirow{2}{*}{\textbf{Circuit}} & \multirow{2}{*}{\textbf{Depth}} & \multirow{2}{*}{\textbf{Paths per instance / total paths}} & \multirow{2}{*}{\textbf{$N_C$}} & \multicolumn{5}{c|}{\textbf{Cores per instance / cores per node}} \tabularnewline
  &  &  &  &  & \textbf{bro} & \textbf{has} & \textbf{ivy} & \textbf{san} & \textbf{sky (Electra)} \tabularnewline
 \hline
 \hline
 1 & Bris.-64  & (1+32+1) & $2/2^8$ & 30 & - & - & - & - & 2/40 \tabularnewline
 2 & Bris.-48 & (1+32+1) & $2/2^8$ & 30 & - & - & - & - & 2/40 \tabularnewline
 2b* & Bris.-48 & (1+32+1) & $2/2^8$ & 256 & 2/28 & 2/24 & 2/20 & 4/16 & 2/40 \tabularnewline
 3 & Bris.-70 & (1+32+1) & $1/2^{16}$ & 512 & & 2/24 & 4/20 & 8/16 & 2/40 \tabularnewline
 3b* & Bris.-70 & (1+32+1) & $1/2^{16}$ & 256 & 2/28 & 2/24 & 2/20 & 4/16 & 2/40 \tabularnewline
 4 & Bris.-70 & (1+24+1) & $1/2^3$ & 62 & - & 8/24 & 20/20 & - & - \tabularnewline
 5 & Bris.-70 & (1+24+1) & $2^3/2^3$ & 62 & - & 8/24 & 20/20 & - & - \tabularnewline
 6 & Bris.-60 & (1+32+1) & $21/2^{12}$ & 30 & 2/28 & 2/24 & 4/20 & 8/16 & 2/40 \tabularnewline
 \hline
\end{tabular}\\
\vspace{0.4cm}
\begin{tabular}{ |c||c|ccccc| }
 \hline
 \multirow{2}{*}{\textbf{Run}} & \multirow{2}{*}{\textbf{Memory footprint (GB)}} & \multicolumn{5}{c|}{\textbf{Num. instances per node}} \tabularnewline
  &  & \textbf{bro} & \textbf{has} & \textbf{ivy} & \textbf{san} & \textbf{sky (Electra)} \tabularnewline
 \hline
 \hline
 1 & 10 & - & - & - & - & 19 \tabularnewline
 2 & 10 & - & - & - & - & 19 \tabularnewline
 2b* & 7 & 14 & 12 & 8 & 4 & 20 \tabularnewline
 3 & 11 & - & 11 & 5 & 2 & 16 \tabularnewline
 3b* & 7 & 14 & 12 & 8 & 4 & 20 \tabularnewline
 4 & 36 & - & 3 & 1 & - & - \tabularnewline
 5 & 36 & - & 3 & 1 & - & - \tabularnewline
 6 & 10 & 11 & 11 & 5 & 2 & 17 \tabularnewline
 \hline
\end{tabular}\\
\vspace{0.4cm}
\begin{tabular}{ |c||ccccc| }
 \hline
 
 \multirow{2}{*}{\textbf{Run}} & \multicolumn{5}{c|}{\textbf{Runtime (s)}} \tabularnewline
  & \textbf{bro} & \textbf{has} & \textbf{ivy} & \textbf{san} & \textbf{sky (Electra)} \tabularnewline
 \hline
 \hline
 1 & - & - & - & - & $335.8\pm 12.2$ \tabularnewline
 2 & - & - & - & - & $246.0\pm 9.4$ \tabularnewline
 2b* & $204.9\pm 6.5$ & $215.6\pm 2.6$ & $256.9\pm 0.5$ & $149.9\pm 0.1$ & $150.0\pm 3.5$ \tabularnewline
 3 & - & $1034.5\pm 91.3$ & $700.9\pm 32.8$ & $386.3\pm 4.0$ & $708.4\pm 79.9$ \tabularnewline
 3b* & $156.3\pm 4.3$ & $161.7\pm 1.7$ & $190.9\pm 1.7$ & $113.3\pm 0.3$ & $113.2\pm 2.8$ \tabularnewline
 4 & - & $226.5\pm 15.8$ & $126.8\pm 5.2$ & - & - \tabularnewline
 5 & - & $1683.2\pm 111.1$ & $885.3\pm 4.4$ & - & - \tabularnewline
 6 & $4555.4\pm 311.7$ & $5092.2\pm 311.7$ & $3606.5\pm 159.8$ & $2040.9\pm 18.9$ & $3054.0\pm 230.3$ \tabularnewline
 \hline
\end{tabular}
\caption{\label{table:times} Number of paths per instance, size of batches of
amplitudes (see Methods~\ref{sec:fast}), number of cores (threads) used per
instance, memory footprint, number of instances fit in a node, and runtime per
instance for all six cases run and for all five node types used on NASA Pleiades
and Electra HPC clusters.  We report single instances of a run, where an
instance corresponds to the computation of a number of paths given a cut
prescription, and the computation of a batch of $N_C$ amplitudes corresponding
to output bit-strings chosen at random over subsystem $C$ (see
Methods~\ref{sec:fast}).  Note that for \rsim3 $N_C=512$, and so computing $N_C$
amplitudes takes about three times the time of computing a single one.  However,
this is strongly mitigated with the contraction used for \rsim{3b}.  Note also
that for \rsim6 we ran 17 jobs per Skylake node, instead of 19, as a
conservative strategy to stay well below the total memory available on these
nodes and hence avoid any unwanted crash in our largest simulation.  Instances
can be collected for a large number of low fidelity amplitudes or for a smaller
number of high fidelity amplitudes at the same computational cost.  *\rsim{2b}
and \rsim{3b} refer to the benchmark of the contraction procedure introduced in
Methods~\ref{sec:cutting} for \Bristlecone{48} and \Bristlecone{70}, respectively;
\rsim{1} and \rsim{3} were run were run using a less performing procedure,
similar to the one used for \Bristlecone{60} (see SI~B).}
\end{table*}

\begin{table*}
\centering
\begin{tabular}{ |c|c|c|ccccc| }
 \hline
 \multirow{2}{*}{\textbf{Circuit}} & \multirow{2}{*}{\textbf{Depth}} & \multirow{2}{*}{\textbf{Fidelity (\%)}} & \multicolumn{5}{c|}{\textbf{Runtime $\times$ num. cores (h)}} \\
  &  &  & \textbf{bro} & \textbf{has} & \textbf{ivy} & \textbf{san} & \textbf{sky (Electra)} \tabularnewline
 \hline
 \hline
 Bris-48 & (1+32+1) & 0.78\% & $1.14\times 10^5$ & $1.20\times 10^5$ & $1.78\times 10^5$ & $1.67\times 10^5$ & $8.33\times 10^{4}$ \tabularnewline
 Bris-64 & (1+32+1) & 0.78\% & - & - & - & - & $1.96\times 10^{5}$ \tabularnewline
 Bris-60 & (1+32+1) & 0.51\% & $3.22\times 10^6$ & $3.09\times 10^6$ & $4.01\times 10^6$ & $4.54\times 10^6$ & $2.00\times 10^6$ \tabularnewline
 Bris-70 & (1+24+1) & 0.50\% & - & $1.87\times 10^4$ & $2.46\times 10^4$ & - & - \tabularnewline
 Bris-70 & (1+32+1) & 0.50\% & $2.85\times 10^7$ & $2.95\times 10^7$ & $4.35\times 10^7$ & $4.13\times 10^7$ & $2.06\times 10^7$ \tabularnewline
 \hline
\end{tabular}
\caption{\label{table:times_sampling} Estimated effective runtimes on a single
core for the computation of $10^6$ amplitudes with a target fidelity of about
$0.5\%$ for the Bristlecone sub-lattices (see Fig.~\ref{fig:bris_x} for
nomenclature).  This is an estimate of the computational cost for the completion
of the RQC sampling task. The estimate is based on the runtimes for single
instances presented in Table~\ref{table:times}. For \Bristlecone{70} with depth
(1+32+1), \rsim{3b} is used, since it offers better performance than \rsim{3}.} 
\end{table*}

\begin{table*}
\centering
\begin{tabular}{|c|c|cc|cc|}
\hline 
\multirow{2}{*}{\textbf{Circuit size}} & \multirow{2}{*}{\textbf{Fidelity (\%)}} & \multicolumn{2}{c|}{\textbf{Runtime (hours)}} & \multicolumn{2}{c|}{\textbf{Energy cost (MWh)}} \tabularnewline
 &  & \textbf{Pleiades} & \textbf{Electra} & \textbf{Pleiades} & \textbf{Electra} \tabularnewline
\hline 
\hline 
$7\times7\times(1+40+1)$  & 100  & $1.22\times10^{-2}$  & $1.16\times10^{-2}$  & $5.35\times10^{-2}$  & $1.89\times10^{-2}$ \tabularnewline
$8\times8\times(1+32+1)$  & 100  & $1.77\times10^{-4}$ & $2.04\times10^{-4}$  & $8.86\times10^{-4}$  & $3.34\times10^{-4}$ \tabularnewline
$8\times8\times(1+40+1)$  & 100  & $2.03$  & $2.12$  & $8.88$  & $3.48$\tabularnewline
$8\times9\times(1+32+1)$  & 100  & $2.90\times10^{-3}$  & $2.98\times10^{-3}$  & $1.45\times10^{-2}$  & $4.89\times10^{-3}$ \tabularnewline
Bris.-70 $\times(1+32+1)$ & 100  & $2.89\times10^{-2}$  & $3.57\times10^{-2}$  & $1.45\times10^{-1}$  & $5.85\times10^{-2}$ \tabularnewline
\hline 
\end{tabular}
\caption{\label{table:verification} Estimated effective runtimes and energy cost
for the computation of a single amplitude with perfect fidelity on NASA HPC
Pleiades and Electra systems.  Note that for the $7\times 7\times (1+40+1)$ and
$8\times 8\times (1+40+1)$ grids, jobs do not fit in Sandy Bridge nodes, due to
their memory requirements; for that reason, the portion of Pleiades with Sandy
Bridge nodes is not considered, and the energy cost estimations of these two
cases do not include those nodes.}
\end{table*}

\begin{table*}
\centering
\begin{tabular}{|c|c|cc|cc|}
\hline 
\multirow{2}{*}{\textbf{Circuit size}} & \multirow{2}{*}{\textbf{Target fidelity (\%)}} & \multicolumn{2}{c|}{\textbf{Runtime (hours)}} & \multicolumn{2}{c|}{\textbf{Energy cost (MWh)}} \tabularnewline
 &  & \textbf{Pleiades}  & \textbf{Electra}  & \textbf{Pleiades}  & \textbf{Electra} \tabularnewline
\hline 
\hline 
$7\times7\times(1+40+1)$  & 0.51 & 62.4  & 59.0  & $2.73\times10^{2}$  & 96.8 \tabularnewline
$8\times8\times(1+32+1)$  & 0.78 & 1.38  & 1.59  & 6.91 & 2.61 \tabularnewline
$8\times8\times(1+40+1)$  & 0.58 & $1.18\times10^{4}$  & $1.23\times10^{4}$  & $5.15\times10^{4}$  & $2.02\times10^{4}$ \tabularnewline
$8\times9\times(1+32+1)$  & 0.51 & 14.8  & 15.2  & 73.9  & 24.9 \tabularnewline
Bris.-70 $\times(1+32+1)$  & 0.50 & 145  & 178  & 723  & 293 \tabularnewline
\hline 
\end{tabular}
\caption{\label{table:classical} Estimated runtimes and energy cost for the
computation of $10^6$ amplitudes with fidelity close to $0.5\%$ on NASA HPC
Pleiades and Electra systems.  Note that for the $7\times 7\times (1+40+1)$ and
$8\times 8\times (1+40+1)$ grids, jobs do not fit in Sandy Bridge nodes, due to
their memory requirements; for that reason, the portion of Pleiades with Sandy
Bridge nodes is not considered, and the energy cost estimations of these two
cases do not include those nodes.} 
\end{table*}

\begin{table*}
\centering
\begin{tabular}{|c|c|c|cc|cc|}
\hline 
\multirow{2}{*}{\textbf{Circuit size}} & \multirow{2}{*}{\textbf{Targ. fidelity (\%)}} & \multirow{2}{*}{\textbf{Num. amps.}} & \multicolumn{2}{c|}{\textbf{Runtime (hours)}} & \multicolumn{2}{c|}{\textbf{Energy cost (MWh)}} \tabularnewline
 &  &  & \textbf{MFIB}  & \textbf{Electra (sky)}  & \textbf{MFIB}  & \textbf{Electra (sky)} \tabularnewline
\hline 
\hline 
 $7\times 7 \times (1+40+1)$ & 0.51 & 1 & $4.96\times 10^{5}$ & $6.63$ & $6.46$ & $8.64\times 10^{-5}$ \tabularnewline
 $7\times 7 \times (1+40+1)$ & 0.51 & $10^5$ & $5.05\times 10^{5}$ & $6.63\times 10^{5}$ & $6.58$ & $8.64$ \tabularnewline
 $7\times 7 \times (1+40+1)$ & 0.51 & $10^6$ & $5.82\times 10^{5}$ & $6.63\times 10^{6}$ & $7.58$ & $86.4$ \tabularnewline
 $7\times 7 \times (1+40+1)$ & 100.0 & 1 & $9.73\times 10^7$ & $1.30\times 10^{3}$ & $1.27\times 10^3$ & $1.69\times 10^{-2}$ \tabularnewline
 $7\times 7 \times (1+40+1)$ & 100.0 & $10^5$ & $9.90\times 10^7$ & $1.30\times 10^{8}$ & $1.29\times 10^3$ & $1.69\times 10^{3}$ \tabularnewline
 $7\times 7 \times (1+40+1)$ & 100.0 & $10^6$ & $1.14\times 10^8$ & $1.30\times 10^{9}$ & $1.49\times 10^3$ & $1.69\times 10^{4}$ \tabularnewline
 \hline
\end{tabular}
\caption{\label{table:igor} 
Estimated runtime and energy cost consumption to compute the specified number of
amplitudes for our simulator on a single processor of the Skylake nodes portion
of the NASA Electra system, compared to Ref.~\cite{markov_quantum_2018} (MFIB).
The energy cost for the MFIB simulations is estimated assuming the same power
consumption per core as the Skylake nodes. In Ref.~\cite{markov_quantum_2018},
the authors report a number of cores $P=625\times 16=10^5$, since they use 625
nodes of 16 cores (32 vCPUs or hyper-threads) each.  For ours, $P=2304\times
40=$ \num{92160} on the Skylake nodes of Electra (note that we consider 40 cores
per node, even though we use only 39 in practice for the $7\times 7\times
(1+40+1)$ simulations; this is due to the ability of modern Intel processors to
``up-clock'' their CPUs in favorable conditions (known as Dynamic Frequency
Scaling), thus consuming a similar amount of energy and achieving a similar
performance as in the case where there are no idle cores.  Note that MFIB's
approach has the advantage to compute a large number of of amplitudes with a
small cost overhead. On the contrary our approach performs much better in the
computation of a smaller subset of amplitudes; both methods use comparable
resources when computing about $10^5$ amplitudes.  The MFIB algorithm becomes
less efficient than our algorithm as the number of qubits grows because of
memory requirements.}
\end{table*}

\clearpage

\appendix
\graphicspath{{figures/si/}}
\renewcommand\thesection{\Alph{section}}
\renewcommand\theequation{\thesection\arabic{equation}}
\renewcommand\thefigure{\thesection\arabic{figure}}
\renewcommand\thetable{\thesection\arabic{table}}
\setcounter{equation}{0}
\setcounter{figure}{0}
\setcounter{table}{0}

\section{Contraction of RQCs on rectangular grids}
\label{sec:grids}

\begin{figure}[t]
\includegraphics[width=0.32\columnwidth]{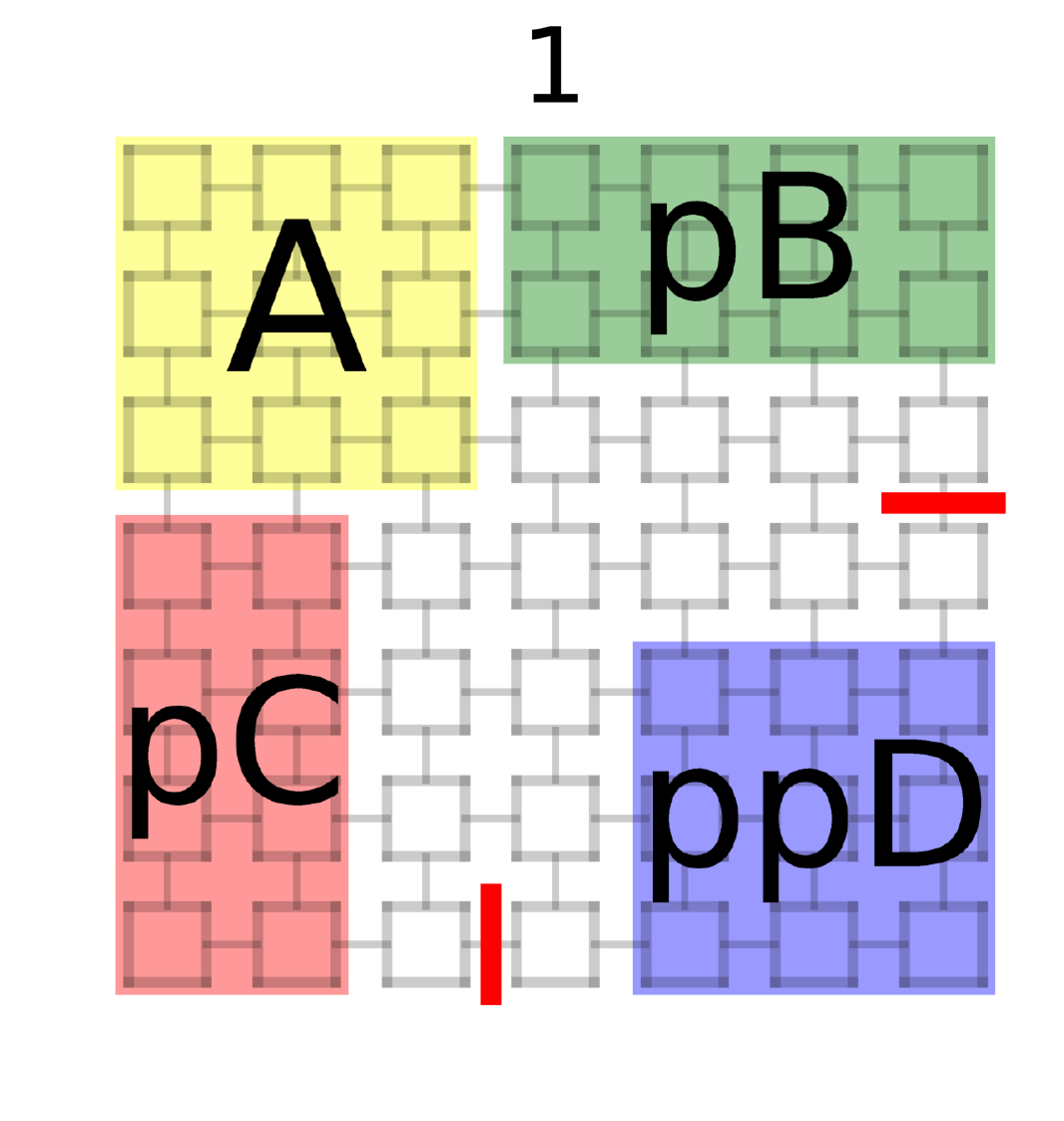}
\includegraphics[width=0.32\columnwidth]{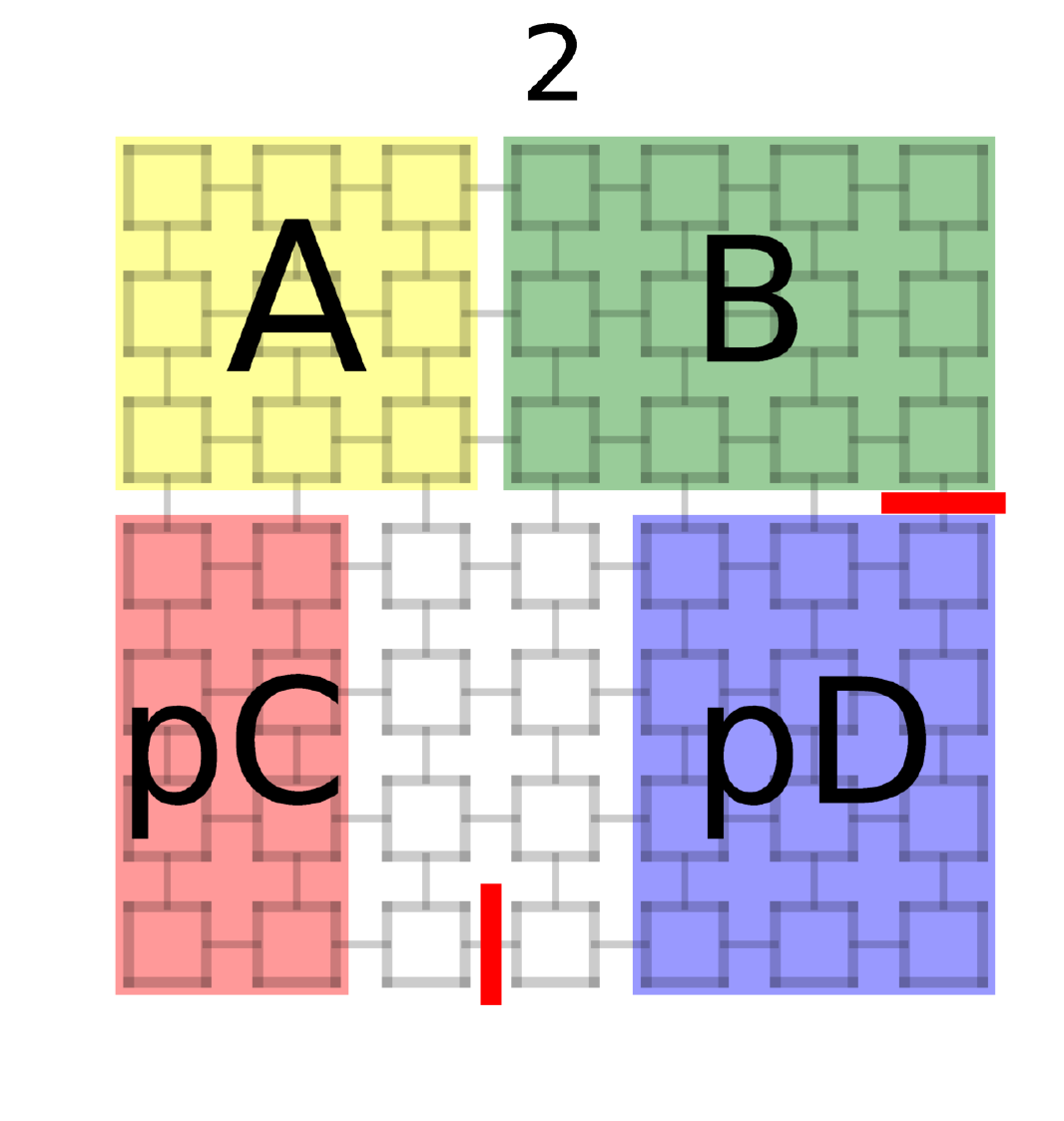}
\includegraphics[width=0.32\columnwidth]{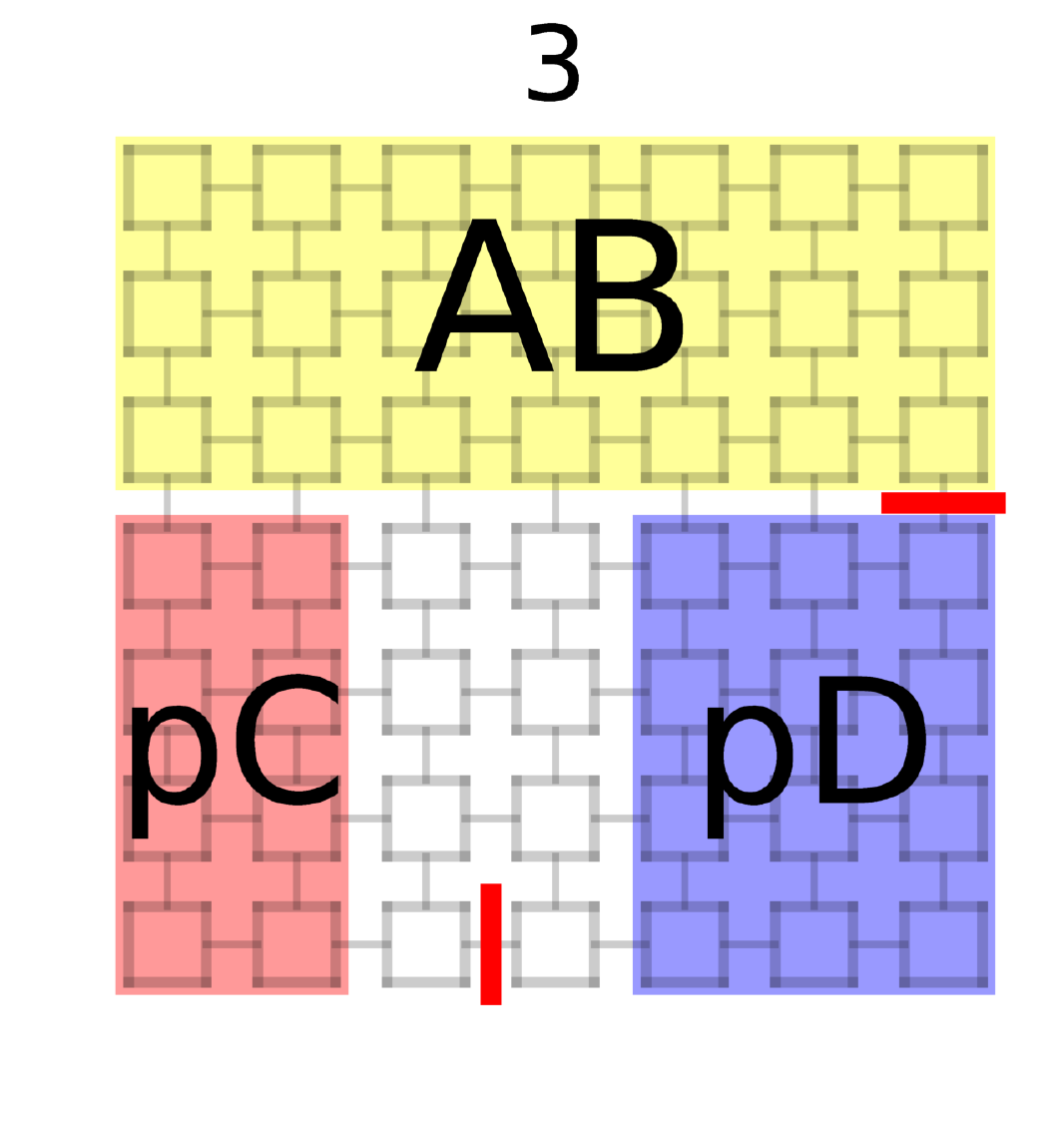}
\includegraphics[width=0.32\columnwidth]{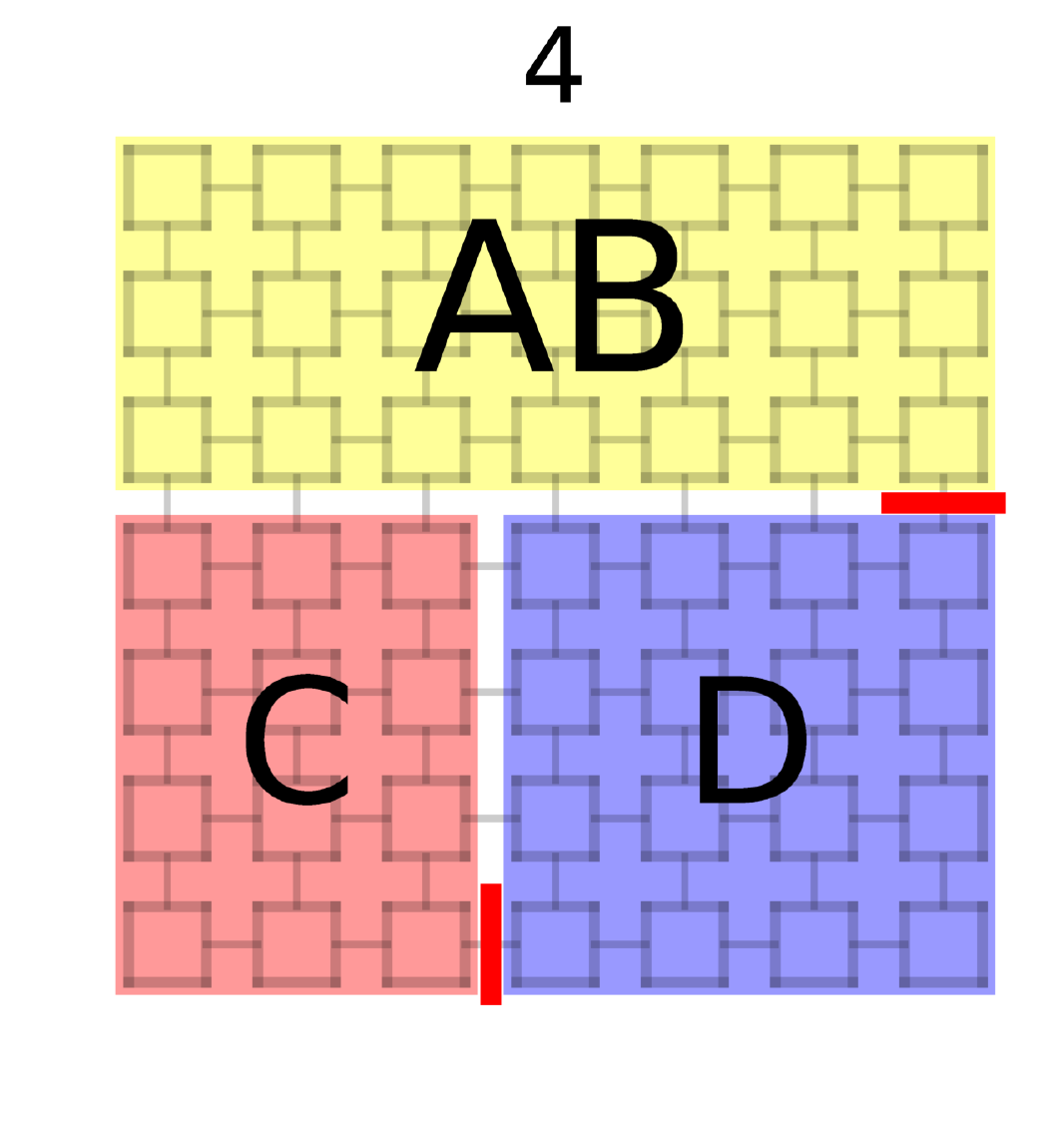}
\includegraphics[width=0.32\columnwidth]{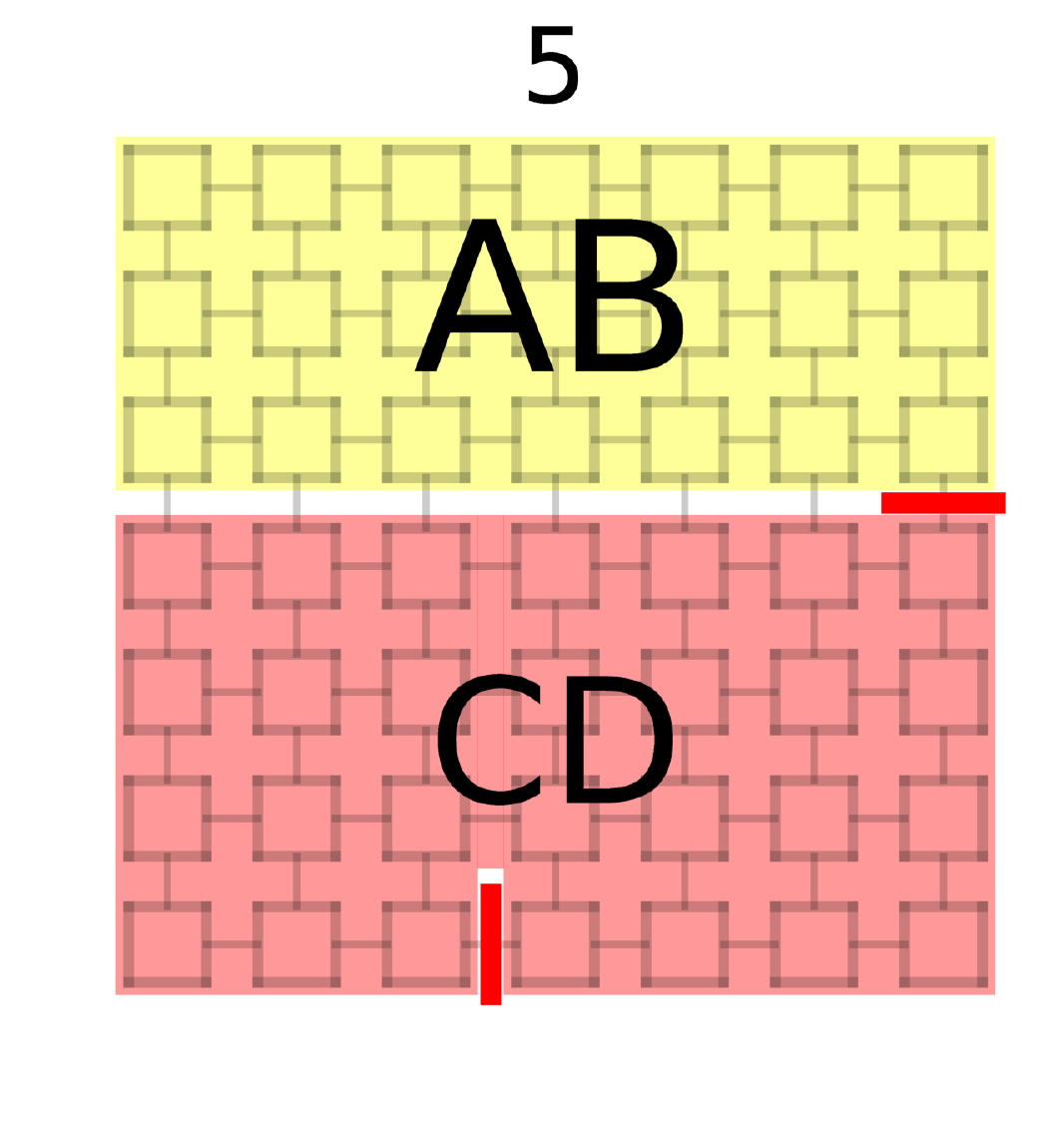}
\includegraphics[width=0.32\columnwidth]{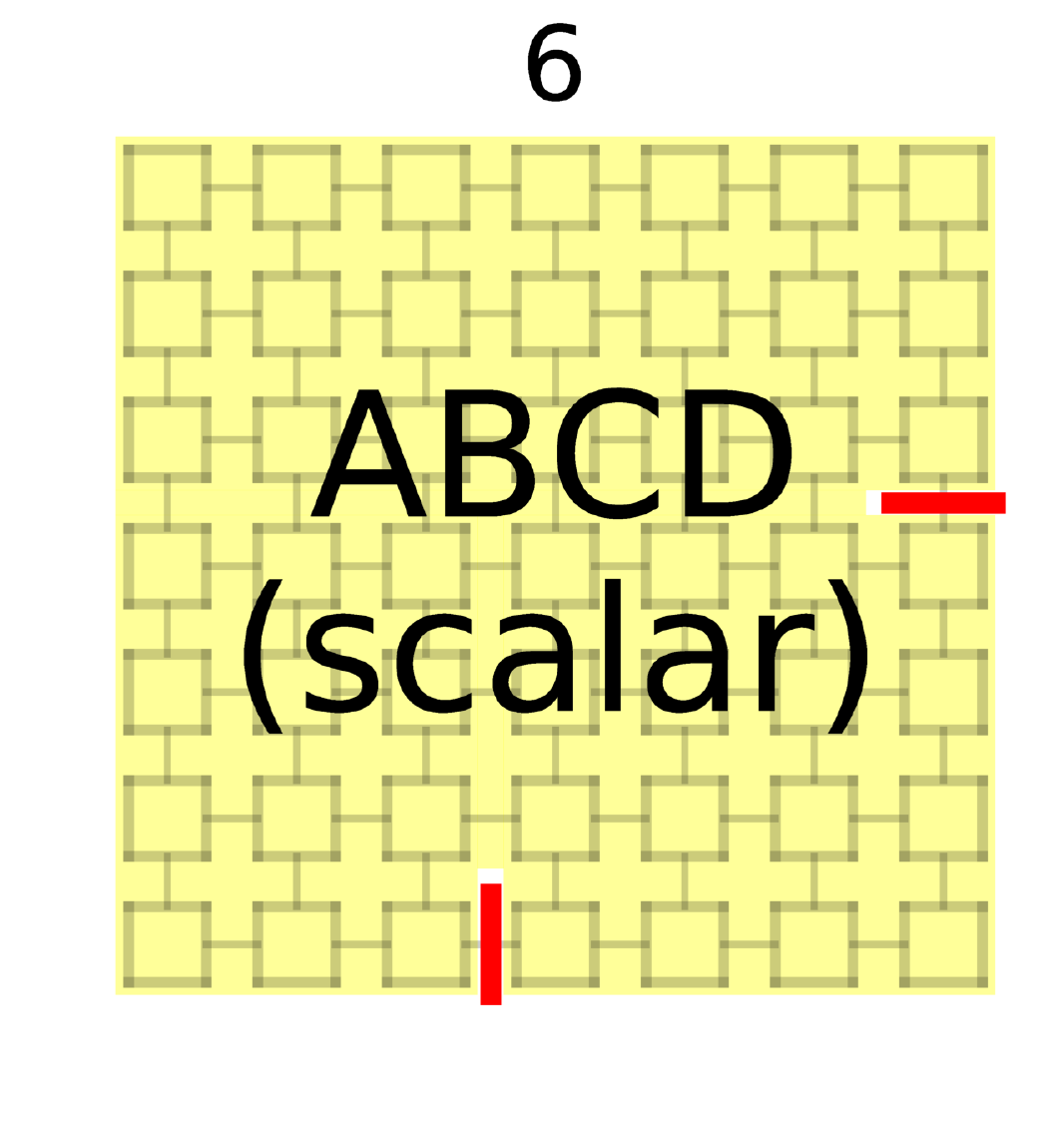}
\caption{\label{fig:7x7_contraction} Sketch of the contraction of the $7\times 7
\times (1+40+1)$ tensor network with two cuts.  The names of the tensors at key
steps shown in the contraction are referred to in Section~\ref{sec:grids}.}
\end{figure}

Here we describe the contraction scheme for $I\times J$ grids. To simplify the
discussion, we consider $7\times7\times(1+40+1)$ RQCs. Nonetheless, the same
procedure applies to other rectangular circuits. For the
$7\times7\times(1+40+1)$ RQCs, we use two cuts ($2^{2\times 5}$ paths) in the
grid (see Fig.~\ref{fig:7x7_contraction}), in order to divide it into four
tensors, $A$, $B$, $C$, and $D$, of dimension $2^{6\times
5}=2^{30}$ each. Then, the contraction proceeds as follows.  
\begin{enumerate}[label={\arabic*)}]
  \item Tensors in region $A$ are contracted onto tensor $A$, which is
        path independent; we do the same for tensors in regions $pB$
        (partial-$B$), $pC$ (partial-$C$), and $ppD$ (partial-partial-$D$),
        which are all path independent; for this reason, all $A$, $pB$, $pC$,
        and $ppD$ are reused over the entire computation.  We now iterate over
        all paths with two nested iterations: the outer one iterates over the
        right cut, while the inner one iterates over the bottom cut.
  \item For each of the $2^5$ right paths, tensors in regions $B$ and $pD$ are
        contracted.
  \item Tensors $A$ and $B$ are contracted onto $AB$; this tensor will be reused
        over all the inner loop iterations.
  \item For each bottom path (given a right path), the tensors on region $C$ and
        $D$ are contracted.
  \item $C$ and $D$ are contracted onto $CD$.
  \item $AB$ and $CD$ are contracted onto a scalar, which is equal to this
        path's contribution to the amplitude.
\end{enumerate}
From here, the iteration over the inner loop takes us back to step 4), for a
different bottom path. After this loop is exhausted, we go back to step 2), for
the next right path. After iterating over both loops, we have obtained all path
contributions.\\

Note that there is a large amount of potential reuse of the tensors built at
each step.  However, taking advantage of already built tensors requires a large
amount of memory in order to store all the tensors to be reused.  While there is
a clear trade-off between tensor reuse and memory usage, in practice we always found
the reuse profitable.\\

\begin{figure}[t]
\includegraphics[width=0.24\columnwidth]{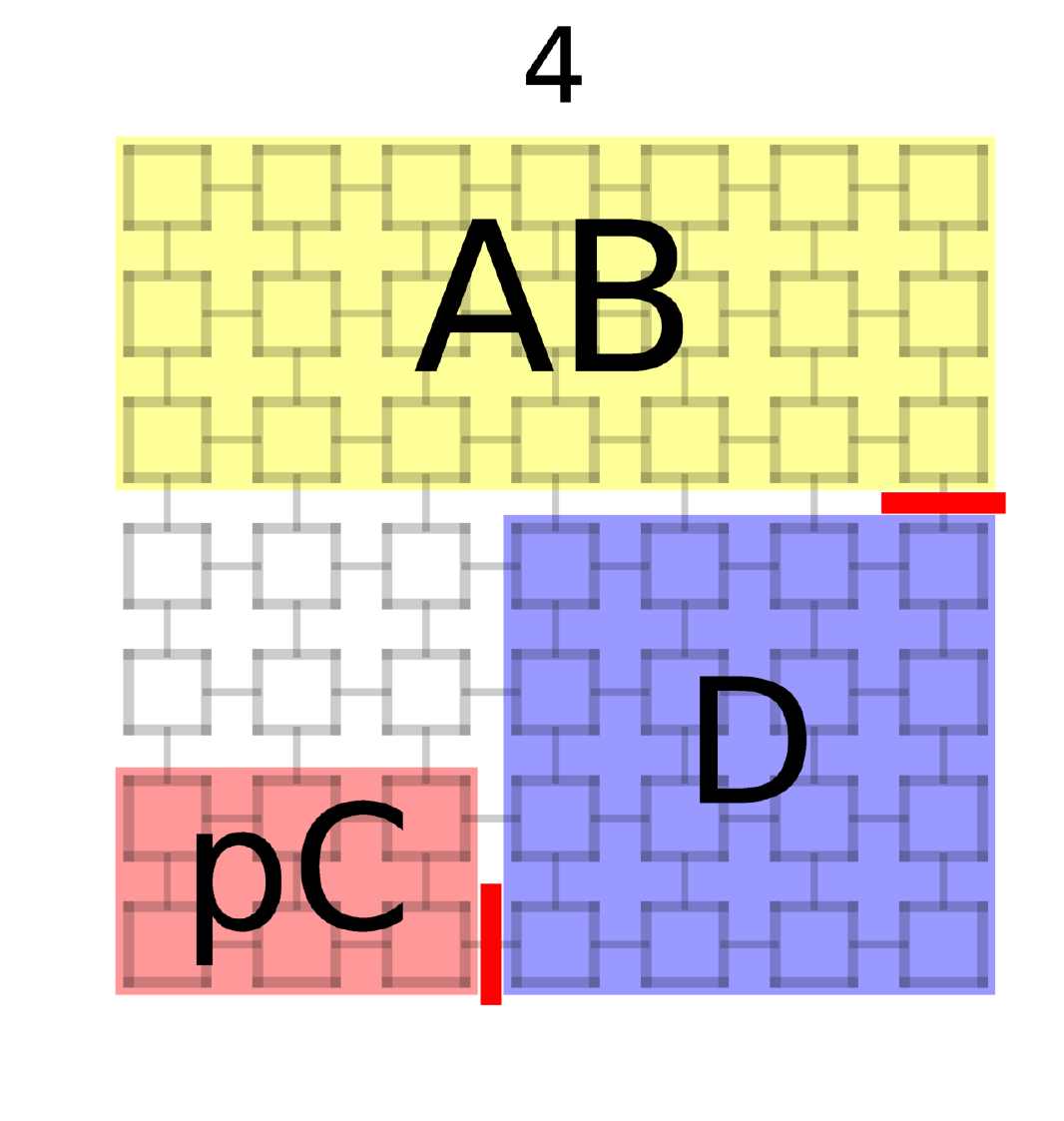}
\includegraphics[width=0.24\columnwidth]{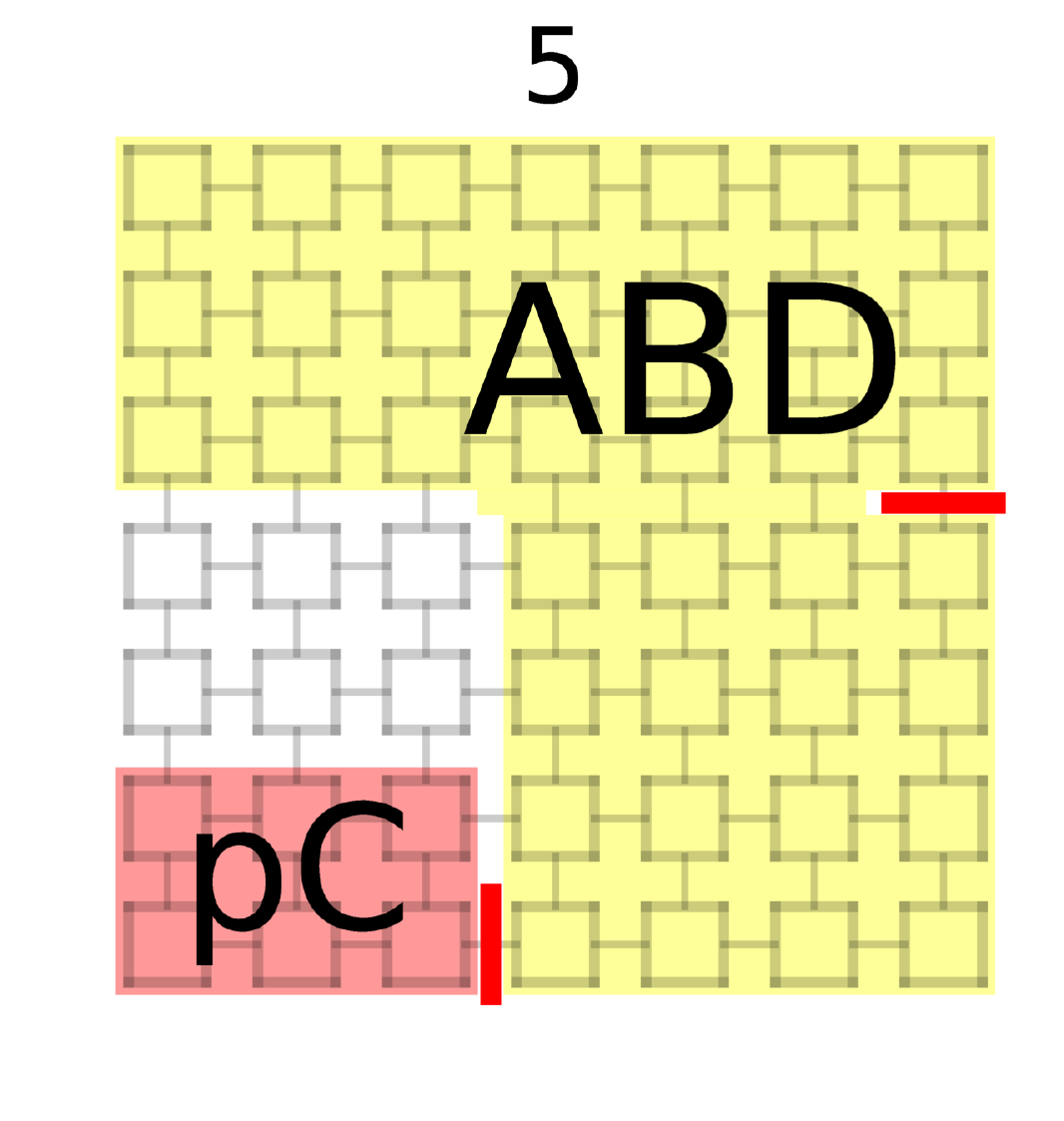}
\includegraphics[width=0.24\columnwidth]{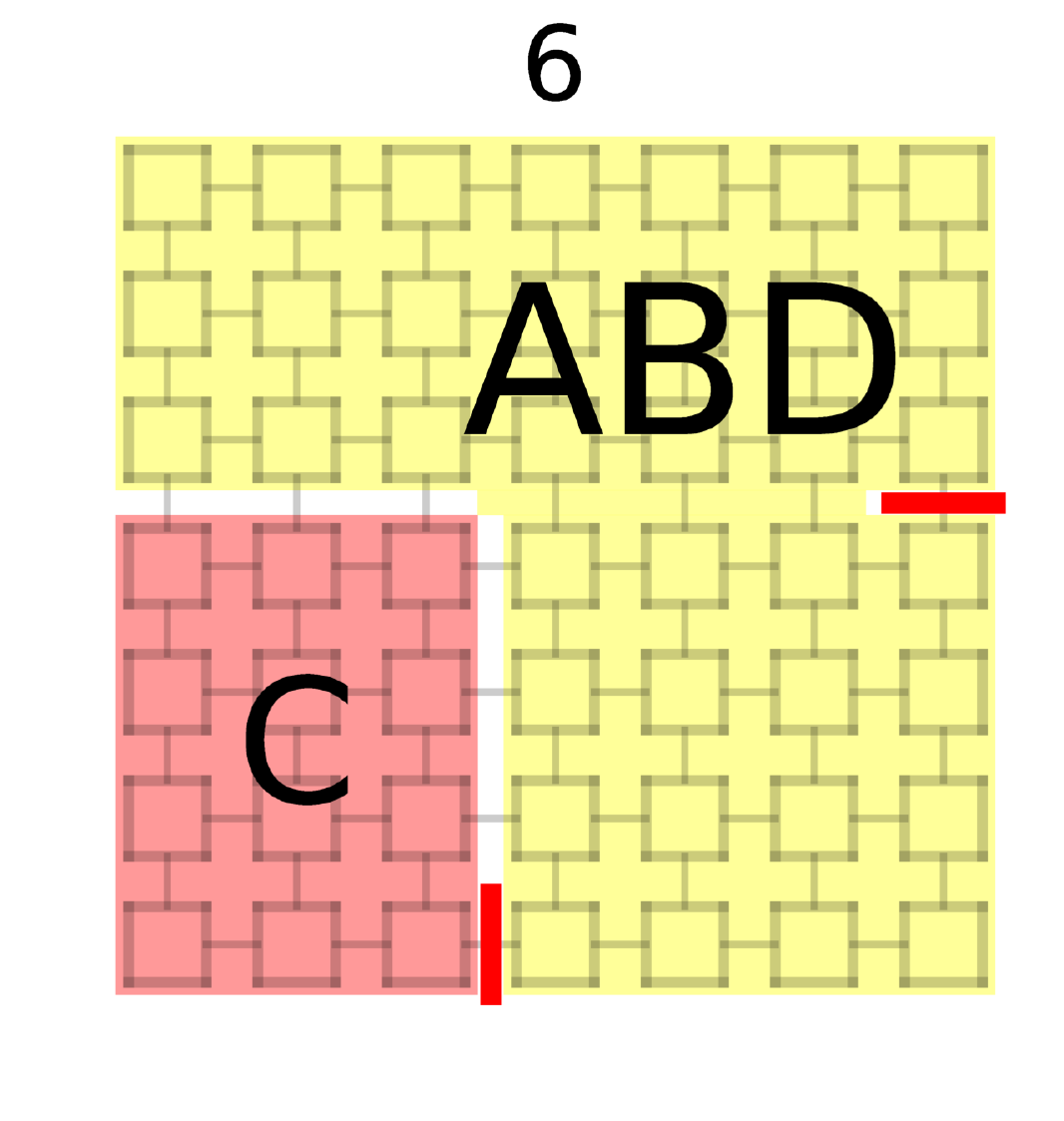}
\includegraphics[width=0.24\columnwidth]{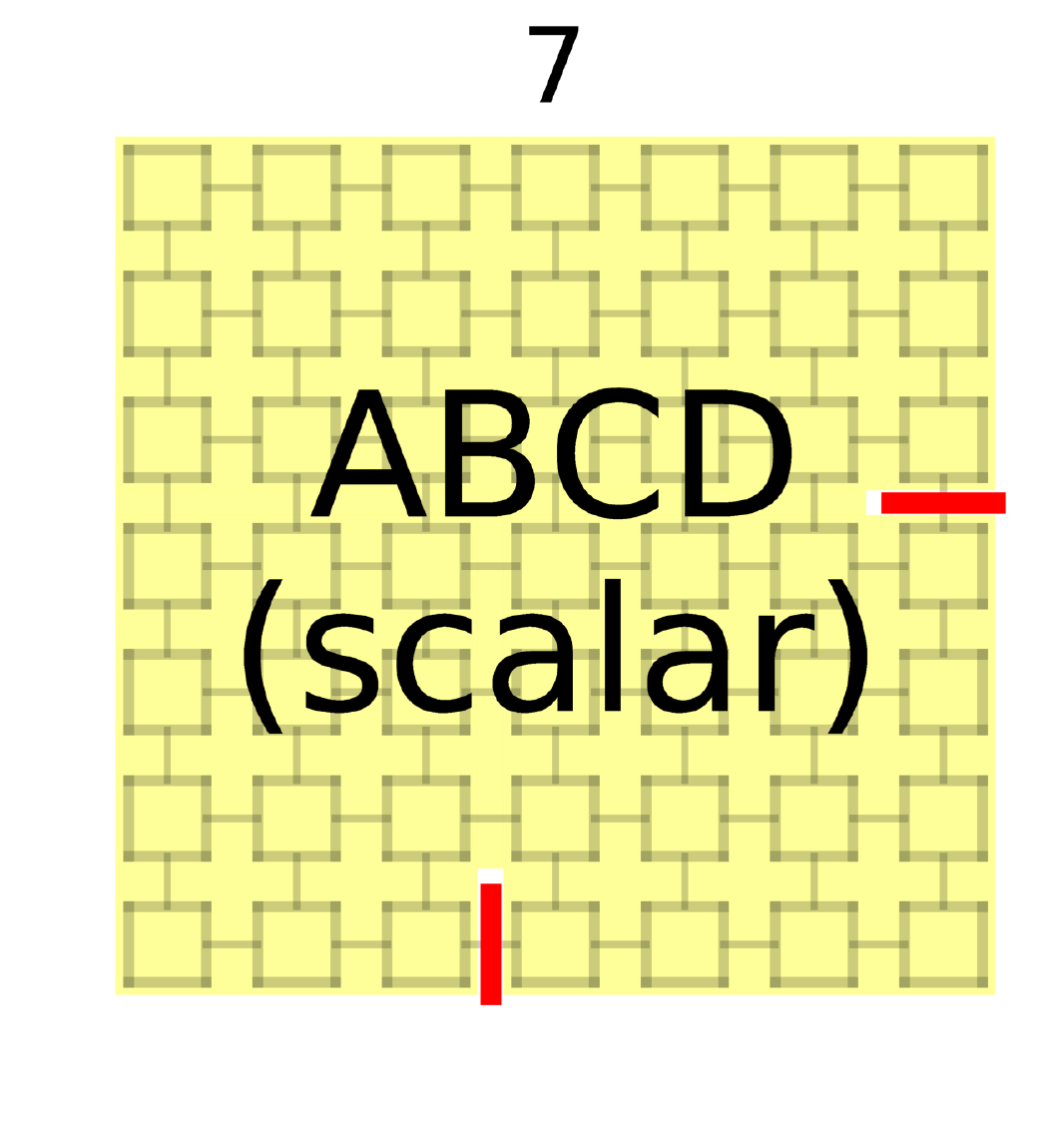}
\caption{\label{fig:7x7_contraction_alternative} Alternative contraction for the
use of fast sampling (see Methods~C.2).}
\end{figure}

Finally, the fast sampling introduced in Methods~C.2 can be also
applied here, by using a slightly different contraction than the one presented
in Fig.~\ref{fig:7x7_contraction}. More precisely, in
Fig.~\ref{fig:7x7_contraction_alternative} we present the final steps of the
alternative contraction. In 4) and 5) the size of the tensor $pC$ is still
small, and we focus, for this particular path, in contracting first $D$ with
$AB$ onto $ABD$.  This leaves six qubits free (above $pC$) for the computation
of batches of as much as $2^6=64$ amplitudes (or more if $pC$ is shrunk
further); for each amplitude, contracting the tensors in region $C$ (where $pC$
is reused) and computing the scalar $ABCD$ is left, \emph{i.e.}, steps 6) and 7)
of Fig.~\ref{fig:7x7_contraction_alternative}.  Once this is done, then we go
back to step 5), in order to loop over all bit-strings in the batch.  After
exhausting this loop, we go back to step 4) to compute the next bottom path.
Note that the contraction following this procedure is dominated by the
contraction of $AB$ with $D$; however, the tensor $ABD$ is reused (for a path)
for all bit-strings in a batch, and so the computation of a batch of amplitudes
adds only a small cost to the computation of a single amplitude.

\section{Old tensor contractions for Bristlecone-48 and Bristlecone-70}
\label{sec:old_contractions}

\begin{figure}[t]
\includegraphics[width=0.49\columnwidth]{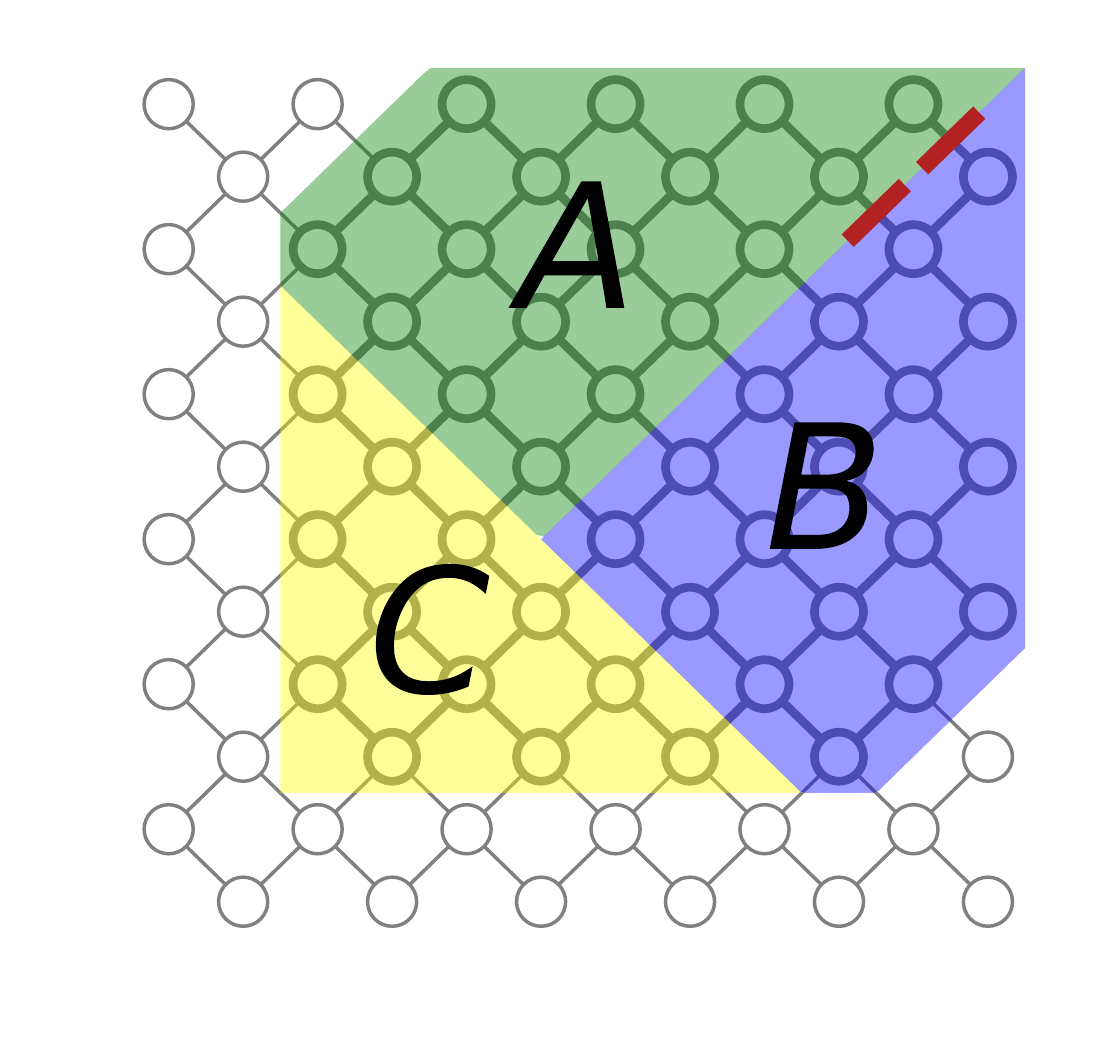}
\includegraphics[width=0.49\columnwidth]{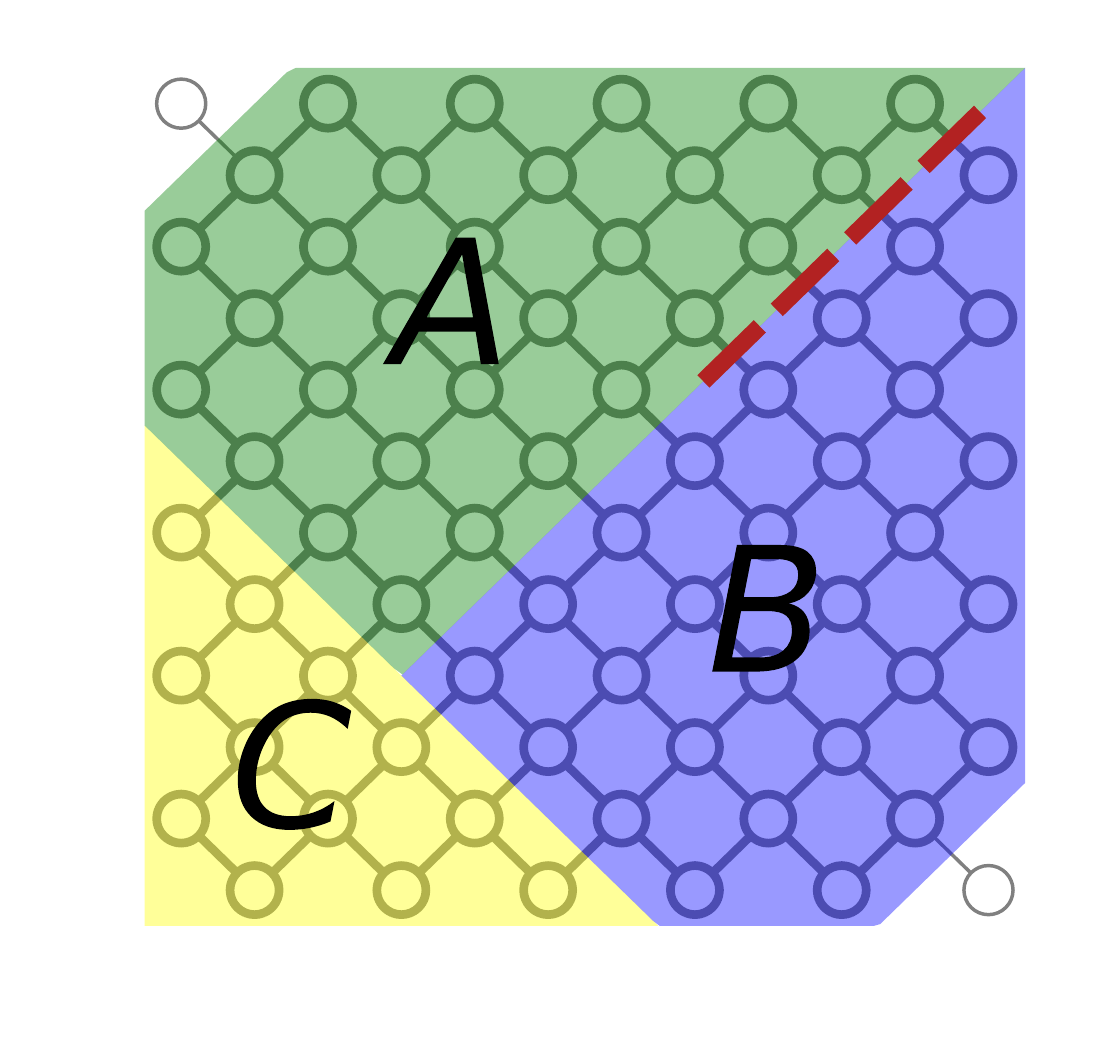}
\caption{\label{fig:old_contractions} Tensors $A$, $B$, and $C$ used in the old
contraction schemes used for \rsim2 and \rsim3 (Bristlecone-48 and -70,
respectively).}
\end{figure}

The contractions procedures followed for \rsim{2-3} are presented in
Fig.~\ref{fig:old_contractions}.  Note that we have identified a faster
contraction (about half the runtime) with less memory requirements for
Bristlecone-48 and -70, which is described in Methods~B.1, and
whose runtimes are reported in the main text (see Results) and referred to as
\rsim{2b} and \rsim{3b}, respectively.  Note also that the new contraction
scheme cannot be adapted to Bristlecone-60, due to its topology. The old
contractions are included here for reference, and apply to our released
simulation data.

\section{Qubit complexity of square grids and Bristlecone sub-lattices for Schro\"dinger-Feynman-type simulators}
\label{sec:qubit_complexity}

To study the time complexity of Schr\"odinger-Feynman-type
simulators~\cite{aaronson2017complexity,chen_64-qubit_2018,markov_quantum_2018},
\emph{i.e.}, those that partition the circuit in sub-circuits and perform a full
wave-function evolution of the sub-circuits for all the paths defined by the
partition, we use the ``qubit complexity''~\cite{chen_64-qubit_2018} for a given depth defined as follows.

\begin{figure}[t]
\includegraphics[width=0.32\columnwidth]{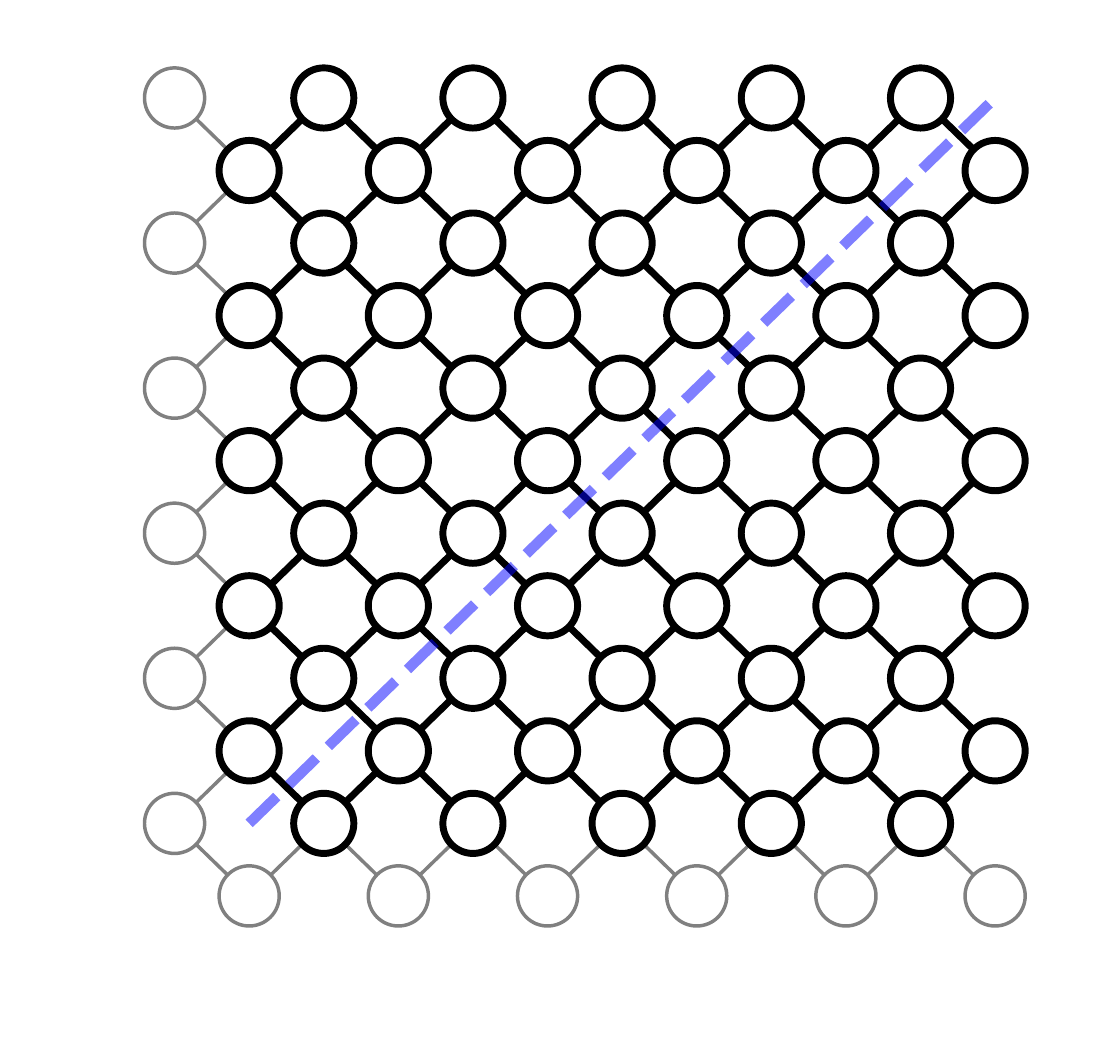}
\includegraphics[width=0.32\columnwidth]{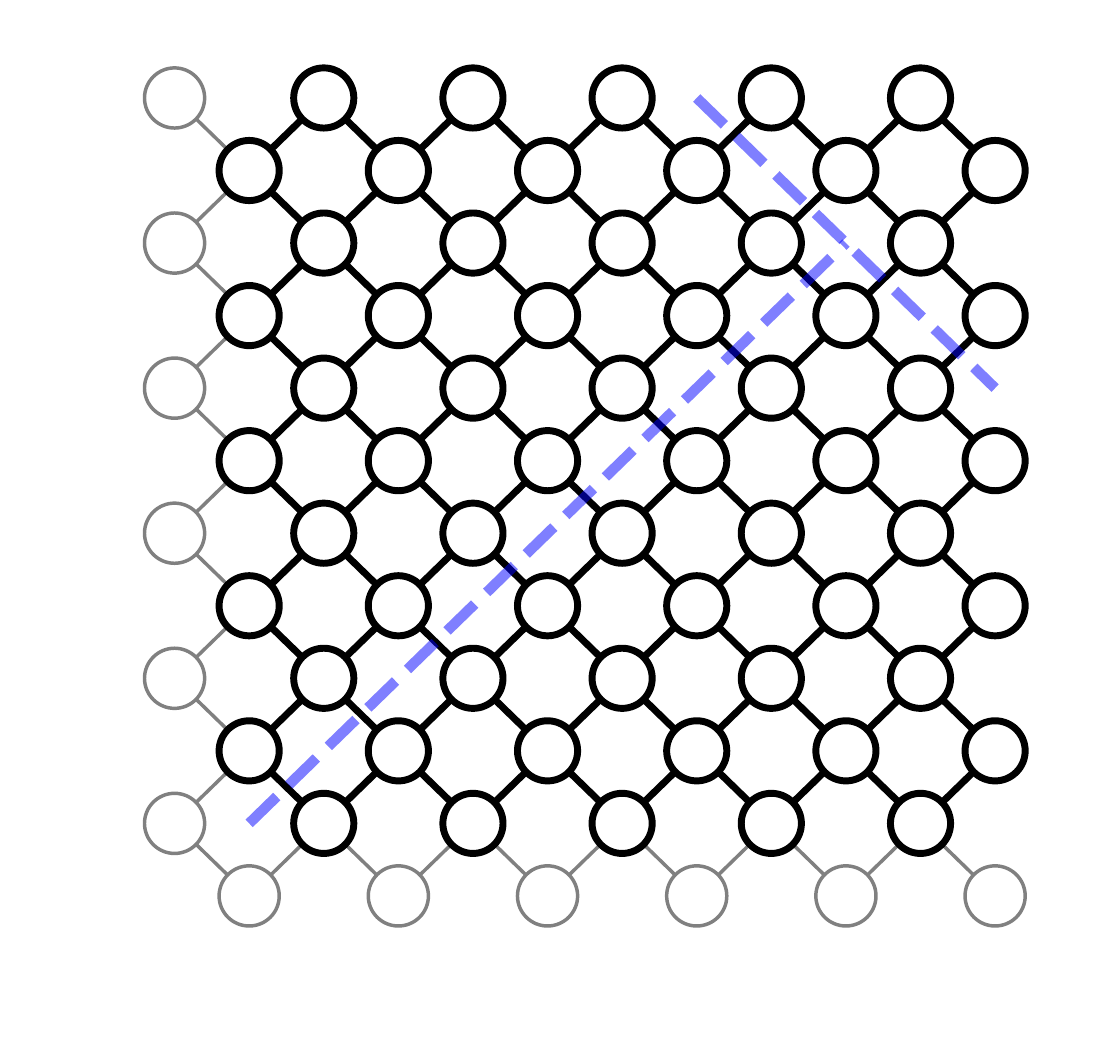}
\includegraphics[width=0.32\columnwidth]{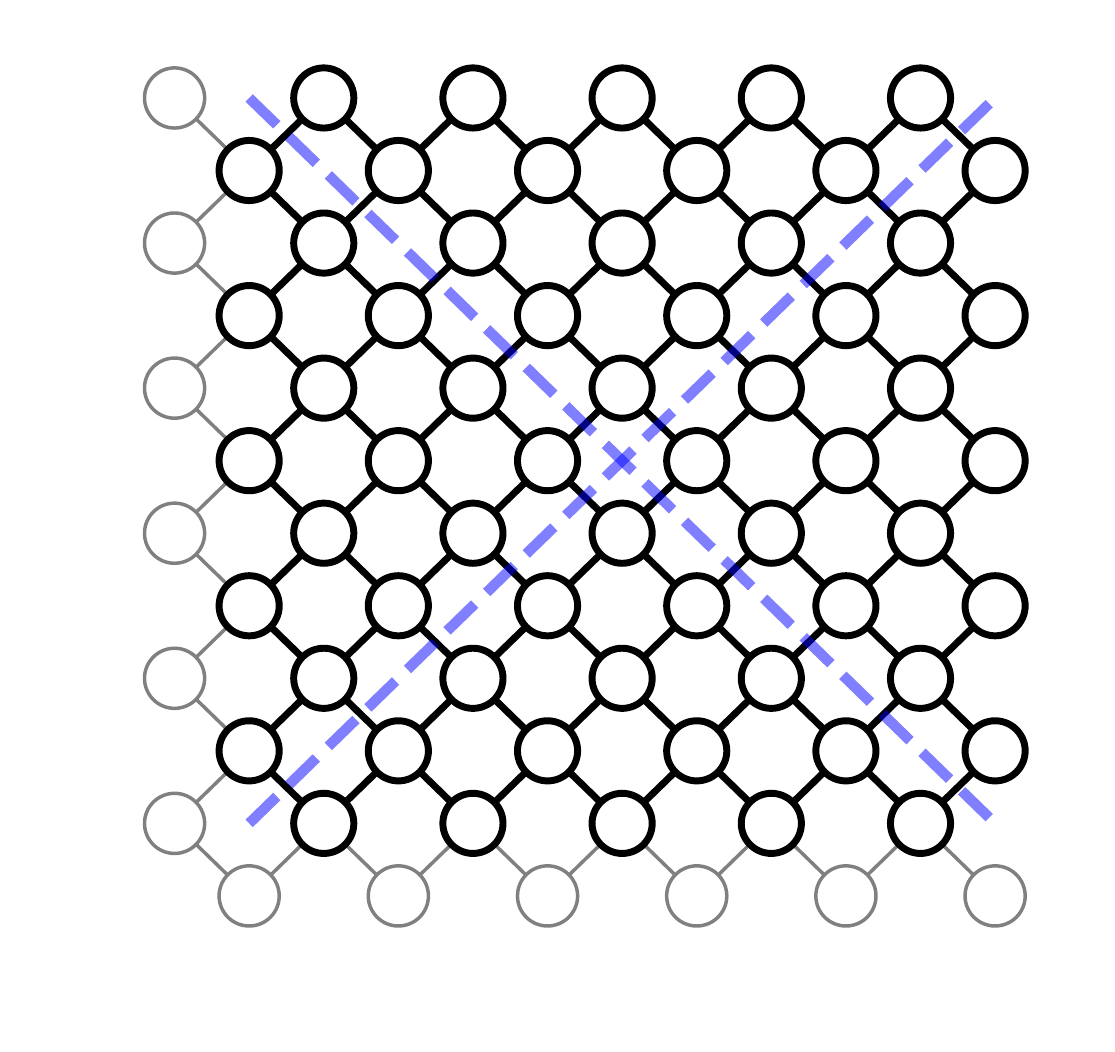}
\includegraphics[width=1.00\columnwidth]{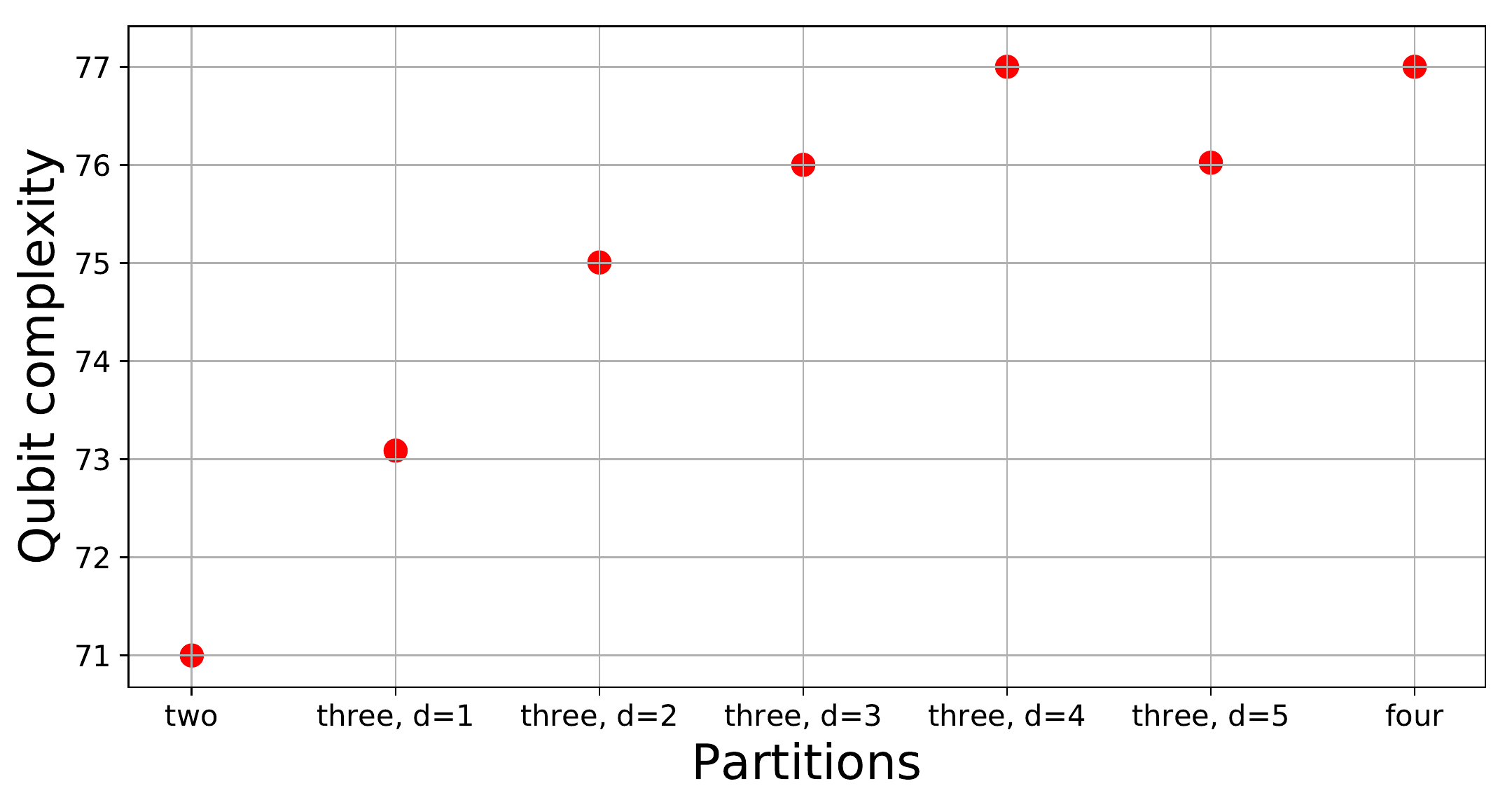}
\caption{\label{fig:qubit_complexity} Qubit complexity of the simulation of
Bristlecone-60 with depth (1+32+1) with Schr\"odinger-Feynman
simulators~\cite{aaronson2017complexity,chen_64-qubit_2018,markov_quantum_2018}
for different partition schemes.  \emph{Top:} From left to right, partition
schemes considered for two partitions, three partitions, and four partitions; in
the three partition case, $d$ is the number of cuts avoided in the diagonal
towards the top-right region, which for the example plotted is $d=2$, and is
extended to $d=1\ldots 5$ in our study.  \emph{Bottom:} Qubit complexity for the
two-partition, three-partition, and four-partition schemes; we can see that the
complexity is best in the two-partition case.}
\end{figure}

For a bipartition, sub-circuits $A$ and $B$ are generated; if sub-circuit $A$
has $n_A$ qubits, $B$ has $n_B$ qubits, and there are $\alpha_{AB}$ ${\rm CZ}$
gates cut, then sub-circuits $A$ and $B$ have to be simulated $2^{\alpha_{AB}}$
times, with complexity $2^{n_A}$ and $2^{n_B}$ each, respectively, for the
computation of one amplitude, or batch of amplitudes at a small overhead cost.
The qubit complexity is in this case $\log_2 \left[ 2^{\alpha_{AB}}
(2^{n_A}+2^{n_B}) \right]$.

For a partition in three regions, sub-circuits $A$, $B$, and $C$ are generated,
with $n_A$, $n_B$, and $n_C$ qubits, respectively.  The number of ${\rm CZ}$
gates cut is now $\alpha_{AB}$ between $A$ and $B$, $\alpha_{AC}$ between $A$
and $C$, and $\alpha_{BC}$ between $B$ and $C$.  In a naive computation, the
qubit complexity would be $\log_2 \left[ \left(2^{\alpha_{AB}} 2^{\alpha_{AC}}
2^{\alpha_{BC}} \right) (2^{n_A}+2^{n_B}+2^{n_C}) \right]$, since each of the
three sub-circuits has to be simulated for each of the $2^{\alpha_{AB}}
2^{\alpha_{AC}} 2^{\alpha_{BC}}$ paths.  However, it is possible to decrease the
complexity by simulating sub-circuit $A$ only once for each particular choice of
path over cuts between $A$ and $B$ and cuts between $A$ and $C$, and iterate
(within this choice) over all paths over cuts between $B$ and $C$.  In that
case, the cost is $\log_2 \left\{ \left(2^{\alpha_{AB}+\alpha_{AC}} \right)
\left[ 2^{n_A} +  2^{\alpha_{BC}} (2^{n_B}+2^{n_C}) \right] \right\}$. The qubit
complexity is in this case minimized when $A$ is the biggest sub-circuit, which
can always be assumed without loss of generality.

For a partition in four sub-circuits ($A$, $B$, $C$, and $D$), where
sub-circuits share ${\rm CZ}$ gates only between $A$ and $B$, $B$ and $C$, $C$
and $D$, and between $D$ and $A$, as in Fig.~\ref{fig:qubit_complexity}, the
naive computation has qubit complexity $\log_2 \left[
\left(2^{\alpha_{AB}+\alpha_{BC}+\alpha_{CD}+\alpha_{AD}} \right) \beta_{ABCD}
\right]$, where $\beta_{ABCD} \equiv 2^{n_A}+2^{n_B}+2^{n_C}+2^{n_D}$.  However,
it is possible, for each combination of paths between $A$ and $B$, and $C$ and
$D$, to iterate over paths between $B$ and $C$, and those between $A$ and $D$,
lowering the qubit complexity to $\log_2 \left[ 2^{\alpha_{AB}+\alpha_{CD}}
\left( 2^{\alpha_{BC}} \beta_{BC} + 2^{\alpha_{AD}} \beta_{AD} \right) \right]$,
where $\beta_{BC} \equiv 2^{n_B}+2^{n_C}$ and $\beta_{AD} \equiv
2^{n_A}+2^{n_D}$.\\

We can see in Fig.~\ref{fig:qubit_complexity} that for Bristlecone-60 with depth
(1+32+1) the best performance is achieved for a partition into two sub-circuits,
as is the case for the rectangular grids considered in
Refs.~\cite{chen_64-qubit_2018,markov_quantum_2018}.  For a square grid $8\times
8\times (1+32+1)$, the qubit complexity is 65, which is lower than the best
complexity found in this section for Bristlecone-60 with depth (1+32+1), \emph{i.e.}, 71, even
though the $8\times 8$ square grid has four more qubits. This suggests that hard
Bristlecone sub-lattices are harder to simulate than square (or rectangular)
grids of the same (or smaller) number of qubits. Similar arguments apply to
Bristlecone-70.

For the square grid $7\times 7\times (1+40+1)$ split into two
sub-circuits~\cite{markov_quantum_2018}, the qubit complexity is slightly over
63, and for the $7\times 8\times (1+40+1)$, the qubit complexity is 64.

\section{Supplementary details on hardware usage and runtimes}
\label{sec:runtimes}

Here we discuss the split in runtimes as reported in the main text (see Results), as well as
the estimation of core-hours necessary for the simulations:

\textbf{Split in runtimes.} Investigation showed that the split in runtimes
discussed in the main text (see Results) appears to be due to differing
amounts of contention on the two sockets within a node (all the
Pleiades and Electra nodes are dual-socket).  We enforced the rule
that all the jobs running on a particular node had the same number of
threads.  This could lead to there being a few unused cores.
Individual threads were ``pinned'' to run on individual cores to avoid
interference.  However, the pinning strategy caused the unused cores
to always be the lowest numbered cores (which are all on the first
socket), and so fewer jobs ran concurrently on the first socket,
causing them to see less contention, and therefore slightly higher 
performance than jobs that ran on the second socket.  Unfavorable
thread counts and core counts could also lead to one job per node
having it's threads split across the two sockets; this creates yet
another source of anomalous timings.

\textbf{Core hour estimation.} In our simulations on the Skylake nodes of
Electra we used \mbox{$P=2304\times40=92160$} cores. Note
that we consider 40 cores per node, even though we use only 39 in practice for
the $7\times 7\times (1+40+1)$ simulations and 36 in the
$8\times 8\times (1+40+1)$ simulations; this is due to the ability of modern
Intel processors to ``up-clock'' their CPUs in favorable conditions (known as
Dynamic Frequency Scaling), thus achieving a performance similar to the case
where there are no idle cores. Similar estimates apply to the benchark of the
other types of nodes used. Regarding our comparisons with
Alibaba~\cite{chen_classical_2018}, we take into account that the authors
report the usage of \mbox{$P=2^{17}=131072$} for their benchmark.

\bibliographystyle{apsrev4-1}
\bibliography{refs}

\end{document}